\documentclass[preprint]{aastex7}

\usepackage{hyperref} 
\usepackage{graphicx} 
\usepackage{subfig} 
\usepackage{multirow}

\begin{document}

\title{On the contradictory case of the binary system HD~81809 hosting two pulsating solar-like stars observed by TESS}

\author[orcid=0000-0001-7801-7484, sname='Di Mauro']{Maria Pia Di Mauro}
\affiliation{INAF-IAPS, Via del Fosso del Cavaliere 100, Roma, Italy}
\email[show]{maria.dimauro@inaf.it}

\author[orcid=0009-0009-6066-0194, sname='Pezzotti']{Camilla Pezzotti}
\affiliation{STAR Institute, Universit\`e de Li\`ege, Li\`ege, Belgium}
\email{camilla.pezzotti@uliege.be}

\author[orcid=0000-0002-2087-6427, sname='Moedas']{Nuno Moedas}
\affiliation{INAF-IAPS, Via del Fosso del Cavaliere 100, Roma, Italy}

\email{nuno.martinsmoedas@inaf.it}

\author[orcid=0000-0003-4337-8612, sname='Catanzaro']{Giovanni Catanzaro}
\affiliation{INAF-Osservatorio Astrofisico di Catania, via S. Sofia 81, Catania, Italy }
\email{giovanni.catanzaro@inaf.it}

\author[orcid=0000-0003-3794-1317, sname='Maxted']{Pierre F. L. Maxted}
\affiliation{Astrophysics Group, Keele University, Staffordshire, ST5 5BG, UK}
\email{p.maxted@keele.ac.uk}

\author[orcid=0000-0001-8835-2075, sname='Corsaro']{Enrico Corsaro}
\affiliation{INAF-Osservatorio Astrofisico di Catania, via S. Sofia 81, Catania, Italy }
\email{}

\author[orcid=0000-0001-8623-5318, sname='Reda']{Raffaele Reda}
\affiliation{Dipartimento di Fisica, Università di Roma Tor Vergata, Roma, Italy} 
\email{raffaele.reda@roma2.infn.it}

\author[orcid=0009-0000-8017-7962, sname='Scuflaire']{Richard Scuflaire}
\affiliation{STAR Institute, Universit\`e de Li\`ege, Li\`ege, Belgium}
\email{r.scuflaire@uliege.be}

\author[orcid=0000-0003-3175-9776, sname='Bonanno']{Alfio Bonanno}
\affiliation{INAF-Osservatorio Astrofisico di Catania, via S. Sofia 81, Catania, Italy }
\email{alfio.bonanno@inaf.it}

\author[orcid=0000-0001-7369-8516, sname='Giovannelli']{Luca Giovannelli}
\affiliation{Dipartimento di Fisica, Università di Roma Tor Vergata, Roma, Italy}
\email{luca.giovannelli@roma2.infn.it}

\author[orcid=0000-0003-4745-2242, sname='Beck']{Paul G. Beck}
\affiliation{ Universidad de La Laguna, Santa Cruz de Tenerife, Santa Cruz de Tenerife, Spain}
\email{paul.beck@iac.es}

\begin{abstract}

We present a new comprehensive study of HD~81809, a nearby binary system 
of two solar-like stars showing high-amplitude X-ray emission and a well-defined 8-year solar-like magnetic cycle.

By analyzing high-resolution spectroscopy, alongside DR3 Gaia astrometry, and bolometric fluxes, we derive updated fundamental parameters for both components.\\
 In particular, we uncover a significant
chemical difference: the primary is metal-poor ([Fe/H]$ \simeq - 0.57$), while the secondary shows solar-like metallicity ([Fe/H]$=0.00$). 
This suggests that the system originated in a mildly metal-poor environment, consistent with the Galactic thick disk population, and that the secondary's surface composition has been altered by a recent accretion event.
 
Using multi-sector TESS photometry,
we detected solar-like oscillations in both components,
deriving global asteroseismic parameters 
$\Delta\nu = 43.32 \pm 3.91~\mu$Hz, $\nu_{\rm max} = 708.74^{+3.23}_{-3.74}~\mu$Hz 
 for HD81809~A, and $\Delta\nu = 97.75 \pm 4.49~\mu$Hz, $\nu_{\rm max} = 2098.07^{+3.07}_{-2.83}~\mu$Hz for HD81809~B.

By combining all the observational constraints
with stellar evolutionary models computed using  CLES and MESA codes, we reconstructed 
the evolutionary scenario of the system.

Our results indicate  that HD~81809 is an old system with an age of $\sim 10\, \mathrm{Gyr}$, composed of a subgiant primary with mass $\sim 0.87M_{\odot}$ and radius $\sim1.96R_{\odot}$ — likely responsible for the reactivated dynamo cycle  — and a main sequence secondary with mass $ M=0.85M_{\odot}$ and radius $R=1.10R_{\odot}$.

This system represents a benchmark for studying stellar evolution, magnetic activity, and the physics of old, metal-poor stars in the Galactic thick disk.

%Thedefinitely appears to be
%belong to the thick disk.  

\end{abstract}

%\keywords{Asteroseismology, stellar modeling, stars, binary stars, magnetic activity, solar-like stars}


\section{Introduction} 

\label{sec:intro}

The system HD~81809 (HIP 46404, TIC 46802551, HR 3750) is a close 
%($\simeq 30.96\, pc$) 
visual binary characterized by a maximum separation $\rho=0.487 $ arcsec, an orbit with a semi-major axis of $\alpha=(0.428\pm 0.001) $ arcsec, an orbital period of $P=34.8\pm0.06$ years and an inclination $i_{ orb} = 85^{\circ}.4\pm 0^{\circ}$, as obtained from SOAR speckle photometry by \citet{Tokovinin}. The distance  according to the \citet{BailerJones2021}
catalog
is  $d=(30.907\pm 0.335)$ pc, as estimated from  the Gaia DR3 parallax $\pi = (32.29 \pm 0.36)$ mas.

The two components show an apparent magnitude of $V_A=5.56\pm0.01$ and $V_B=7.45\pm0.01$, respectively \citep{Egeland2018}.
%The large physical separation of the two stars makes this a non-interacting system.
The primary HD~81809A (HR~3750, TIC~46802551) is a G1.5 solar-type star
with $T_{\rm eff, A}= (5757\pm 57)$ K from two-color photometry  \citep{Egeland2018},
and $T_{\rm eff, A}= (5620\pm 80)$ K from high-resolution spectroscopy \citep{Fuhrmann2018}. 
Its rotational period is $P_{rot, A}=(40.2\pm2.3)$~d as derived by \citet{Donahue1993, Donahue1996, Egeland2018}. The secondary seems to be a G type star too,
with $T_{\rm eff, B}= (5705\pm 73)$ K as obtained from two-color photometry  \citep{Egeland2018}, quite in agreement with
 $T_{\rm eff,B}= (5730\pm 100)$ K deduced from high-resolution spectroscopy by \citet{Fuhrmann2018}. 
%\textcolor{blue}{CAMI: maybe I would remove this line on the rotational period of the secondary, or specify that the value does not come from direct observational data, but rather from indirect analysis of activity index of the system} The rotational period is $P_{rot, B}\simeq28~d$ \citep{Donahue1993, Donahue1996, Egeland2018}. 

%Furthermore, the galactic space velocities with respect the Sun have been measured by \citet{Soubiran2005} as respectively $U=	-42.9 	$\,km/s , 
%$V=	-46.8$\, 	km/s,    
%$W=	-2.0 $ \, 	km/s.

Binary stars in which one or both components can be detected asteroseismically provide powerful constraints on our understanding
of stellar physics.
The HD~81809 system appears particularly interesting, because
it has been observed with different instruments at different epochs, which allowed to collect several information.
Among these, it should be mentioned  a well-defined magnetic activity cycle similar to that of the Sun with a period $\simeq8.2$ years \citep{Egeland2018} detected through the regular modulation of chromospheric indicators. 
%such as the Ca ii H\&K index, with measurements starting from 1975, which allowed to reveal a periodical magnetic variability similar to that of the Sun.

The system is also characterized by high amplitude and cyclical emission of X-ray flux, monitored by the XMM-Newton program, which seems to originate from the more massive of the two components \citep{Favata2004, Favata2008, Orlando2017}. 
%Although \citep{Radick2018} did not find this hypothesis so convincing,
%pointing out that solar-like younger, fast-rotating main-sequence stars also might show X-ray luminosities well exceeding those of the Sun \cite[e.g.,][]{Wright2011},
%with a similar surface area. However, 
% noticing that
%the Ca ii H\&K and X-ray emission vary in phase, it is reasonable to assume that the primary 
%with a $\simeq$ 40-day rotational
%period 
%is the active component.

%This is certainly not the rotation period of a young star, and therefore 
%Hence, it is reasonable to assume, as supposed by \citet{Favata2004} and confirmed by %\citet{Orlando2017},  that the primary is the active component.

Even though this object has been largely observed, the question of the characterization of the two components is still quite debated and the estimate of the age and masses appears to be quite crucial.
In particular considering that the age of a star is one of the most challenging parameter to determine because it  is usually inferred through indirect methods \citep{Lebreton2014}.
%isochrone fittings when atmosphere parameters are known \citep{Soderblom2010}.
In this regard, by using stellar evolutionary models, \cite{Fuhrmann2018} estimated for the primary component a value of age
$\rm \tau = 3.2$ Gyr for a mass $M_A = (1.39 \pm 0.09) M_{\odot }$ in contradiction with the long rotational period resulted by the action of magnetic braking and the estimated low spectroscopic iron-to-magnesium abundance $\rm{[Fe/Mg]}=-0.35\pm0.05$, which classifies
this target as a Population II star (thick disk). 
In order to reconcile this contradictory scenario,  \citet{Fuhrmann2018} hypothesized that the young massive primary subgiant could be the result of a fossil  merger between an old progenitor
with mass $\simeq 1.0~M_{\odot}$ and a former tertiary component with  mass of $\simeq0.39~M_{\odot}$.

%Thanks to the successful photometric space mission CoRoT (Baglin et al., 2006), Kepler/K2 (Borucki et al., 2010) and TESS (Ricker et al., 2014),
%steroseismology has proven to be a formidable mean to characterize stars which, similarly to the Sun, are characterized by the presence of an external convective envelope able to excite normal modes of oscillations.
The purpose of this manuscript is to draw  new conclusions on the fundamental parameters of HD~81809, taking advantage of the photometric time series and asteroseismic measurements provided by the TESS \citep{Ricker} space mission, which, combined with the spectroscopic atmospheric results, allow us to define a new evolutionary scenario, confirmed by all the available observational data.

%\textcolor{blue}{CAMI: maybe it is possible to merge this part about the description of Section 2, with the part below ("This manuscript is organised as follow ...")}
The manuscript is organized as follows:
in Section~\ref{sec:fund_params} we will present the revised orbital parameters, masses, luminosities and the new analysis of HERMES high-resolution spectra to obtain a complete set of atmospheric parameters and the elements abundance for the two stars; in Section~\ref{sec:osc} the analysis of the TESS photometric light curve will be presented in order to extract the global oscillation parameters; in Section~\ref{sec:models} we will characterize the properties of the two stars of HD~81809 based on the spectroscopic atmospheric parameters and the
TESS asteroseismic data, supplemented by stellar masses and luminosities;
in Section \ref{sec:magnetic} we will present the analysis of the magnetic activity measurements to confirm the evolutionary stage;
in Section \ref{sec:spi} we will present the rotational evolutionary model of the primary component and compare with observed X-ray flux;
in Section \ref{sec:conclus} we will make our considerations and conclusion.

\section{Fundamental parameters of HD~81809}
\label{sec:fund_params}

\subsection{Radial velocity measurements}
\label{sec:rv}

\begin{figure}
\centering
\includegraphics[width=0.49\textwidth]{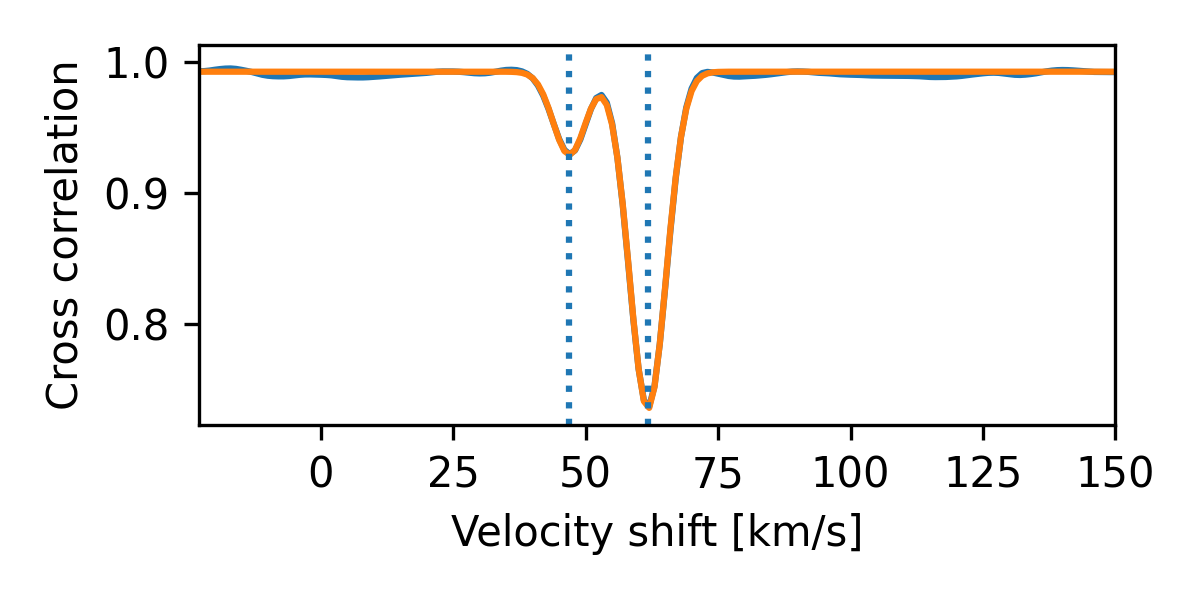}
\caption{The cross-correlation function for one HERMES spectrum of HD\,81809 with the double Gaussian fit used to measure the radial velocities of the two stars.}
\label{fig:ccf}
\end{figure}

We analyzed high-resolution spectroscopic observations of HD~81809 obtained with several instruments spanning more than two decades.
The largest and most precise contribution comes from 12 spectra acquired between 2012 January 10 and 2013 January 18 with the HERMES spectrograph on the MERCATOR 1.2-m telescope, retrieved from the Mercator library of stellar spectroscopy \citep{2024A&A...681A.107R}.
HERMES operates at a resolving power of $R \approx 85\,000$.
The cross-correlation functions (CCFs), computed using a numerical mask based on the solar spectrum, exhibit two well-separated peaks corresponding to the two stellar components (Figure~\ref{fig:ccf}).

Radial velocities were extracted by performing a simultaneous least-squares fit of two Gaussian profiles to each CCF.
The resulting measurements are listed in Table~\ref{tab:rv}.
The mean full width at half-maximum (FWHM) of the best-fit Gaussian profiles are $8.15 \pm 0.01$\,km/s for HD\,81809\,A and $7.26 \pm 0.09$\,km/s for  HD\,81809\,B.
%It is difficult to make an accurate estimate of $v\,\sin i$ from these FWHM measurements because of uncertainties in the intrinsic line width due to macro-turbulence, but the slightly larger width of the profile for HD\,81809\,A compared to  HD\,81809\,B is consistent with the expectation that it has a projected  rotational velocity $v_{\rm A}\,\sin i\approx 2.6$\,km/s based on its estimated radius and the rotational period from \citet{Egeland2018}.}

We complemented the HERMES dataset with several archival observations.
A spectrum obtained on 1996 December 01 with the HIRES spectrograph on the Keck~I telescope (nominal resolving power $R \approx 71\,700$) was retrieved from the Keck Observatory Archive.
The CCF for this spectrum appeared clearly asymmetric, so we also performed a least-squares fit of two Gaussian functions to measure the RVs of both stars with the FWHM of the two Gaussian profiles set equal to one another.
% A more complicated fit with the FWHM of B computed from the FWHM of star A and the FWHM measured from the HERMES spectra gives almost exactly the same result.

Two spectra of HD\,81809 observed on the night 1998 January 14 with the 1.93-m telescope of Observatoire de Haute Provence with the ELODIE spectrograph were retrieved from the ELODIE data archive. 
These spectra have a nominal resolving power of $R\approx 42\,000$. 
These spectra also produced asymmetric CCFs and so we fit them in the same way as we did for the HIRES spectra.

Finally, two spectra observed on 2017 May 09 and 2018 January 12 with the FEROS spectrograph mounted on the ESO/MPG 2.3-m telescope at La Silla \citep{kaufer99} were included.
These spectra, with resolving power $R \approx 48\,000$, display symmetric CCFs and were therefore fitted with a single Gaussian profile to derive radial velocities for the primary component only.
All radial velocity measurements and residuals from the best-fit orbit are reported in Table~\ref{tab:rv}.
\begin{table}
\centering
\caption{Radial velocity measurements for HD\,81809\,A and  HD\,81809\,B and residuals from the best-fit orbit.}
\begin{tabular}{rrrrrrl}
\hline
\multicolumn{1}{l}{MJD} &
\multicolumn{1}{l}{V$_{\rm r,A}$} &
\multicolumn{1}{l}{$({\rm O}-{\rm C})_{\rm A}$} &
\multicolumn{1}{l}{V$_{\rm r,B}$} &
\multicolumn{1}{l}{$({\rm O}-{\rm C})_{\rm B}$} &
\multicolumn{1}{l}{Phase} &
Source \\
&
\multicolumn{1}{l}{[km s$^{-1}$]} &
\multicolumn{1}{l}{[km s$^{-1}$]} &
\multicolumn{1}{l}{[km s$^{-1}$]} &
\multicolumn{1}{l}{[km s$^{-1}$]} &
 &
 \\
\hline
21615.0 &    54.65 &    +0.68 &         &          &  0.348 & SB9       \\ 
22338.4 &    51.50 &  $ -1.38$&         &          &  0.405 & SB9       \\ 
22631.0 &    52.67 &    +0.13 &         &          &  0.428 & SB9       \\ 
24192.0 &    52.70 &    +1.05 &         &          &  0.552 & SB9       \\ 
39907.0 &    54.23 &   $-1.18$&         &          &  0.801 & SB9       \\ 
40780.0 &    56.89 &   $-0.98$&         &          &  0.870 & SB9       \\ 
41338.0 &    58.28 &   $-1.34$&         &          &  0.915 & SB9       \\ 
43510.0 &    60.71 &   $-0.80$&    47.14 &  $ -0.18$&  0.087 & SB9       \\ 
45269.0 &    57.24 &    +0.02&    51.42 &   $-1.11$&  0.227 & SB9       \\ 
45911.0 &    55.53 &   $-0.18$&    56.51 &    +2.15 &  0.278 & SB9       \\ 
46635.0 &    54.00 &   $-0.25$&    59.65 &    +3.52 &  0.335 & SB9       \\ 
55964.1 &    61.81 &    +0.06&    47.09 &    +0.06 &  0.076 & HERMES    \\ 
55976.1 &    61.77 &    +0.04&    47.10 &    +0.05 &  0.077 & HERMES    \\ 
55938.2 &    61.82 &    +0.03&    47.01 &    +0.03 &  0.074 & HERMES    \\ 
55998.0 &    61.61 &   $-0.08$&    47.01 &   $-0.09$&  0.079 & HERMES    \\ 
56310.1 &    61.11 &    +0.03&    47.85 &    +0.02 &  0.104 & HERMES    \\ 
56062.9 &    61.53 &   $-0.04$&    47.14 &  $ -0.10$&  0.084 & HERMES    \\ 
55938.2 &    61.82 &    +0.04&    47.02 &    +0.04 &  0.074 & HERMES    \\ 
55983.0 &    61.70 &   $-0.01$&    47.03 &  $ -0.03$&  0.078 & HERMES    \\ 
55976.1 &    61.76 &    +0.03&    47.09 &    +0.03 &  0.077 & HERMES    \\ 
55937.1 &    61.74 &   $-0.05$&    46.93 &  $ -0.04$&  0.074 & HERMES    \\ 
56310.1 &    61.10 &    +0.02&    47.86 &    +0.03 &  0.104 & HERMES    \\ 
55983.0 &    61.70 &   $-0.01$&    47.05 &  $ -0.02$&  0.078 & HERMES    \\ 
50418.6 &    50.49 &   $-1.54$&    60.19 &    +1.35 &  0.636 & HIRES     \\ 
50828.1 &    51.45 &   $-0.97$&    60.24 &    +1.87 &  0.668 & ELODIE    \\ 
50828.6 &    51.46 &   $-0.96$&    60.28 &    +1.91 &  0.668 & ELODIE    \\ 
57858.5 &    56.20 &   $-1.01$&         &          &  0.227 & FEROS     \\ 
58091.9 &    56.15 &   $-0.50$&         &          &  0.245 & FEROS     \\  
\hline
\end{tabular}
\label{tab:rv}
\end{table}

\subsection{Orbital solution and stellar masses}
\label{sec:orbit}

The orbital solution was obtained by fitting a Keplerian model simultaneously to the radial velocities and to measurements of the relative astrometric positions of the two stars.
In addition to our new RV measurements, we included published velocities from the $9^{\rm th}$ Catalogue of Spectroscopic Binary Orbits \citep[SB9;][]{Pourbaix2004}.
The astrometric dataset consists of angular separations and position angles compiled from the literature (See Appendix Table~\ref{tab:rhopa}).

The posterior probability distributions of the orbital parameters were sampled using the affine-invariant Markov chain Monte Carlo ensemble sampler implemented in the {\sc emcee} package \citep{2013PASP..125..306F}.
We assumed Gaussian uncertainties for all observables and adopted broad, uniform priors on all model parameters.
Position angles obtained from speckle interferometry suffer from a 180$^\circ$ ambiguity, which we resolved by comparing their temporal evolution with the astrometric orbit published by \citet{Egeland2018}.

\begin{table}
\caption{Orbital parameters of the combined spectroscopic and astrometric solution for HD\,81809.
The fit uses measured radial velocities of both components together with relative astrometric positions.
Angles follow the standard visual-orbit convention: $\omega$ refers to the primary star, and $\Omega$ is the position angle of the ascending node.
$T$ is given as the JD$_{\rm (TDB)}$ time of periastron passage. 
Quoted $\sigma$ values represent additional Gaussian error terms fitted for each dataset.}
%\caption{Parameters of the spectroscopic and astrometric orbit model fit to measured radial velocities and relative positions of the two stars.
\label{tab:orbit}

\centering
\begin{tabular}{lrcl}
\hline
Parameter & \multicolumn{1}{l}{Value} & Units & Notes \\
\hline
$T $&    2455083  $\pm$      211 &            &  Time of periastron, JD \\
$P $&    34.53  $\pm$       0.16  &  yr        &  Orbital period \\
$K_A $&     5.42  $\pm$     0.17 & km/s     &  Semi-amplitude of primary spectroscopic orbit \\
$K_B $&     6.30  $\pm$     0.25 & km/s     &  Semi-amplitude of secondary spectroscopic orbit \\
$\gamma $&    54.93  $\pm$     0.21 & km/s       &  RV of system barycentre \\
$\alpha $&   0.4089  $\pm$   0.0079 &  $^{\prime\prime}$    &  Semi-major axis of the visual orbit \\
$e $&    0.366  $\pm$    0.022   &          &  Orbital eccentricity \\
$\omega $&    347.5  $\pm$      6.7 & $^{\circ}$   &  Longitude of periastron for the primary star orbit \\
$\Omega $&   150.42  $\pm$     0.39  &$^{\circ}$   &  Longitude of the ascending node \\
$\sin i $&  0.9968  $\pm$  0.0005    &      &  sin(orbital inclination) \\
$\sigma_A $&     0.94  $\pm$     0.21  & km/s   &   Std. err. on primary star RVs excluding HERMES  \\
$\sigma_B $&     2.38  $\pm$     0.77  & km/s   &  Std. err on secondary star RVs excluding HERMES  \\
$\sigma_{A,H} $&    0.049  $\pm$    0.012 & km/s   &  Std. err. on HERMES primary star RVs \\
$\sigma_{B,H} $&    0.061  $\pm$    0.015 & km/s   &  Std. err. on HERMES secondary star RVs \\
$\sigma_{\rho,S} $&   0.0361  $\pm$   0.0054 & $^{\prime\prime}$ &  Std. err. on speckle $\rho$ measurements \\
$\sigma_{\rho,M} $&    0.080  $\pm$    0.017 & $^{\prime\prime}$ & Std. err. on micrometer $\rho$ measurements \\
$\sigma_{\theta,S} $&     2.25  $\pm$     0.38 &  $^{\circ}$ & Std. error on speckle PA measurements \\
$\sigma_{\theta,M} $&     3.57  $\pm$     0.72 & $^{\circ}$ & Std. error on micrometer PA measurements \\
\hline
\end{tabular}
\end{table}

\begin{figure}[h]
\centering
\includegraphics[width=0.9\textwidth]{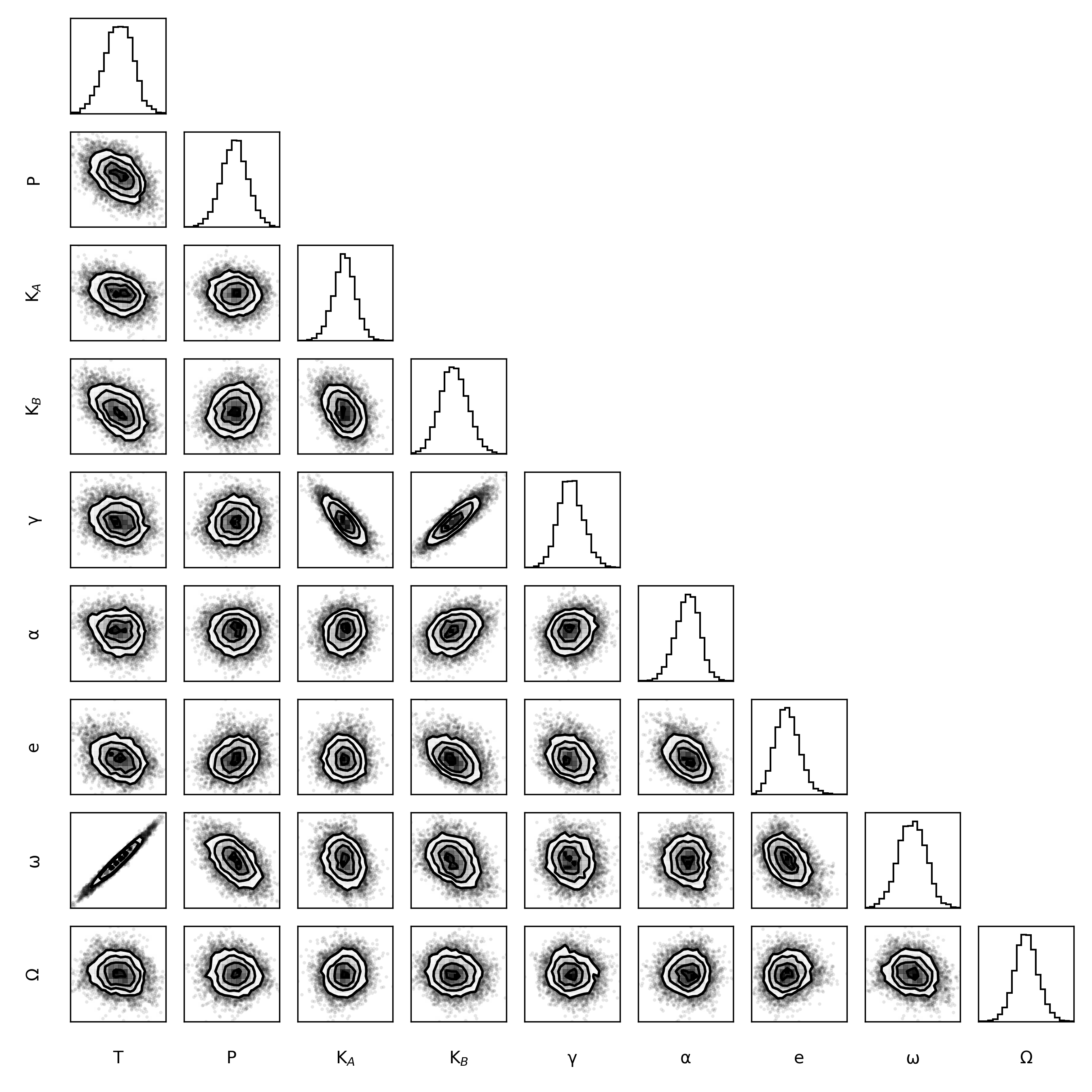}
\caption{Parameter correlation plot for parameters of interest given in Table \ref{tab:orbit} in the fit of the visual and spectroscopic orbits to radial velocity, position angle and angular separation measurements for HD\,81909.}
\label{fig:corner}
\end{figure}

\begin{figure}
\centering
\includegraphics[width=0.49\textwidth]{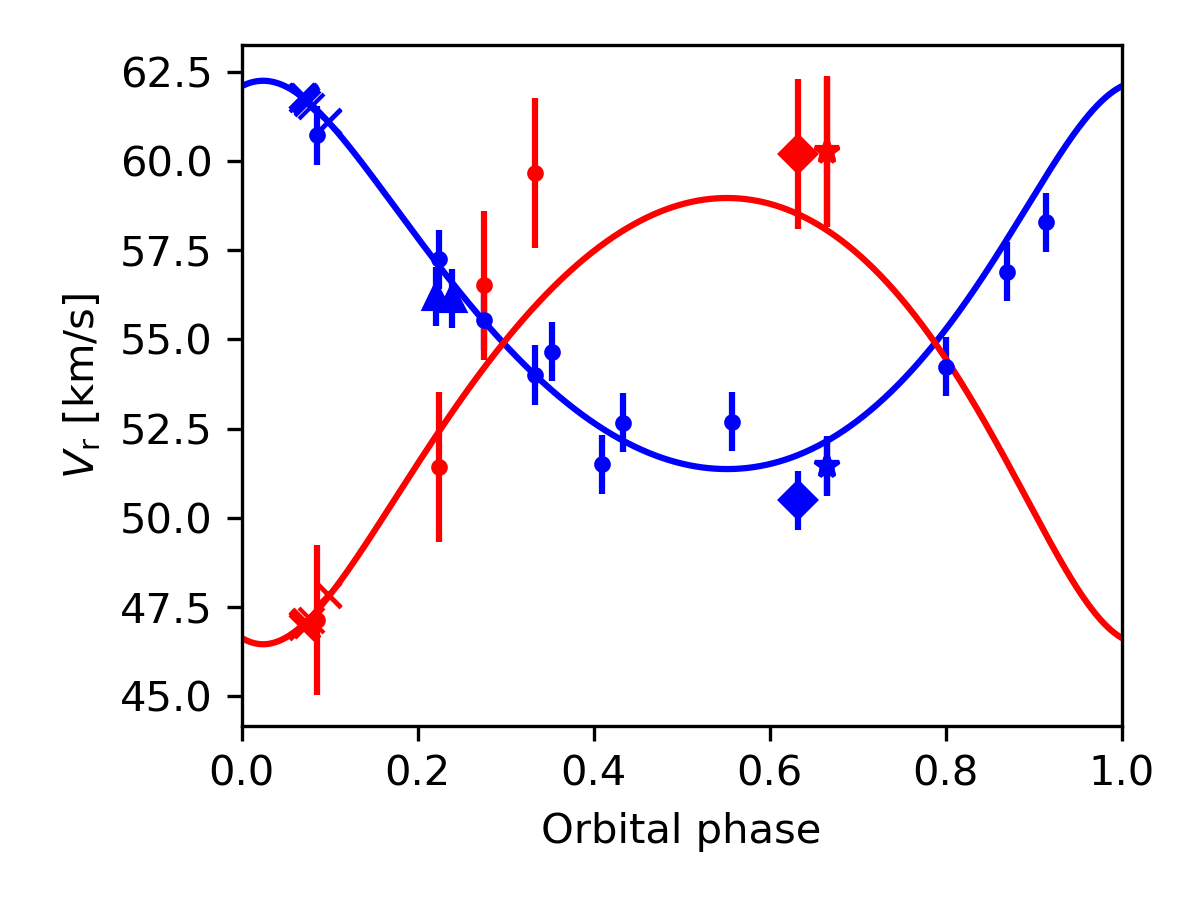}
\includegraphics[width=0.49\textwidth]{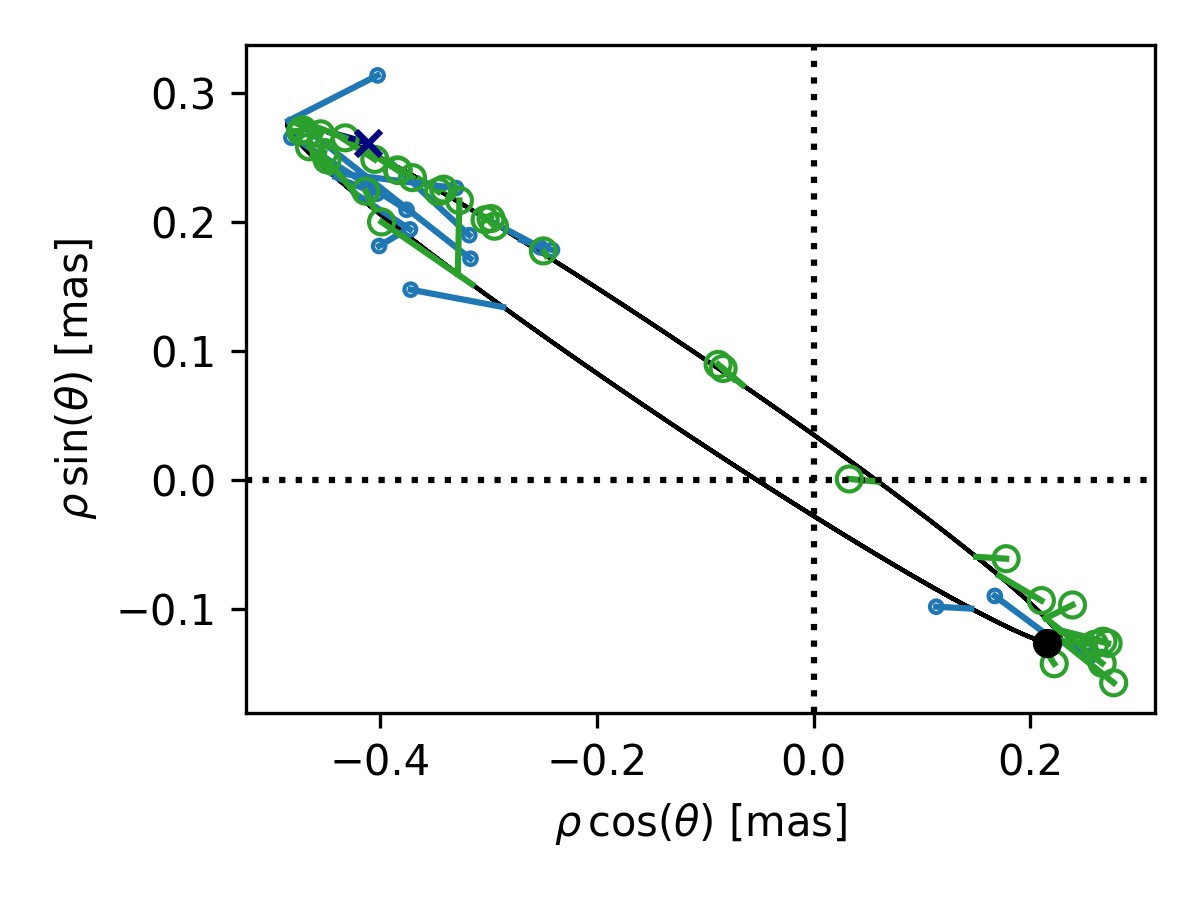}
\caption{Left panel: 
Radial velocity measurements for HD~81809\,A (blue points) and HD~81809\,B (red points) with the best-fit spectroscopic orbit (solid lines).
Symbols denote the source of the radial velocity measurements as follows: cross -- HERMES; diamond -- HIRES; triangle -- FEROS; star -- ELODIE; dots -- SB9.
Right panel: 
Measurements of the relative positions of the two stars with the best-fit astrometric orbit. 
Points measured by speckle interferometry are plotted with open circles, micrometer measurements are plotted with dots, and the Hipparcos measurement is shown with a cross. 
Lines connect measurements to the point on the visual orbit corresponding to their time of observation. 
The point of periastron on the orbit is marked with a filled circle.}
\label{fig:orbit}
\end{figure}

The orbital parameters derived from our analysis are listed in Table~\ref{tab:orbit}. Correlations among selected parameters are presented in Figure~\ref{fig:corner}, while the best-fit model is overplotted on the radial velocity and astrometric data in Figure~\ref{fig:orbit}.
The orbital parallax derived from the joint solution, $\pi = 32.3 \pm 0.8$\,mas, is in excellent agreement with the Gaia DR3 value of $\pi = 32.3 \pm 0.4$\,mas.

%\subsection{Stellar mass determination}
%\label{sec:masses}

%The maximum-likelihood fit to the measurements of $\rho$, PA, and the radial velocities of the two stars are shown in Figure~\ref{fig:orbit}. 

The masses of HD~81809\,A and HD~81809\,B were determined through a joint analysis of radial velocity measurements and relative astrometry.
We performed a simultaneous fit of a Keplerian orbit to the spectroscopic and visual data, allowing us to derive a fully constrained orbital solution.
The resulting mass estimates represent a significant improvement over previous determinations, primarily due to the inclusion of high-resolution spectra that clearly resolve both components of this SB2 system and provide radial velocities for each star.
The stellar masses inferred are reported in Table~\ref{tab:atmos}.

\subsection{Atmospheric Parameters from Spectroscopic Analysis}
\label{Sec:SpectAtmo}
We determined the overall iron abundance and derived detailed abundances for several other chemical elements from a high-resolution HERMES spectrum obtained by averaging nine individual spectra, all taken near orbital quadrature (phases between 0.074 and 0.079, see Table~\ref{tab:rv}) to maximize the separation between the components and enhance the visibility of lines from both the primary and the secondary, achieving a signal-to-noise ratio of $\sim 500$ at $\lambda = 6000$~\AA, thus providing a comprehensive picture of the stellar chemical composition.

The determination of chemical abundances began with the definition of the primary atmospheric parameters: effective temperature ($T_{\rm eff}$), surface gravity ($\log g$), microturbulent velocity ($\xi$), and projected rotational velocity ($v \sin i$).

To estimate the effective temperature, we applied to both components the line depth ratio (LDR) method \citep{kot2000}, which is sensitive to temperature variations and largely unaffected by abundance differences or interstellar reddening. For our spectrum, we measured approximately 32 LDRs using the calibration lines listed in \citet{kot2000}. The final temperature and its uncertainty were derived as the mean value and the weighted standard deviation of these measurements.

The remaining parameters ($\xi$ and $\log g$) were derived through an iterative process. The microturbulent velocity was obtained by requiring the iron abundance to be independent of line strength, i.e., ensuring a null slope in the [Fe/H] versus equivalent width (EW) diagram. We measured the EWs of 145 Fe\,\textsc{i} lines using a custom semi-automatic routine written in \textit{IDL}, with line data taken from \citet{romaniello2008}. These EWs were converted into abundances using the WIDTH9 code \citep{kur81}, in combination with model atmospheres computed with ATLAS9 \citep{kur93}. Since Fe\,\textsc{i} lines are insensitive to surface gravity, $\log g$ was determined by imposing ionization equilibrium between Fe\,\textsc{i} and Fe\,\textsc{ii}, using 24 Fe\,\textsc{ii} lines from the same reference. Errors were estimated by propagating the uncertainties from the linear fits.

The final atmospheric parameters, adopted for the two components, are reported in Table~\ref{tab:atmos}. The values result in good agreement, within the errors, with previous measurements.

\begin{table}[h]
\begin{center}
\caption{The stellar parameters of HD\,81809\,A and HD\,81809\,B as deduced from the orbital and spectroscopic analysis.
\label{tab:atmos}}
\renewcommand{\tabcolsep}{0mm}
\begin{tabular}{l c c}

\tableline\tableline
\noalign{\smallskip}
                          & HD~81809\,A\,  \, & HD~81809\,B \\
\hline
$M/{\mathrm M_{\odot} } $\,$^a$    & $ 0.92 \pm 0.09  $ & $ 0.79 \pm 0.06  $ \\
$M_A/M_B$  & \multicolumn{2}{c}{$1.16 \pm 0.07$} \\
$T_{\mathrm{eff}}$  (K)   & $5580 \pm 140 $  & 5520 $\pm$ 150 \\
$v \sin i$ (km\,s$^{-1}$) &  4.0 $\pm$ 0.5  &  4.0 $\pm$ 0.5 \\
$\log g$ (dex)            & 3.8 $\pm$ 0.5   &  4.2 $\pm$ 0.5 \\
 $\xi$  (km\,s$^{-1}$)    &  1.1 $\pm$ 1.7  &  1.0 $\pm$ 1.5 \\
\hline
\noalign{\smallskip}
\multicolumn{3}{l}{$^a$ covariance $cov(M_A/ {\mathrm M_{\odot}}, M_B/{\mathrm M_{\odot} })= 0.0039$ }
%\noalign{\smallskip}

\end{tabular}
\end{center}

\end{table}

Furthermore, we identified spectral lines suitable for the determination of the abundances of twenty-seven chemical elements, including Fe. This list is reported in Table~\ref{tab:lines}. We derived the abundances by spectral synthesis, minimizing the $\chi^2$ difference between the observed and synthetic spectra. Synthetic spectra were generated in three steps: 
(i) plane-parallel local thermodynamic equilibrium (LTE) atmospheric models were computed using the ATLAS9 code \citep{kur93}; 
(ii) stellar spectra were synthesized using SYNTHE \citep{kur81}; 
and 
(iii) the synthetic spectra were convolved with instrumental (R = 85000) and rotational velocity broadening profiles to match observed line profiles.
(iv) the total synthetic spectrum has been computed following \citet{catanzaro2024}, using the equation:
\begin{equation}
\label{flux}
F^{th}_{\text{Tot}} = \frac{l_\lambda F^{th}_A + F^{th}_B}{1+l_\lambda}
,\end{equation}
\noindent
where F$_{A,B}^{th}$ are the synthetic fluxes of the primary and secondary and the luminosity ratio 
$l_\lambda=(T_{\mathrm{eff},A}/T_{ \mathrm{eff},B}) ^4 (R_A/R_B)^2\simeq 3.05 $ is calculated from the 
$T_{\mathrm{eff}}$ and $\log g$ 
of the single stars reported in Table~\ref{tab:atmos}.

\begin{figure*}[h]
    \centering
 \includegraphics[trim=0 0 0 170,width=15cm]{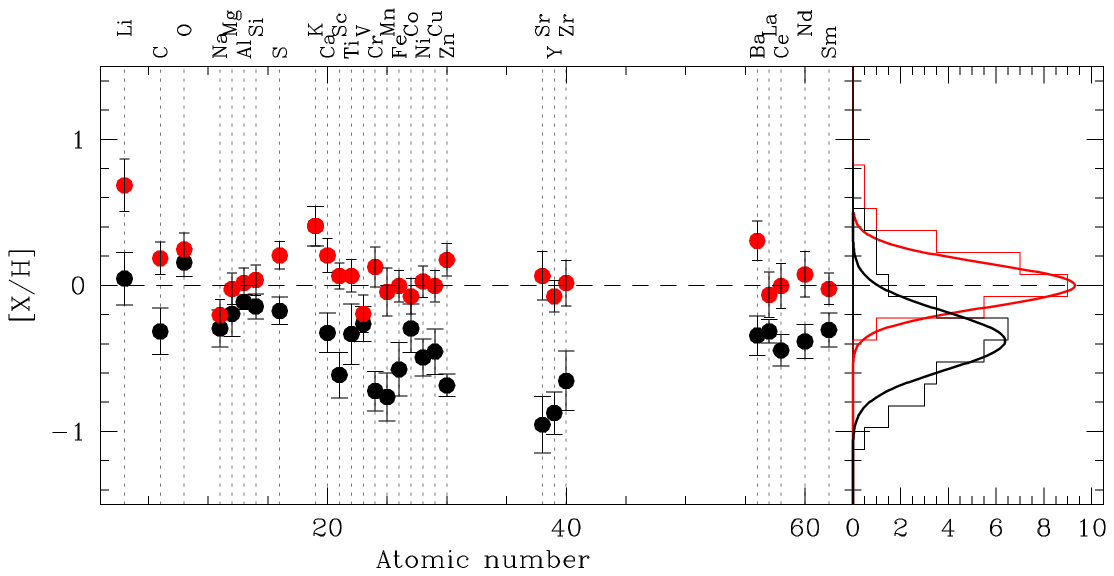}
    \caption{Atmospheric abundances obtained by present analysis for HD\,81809\,A (black dots) and HD\,81809\,B (red dots) compared with those of the Sun \citep{grevesse2011chemical} (dashed line). In the right panel, we show the histogram of the distribution of the abundances around the average with overplotted Gaussian fit, colors as in left panel.}
    \label{fig:pattern}
\end{figure*}

The results are shown in Figure~\ref{fig:pattern} and in Table~\ref{tab:abund} given in terms of ratio both with H and Fe. The chemical properties of HD 81809\,A and B provide important results on the scenario of the formation and evolution of this binary system.

%From inspection of Figure~\ref{fig:pattern}, we can see that the primary star has [Fe/H]$=-0.57\pm0.18$ and is therefore mildly metal-poor. 

The abundance pattern of the two components of HD\,81809 reveals a marked chemical difference. The primary (HD\,81809~A) is metal-poor, with an iron abundance of [Fe/H]~$\simeq -0.57$, while the secondary (HD\,81809~B) has essentially solar metallicity ([Fe/H]~$\simeq 0.00$).

For the primary, most of the measured chemical species - in particular C, Na, and Mg - are underabundant in [X/H] relative to the solar reference values \citep{grevesse2011chemical}, with the notable exception of oxygen, which shows a slight overabundance of $\sim 0.16$~dex. The abundance distribution, represented by the histogram in the right panel of Figure~\ref{fig:pattern}, yields an average metallicity of [M/H]~$=-0.44$.
The $\alpha$-elements such as Mg, Si, S, and Ca show clear enhancements relative to iron, with typical [X/Fe] values in the range $\sim 0.2$–0.4~dex, consistent with an $\alpha$-enhanced, chemically older population.
%The high [Mg/Fe] ratio ([Mg/Fe]$=0.38\pm0.24$), together with consistently elevated [X/Fe] values for O, Si, and S (and even Al, which behaves in a similar way), indicates a clear $\alpha$-enhancement. 
This suggests rapid formation in an environment 
where chemical enrichment occurred on short timescales and was dominated by Type~II supernovae (originating from massive stars), before Type~Ia supernovae (which mainly produce iron) had a major impact on the chemical composition.

These chemical signatures, combined with the star's kinematic properties $U = -42.9$, $V = -46.8$, $W = -2.0$~km\,s$^{-1}$ \citep{Soubiran2005}, which correspond to a total space velocity of $\sim$63~km\,s$^{-1}$ relative to the LSR, further support that the star, and hence the system, belongs to an old stellar population, most likely the Galactic thick disk \citep{anguiano2020,vieira2022}. In addition, the negative $V$ component indicates a lag with respect to the Galactic rotation, which is also characteristic of thick-disk stars.

Iron-peak elements (Cr, Mn, Ni, Cu, Zn) in HD\,81809~A are generally subsolar in [X/H], but their [X/Fe] ratios are near solar values, indicating behaviour similar to iron. Neutron-capture elements (Sr, Y, Zr, Ba, La, Ce, Nd, Sm) are also subsolar in [X/H], but tend to be mildly enhanced in [X/Fe], suggesting a moderate enrichment in s-process material relative to iron.

In contrast, HD\,81809~B displays abundances much closer to solar for most species, with [X/H] and [X/Fe] typically within a few hundredths of a dex of the solar values. The secondary therefore appears chemically normal, while the primary is both more metal-poor and $\alpha$-enhanced. This chemical dichotomy between the two components points to a complex evolutionary history for the system and will be further discussed in the next Sections.

Lithium was also measured from the Li\,\textsc{i} 6707.8~\AA{} feature for both components. The primary (HD\,81809~A) shows a lithium abundance consistent with the solar value, indicating that only modest depletion has occurred despite the star’s evolved state. In contrast, the secondary (HD\,81809~B) exhibits a significantly higher lithium abundance, about 0.69~dex above the solar value. This behavior is noteworthy, as lithium is typically depleted in solar-type stars over Gigayear timescales. The enhanced abundance in HD\,81809~B may therefore reflect either a younger evolutionary status compared to the primary, a different internal mixing history, or possible recent accretion events affecting the surface composition. The marked lithium difference between the two stars adds on to the chemical dichotomy observed in this system and may provide additional constraints on its formation and evolutionary history.

 \begin{table}[h]
 \caption{Atmospheric abundances for both components HD\,81809\,A and B, for all identified elements, including those of the iron group and the neutron-capture elements. Columns list abundances in the standard logarithmic form relative to hydrogen and scaled to solar values ([X/H]), and abundances relative to iron ([X/Fe]).}
    \label{tab:abund}
    \centering
    \begin{tabular}{l|rr|rr}
    \hline
    El  &     [X/H]~~~~    &    [X/Fe]~~~~ &     [X/H]~~~~    &    [X/Fe]~~~~  \\
    \hline
 Li & -0.05~$\pm$~0.18 &     0.62~$\pm$~0.24   &    0.69~$\pm$~0.18 &     0.69~$\pm$~0.24  \\    
  C & -0.31~$\pm$~0.16 &     0.26~$\pm$~0.24   &    0.19~$\pm$~0.11 &     0.19~$\pm$~0.16  \\ 
  O &  0.16~$\pm$~0.09 &     0.73~$\pm$~0.21   &    0.25~$\pm$~0.11 &     0.25~$\pm$~0.16  \\ 
 Na & -0.29~$\pm$~0.13 &     0.28~$\pm$~0.22   & $-$0.20~$\pm$~0.11 &  $-$0.20~$\pm$~0.15  \\ 
 Mg & -0.19~$\pm$~0.16 &     0.38~$\pm$~0.24   & $-$0.02~$\pm$~0.11 &  $-$0.02~$\pm$~0.15  \\ 
 Al & -0.11~$\pm$~0.05 &     0.46~$\pm$~0.19   &    0.02~$\pm$~0.10 &     0.02~$\pm$~0.15  \\ 
 Si & -0.14~$\pm$~0.09 &     0.43~$\pm$~0.20   &    0.04~$\pm$~0.10 &     0.04~$\pm$~0.15  \\ 
  S & -0.17~$\pm$~0.09 &     0.40~$\pm$~0.21   &    0.21~$\pm$~0.09 &     0.21~$\pm$~0.14  \\ 
  K &  0.41~$\pm$~0.13 &     0.98~$\pm$~0.23   &    0.41~$\pm$~0.13 &     0.41~$\pm$~0.17  \\ 
 Ca & -0.32~$\pm$~0.14 &     0.25~$\pm$~0.23   &    0.21~$\pm$~0.12 &     0.21~$\pm$~0.16  \\ 
 Sc & -0.61~$\pm$~0.16 &  $-$0.04~$\pm$~0.24   &    0.07~$\pm$~0.09 &     0.07~$\pm$~0.14  \\ 
 Ti & -0.33~$\pm$~0.21 &     0.24~$\pm$~0.28   &    0.07~$\pm$~0.11 &     0.07~$\pm$~0.16  \\ 
  V & -0.26~$\pm$~0.12 &     0.31~$\pm$~0.22   & $-$0.19~$\pm$~0.13 &  $-$0.19~$\pm$~0.17  \\ 
 Cr & -0.72~$\pm$~0.14 &  $-$0.15~$\pm$~0.23   &    0.13~$\pm$~0.14 &     0.13~$\pm$~0.17  \\ 
 Mn & -0.76~$\pm$~0.16 &  $-$0.19~$\pm$~0.25   & $-$0.04~$\pm$~0.16 &  $-$0.04~$\pm$~0.20  \\ 
 Fe & -0.57~$\pm$~0.18 &     $---$ ~~~~        &    0.00~$\pm$~0.11 &     $---$ ~~~~       \\ 
 Co & -0.29~$\pm$~0.17 &     0.28~$\pm$~0.25   & $-$0.07~$\pm$~0.12 &  $-$0.07~$\pm$~0.16  \\ 
 Ni & -0.49~$\pm$~0.13 &     0.08~$\pm$~0.22   &    0.03~$\pm$~0.11 &     0.03~$\pm$~0.15  \\ 
 Cu & -0.45~$\pm$~0.16 &     0.12~$\pm$~0.24   & $-$0.00~$\pm$~0.11 &  $-$0.00~$\pm$~0.15  \\ 
 Zn & -0.68~$\pm$~0.08 &  $-$0.11~$\pm$~0.20   &    0.18~$\pm$~0.11 &     0.18~$\pm$~0.16  \\ 
 Sr & -0.95~$\pm$~0.19 &  $-$0.38~$\pm$~0.27   &    0.07~$\pm$~0.17 &     0.07~$\pm$~0.20  \\ 
  Y & -0.87~$\pm$~0.15 &  $-$0.30~$\pm$~0.23   & $-$0.07~$\pm$~0.11 &  $-$0.07~$\pm$~0.15  \\ 
 Zr & -0.65~$\pm$~0.20 &  $-$0.08~$\pm$~0.27   &    0.02~$\pm$~0.16 &     0.02~$\pm$~0.19  \\ 
 Ba & -0.34~$\pm$~0.13 &     0.23~$\pm$~0.23   &    0.31~$\pm$~0.13 &     0.31~$\pm$~0.17  \\ 
 La & -0.31~$\pm$~0.08 &     0.26~$\pm$~0.20   & $-$0.06~$\pm$~0.16 &  $-$0.06~$\pm$~0.19  \\ 
 Ce & -0.44~$\pm$~0.11 &     0.13~$\pm$~0.21   & $-$0.00~$\pm$~0.16 &     0.00~$\pm$~0.19  \\ 
 Nd & -0.38~$\pm$~0.12 &     0.19~$\pm$~0.22   &    0.08~$\pm$~0.16 &     0.08~$\pm$~0.19  \\ 
 Sm & -0.30~$\pm$~0.12 &     0.27~$\pm$~0.22   & $-$0.02~$\pm$~0.11 &  $-$0.02~$\pm$~0.15  \\ 

   \end{tabular}
\end{table}

\subsection{Luminosities from bolometric fluxes}
The luminosities of the two components were determined from their bolometric fluxes using the {\sc teb} software package \citep{teb,teb2}.
This tool allows a self-consistent analysis of multi-band photometry and flux ratios to recover the bolometric fluxes, effective temperatures, and luminosities of the individual stars in binary systems.

%This package is designed to measure the effective temperature  ($T_{\rm eff}$) of stars in eclipsing binary systems directly from the bolometric fluxes of the two stars and their angular diameters.}

We fitted the observed broadband photometry of the unresolved system together with the measured flux ratios in the Tycho $B_{\rm T}$ and $V_{\rm T}$ bands (Table~\ref{tab:mags}).
The spectral energy distributions (SEDs) used in the analysis are semi-empirical, consisting of BT-Settl model atmospheres \citep{2013MSAIS..24..128A} multiplied by a smooth function.
This approach allows for the SEDs to contain realistic stellar absorption features, but the overall shape of the SEDs is determined by the observed magnitudes and flux ratios of the binary system.
In addition, we used empirical colour -- $T_{\rm eff}$ relations to ensure that the optical -- infrared and near-UV -- optical colours computed from the SEDs are similar to those of others stars of the same effective temperature. 
The derived bolometric fluxes and luminosities are largely insensitive to the assumed stellar radii and effective temperatures, owing to the strong observational constraints imposed by the multi-wavelength photometry.

%% Observed and computed magnitudes, colors and flux ratios %%

\begin{table*}
\caption{Observed magnitude and flux ratios for HD\,81809 and predicted values based on our synthetic photometry. The
predicted magnitudes are shown with error estimates from the uncertainty on the zero-points for each photometric system.  The
pivot wavelength for each band pass is shown in the column headed  $\lambda_{\rm pivot}$.  
U$_{G}$, B$_{G}$ etc. are magnitudes in the Geneva 7-colour photometry system. 
TD1 fluxes have been converted to AB magnitudes, e.g. F2365 to m$_{2365}$.}
\label{tab:mags}
\centering
\begin{tabular}{@{}lrrrr}
\hline

Band &  $\lambda_{\rm pivot}$ [nm]& \multicolumn{1}{c}{Observed} &\multicolumn{1}{c}{Computed} &
\multicolumn{1}{c}{$\rm O-\rm C$} \\
\hline
\noalign{\smallskip}
UVW2  &   213.6 & $11.018\pm 0.010 $& $10.974\pm 0.049 $& $+0.043 \pm 0.050 $ \\
UVM2  &   232.7 & $10.780\pm 0.072 $& $10.546\pm 0.062 $& $+0.234 \pm 0.095 $ \\
m$_{2365}$ &   236.3 & $10.560\pm 0.160 $& $10.858\pm 0.136 $& $-0.298 \pm 0.210 $ \\
m$_{2740}$ &   273.9 & $ 8.930\pm 0.030 $& $ 9.174\pm 0.085 $& $-0.244 \pm 0.090 $ \\
U$_G$  &   345.5 & $ 6.583\pm 0.010 $& $ 6.552\pm 0.017 $& $+0.031 \pm 0.019 $ \\
B1$_G$ &   401.9 & $ 6.277\pm 0.010 $& $ 6.286\pm 0.018 $& $-0.009 \pm 0.021 $ \\
B$_{\rm T}$    &   421.2 & $ 6.196\pm 0.014 $& $ 6.194\pm 0.014 $& $+0.002 \pm 0.020 $ \\
B$_G$  &   423.9 & $ 5.208\pm 0.009 $& $ 5.220\pm 0.017 $& $-0.012 \pm 0.019 $ \\
B     &   439.8 & $ 6.040\pm 0.010 $& $ 6.050\pm 0.008 $& $-0.010 \pm 0.013 $ \\
V$_{\rm T}$    &   533.5 & $ 5.483\pm 0.009 $& $ 5.466\pm 0.014 $& $+0.017 \pm 0.017 $ \\
V1$_G$ &   540.7 & $ 6.122\pm 0.010 $& $ 6.130\pm 0.008 $& $-0.008 \pm 0.013 $ \\
V     &   550.0 & $ 5.390\pm 0.010 $& $ 5.389\pm 0.004 $& $+0.001 \pm 0.011 $ \\
V$_G$  &   550.2 & $ 5.377\pm 0.009 $& $ 5.391\pm 0.012 $& $-0.014 \pm 0.015 $ \\
Hp    &   550.7 & $ 5.503\pm 0.005 $& $ 5.533\pm 0.011 $& $-0.029 \pm 0.012 $ \\
G$_G$  &   581.3 & $ 6.418\pm 0.010 $& $ 6.413\pm 0.016 $& $+0.005 \pm 0.019 $ \\
R     &   655.6 & $ 5.020\pm 0.010 $& $ 5.006\pm 0.004 $& $+0.014 \pm 0.011 $ \\
I     &   804.6 & $ 4.660\pm 0.010 $& $ 4.640\pm 0.009 $& $+0.020 \pm 0.014 $ \\
J     &  1240.6 & $ 3.984\pm 0.180 $& $ 4.130\pm 0.015 $& $-0.146 \pm 0.181 $ \\
H     &  1649.0 & $ 3.586\pm 0.176 $& $ 3.818\pm 0.019 $& $-0.232 \pm 0.177 $ \\
K$_{s\rm }$    &  2162.9 & $ 3.568\pm 0.176 $& $ 3.766\pm 0.030 $& $-0.198 \pm 0.179 $ \\
S9W   &  9097.2 & $ 8.268\pm 0.002 $& $ 8.271\pm 0.053 $& $-0.003 \pm 0.053 $ \\
L18W  & 19507.0 & $ 9.788\pm 0.003 $& $ 9.778\pm 0.070 $& $+0.010 \pm 0.070 $ \\
\noalign{\smallskip}
\hline
\multicolumn{5}{@{}l}{Flux ratios} \\
\hline
\noalign{\smallskip}
$(F_B/F_A)_{{\rm B_T}}$    &   421.2 & $ 0.172 \pm 0.004 $& $ 0.172 $& $+0.000 \pm 0.004 $ \\
$(F_B/F_A)_{{\rm V_T}}$    &   533.5 & $ 0.175 \pm 0.002 $& $ 0.176 $& $-0.000 \pm 0.002 $ \\
\noalign{\smallskip}
\hline
\end{tabular}
\end{table*}

The used photometric dataset given in Table~\ref{tab:mags}  includes measurements from the near-UV to the mid-infrared: 
 J, H and Ks magnitude from the 2MASS survey \citep{2006AJ....131.1163S}; 
B$_{\rm T}$ and V$_{\rm T}$ magnitudes from the Tycho-2 catalogue \citep{2000A&A...355L..27H}; H$_{\rm p}$ from the Hipparcos mission \citep{2007A&A...474..653V}; Geneva 7-colour photometry from \citet{2022A&A...661A..89P}; Akari S9W and L18W mid-infrared fluxes from \citet{2010A&A...514A...1I}; UVM2 and UVW2 magnitudes from the XMM-Newton Optical Monitor Serendipitous Source Survey Catalogue, Version 6.0 \citep{2012MNRAS.426..903P}; BVRI photometry from \citet{1990A&AS...83..357B}; and near-UV F2365 and F2740 fluxes from the TD1 mission \citep{1973MNRAS.163..291B}.
%Here we decided to not use the GAIA magnitude as input since 
%relative RUWE flag ${(\rm > 1.5)} $
%used to validate the measurement indicates that it is not fully trustful. 
It is worth noticing that 
the two stars of the HD 81809 system have not been resolved in Gaia DR3, and the corresponding flag (RUWE$=4.91$) is above the recommended threshold for a trustworthy astrometric solution. Therefore, we decided not to use the Gaia G magnitude as input.

We adopted the Gaia DR3 parallax
 ($\pi = 32.32 \pm 0.36$\,mas) corrected for the zero-point offset following \citet{2021A&A...654A..20G}.
For the colour excess we used the value $E(B-V) = 0.0054 \pm 0.0011$ from  the {\tt bayestar17} Galactic reddening 3D maps \citep{Green2018}.
This negligible reddening value is consistent with the lack of interstellar Na\,I absorption in the spectra \citep{2025RNAAS...9..146M} and is expected for stars less than 40\,pc from the Sun \citep{gcs2}.

\begin{figure}[ht]
\centering
\includegraphics[width=0.9\textwidth]{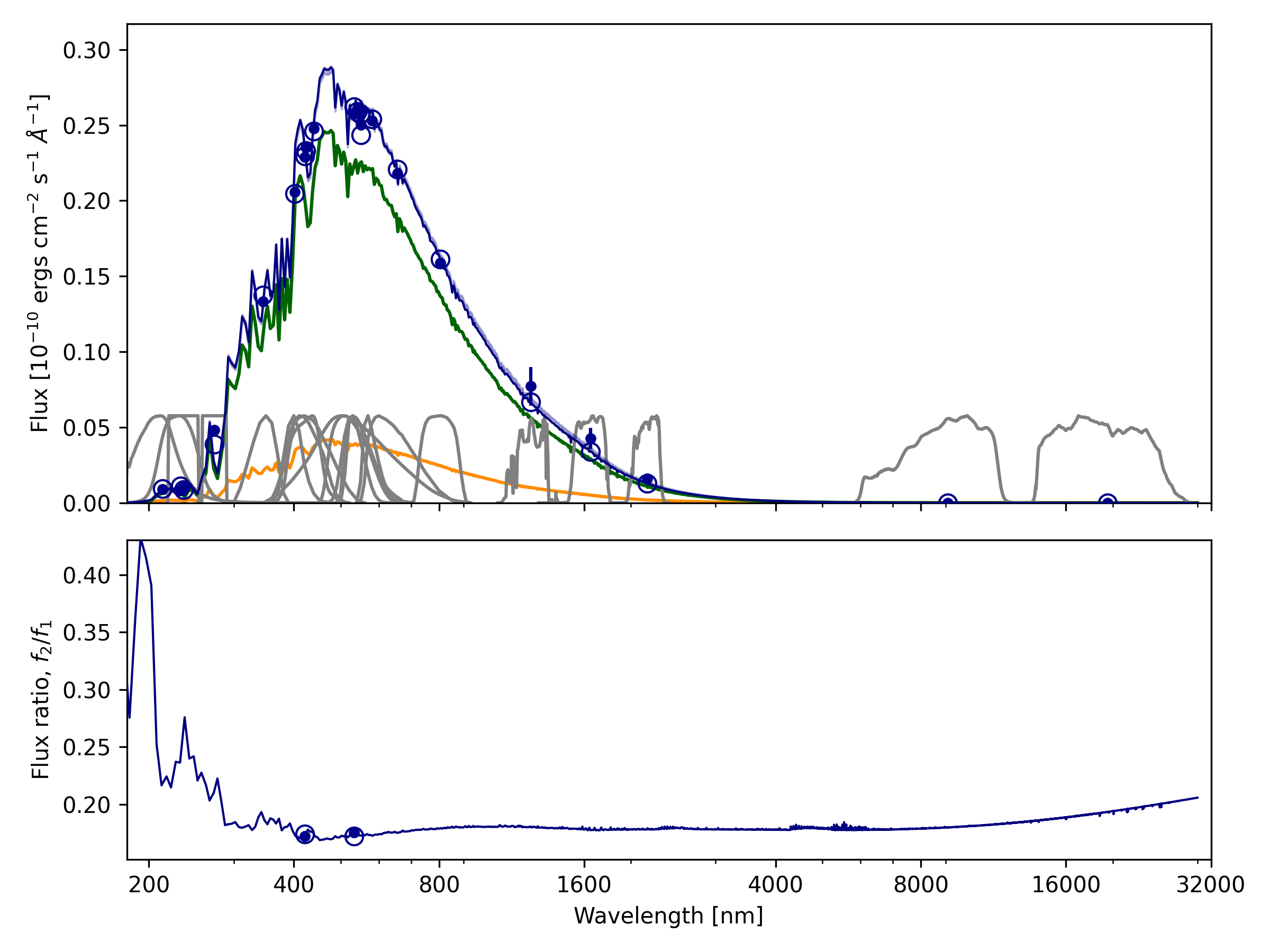}
\caption{In the upper panel the SED of HD~81809 is shown. 
The best-fit SED is plotted as dark blue line.
%and the mean SED $\pm 1$ standard deviation is plotted as a filled region. 
The observed fluxes are plotted as blue points with error bars while predicted fluxes for the best-fit SED integrated over the response functions shown in grey are plotted with open circles.
The SEDs of the two stars are also plotted (green -- primary, orange -- secondary). 
Lower panel: Flux ratio as a function of wavelength for the best-fit SEDs. 
The observed flux ratios are plotted as points
%with error bars 
and the predicted flux ratios in the Tycho $B_T$ and $V_T$ bands are plotted as open circles.}
\label{fig:teb}
\end{figure}

\begin{table}[h]
\centering
\caption{Stellar parameters from photometric analysis.}
\label{tab:params}
\begin{tabular}{lrr}
\hline
Parameter & HD 81809 A  & HD 81809  B \\
\hline

%$R/R_{\odot}$     & $ 2.07 \pm 0.14  $ & $ 1.145 \pm 0.05 $ \\
%$\log g$          & $ 3.77 \pm 0.05  $ & $ 4.22 \pm 0.03  $ \\
$T_{\rm eff}$ (K) & $ 5554 \pm 120   $ & $ 5481 \pm 120   $ \\
$L/L_{\odot}$     & $ 5.10 \pm 0.14  $ & $ 0.92 \pm 0.05  $ \\
\hline

\end{tabular}
\end{table}
The observed and synthetic fluxes and the best-fit spectral energy distributions are shown in Figure~\ref{fig:teb}. The luminosities and effective temperatures obtained from this analysis are given in Table~\ref{tab:params}. It should be pointed out that while
the effective temperatures results in good agreement with the spectroscopic values of Table \ref{tab:atmos},
the ratio between the luminosities $ L_A/L_B\sim 5.54$ is higher than the ratio $l_\lambda\sim3.05$ deduced spectroscopically. 

%\subsection{Questions on the stellar luminosity}
% With the aim to properly characterize the fundamental parameters of this system we  performed an analysis of the broadband Spectral Energy Distribution (SED) together with the {\it Gaia/}EDR3 \citep{Gaia2020} parallax measurement. 
 %The observed SED was assembled using the VOSA tool \citet{VOSA}. Given the proximity of the star,
%interstellar reddening was neglected. In Figure \ref{fig:sed}, we show the comparison between the synthetic flux and the observed photometric data. 
%Table \ref{tab:sed} summarizes the input parameters and the results obtained as convolved measurements. 

The spectral energy distribution (SED) analysis reveals excellent agreement between the observed fluxes and the photospheric models from the UV down to the mid-infrared (AKARI L18W, $19.5\mu m$), as shown in Table \ref{tab:mags}.
However, a strong infrared excess becomes evident at longer wavelengths $\lambda>30 \mu$m in the flux ratio profile, which may be interpreted as evidence for the presence of a cold debris disk which can surround one of the two stars (S-type orbits circumprimary/circumsecondary) or
can orbit around the center of mass (P-type orbits circumbinary).
We used data from the
IRAS Point Source Catalog \citep{Helou1988},
 to pinpoint the thermal properties of the dust (see Table \ref{tab:iras_excess}).
At $60\,\mu\mathrm{m}$, the observed flux  is approximately three times higher than the predicted photospheric value, while at $100\,\mu\mathrm{m}$, where the stellar contribution is negligible.
Fitting these infrared residuals with a blackbody function yields a characteristic dust temperature of $T_{\rm BB} \simeq 96.6$\,K.

\begin{table}
\caption{IRAS flux densities and infrared excess analysis. The observed fluxes ($F_{\rm obs}$) are compared with the photospheric model predictions ($F_{\rm mod}$). The excess ratio ($F_{\rm obs}/F_{\rm mod}$) clearly indicates the presence of cold dust radiating at $\lambda \ge 60\,\mu$m.
Flux values are scaled by $10^{16}$ (25$\mu$m), $10^{18}$ (60$\mu$m), and $10^{19}$ (100$\mu$m) for readability. The uncertainties on the excess ratio include the observational errors.
}
\label{tab:iras_excess}
\centering
\begin{tabular}{l c c c c}
\hline

Band & $\lambda_{pivot}$ & $F_{\rm obs}$ & $F_{\rm mod}$ & Excess \\
     & [$\mu$m]  & [arbitrary units] & [arbitrary units] & ($F_{\rm obs}/F_{\rm mod}$) \\
\hline
\noalign{\smallskip}
IRAS 25  & 25 & $2.47 \pm 0.26$ & $2.32$ & $1.07 \pm 0.11$ \\
IRAS 60  & 60 & $22.6 \pm 9.0$  & $7.42$ & $3.05 \pm 1.21$ \\
IRAS 100 & 100& $191.3 \pm 67.7$& $3.89$ & $49.1 \pm 17.4$ \\
\hline
\end{tabular}

\end{table}

Belts of rocks and dust, known as debris disks are commonly detected around main-sequence stars. However, very little is known regarding the debris disks around subgiants, as for the evolutionary stage of the primary. It is not clear 
whether the presence of debris disks correlates with the presence of planets \citep{MoroMartin2007,Bryden2009,Kospal2009}.

The presence of this debris disk provides a plausible explanation for the observed metallicity difference between the two stars. While the primary remains metal-poor, the secondary exhibits nearly solar metallicity, which could result from selective accretion of metal-rich solids from the disk during past dynamical evolution. Such accretion is more effective for stars with shallower convective envelopes and could naturally enhance the photospheric metallicity of the secondary, leaving the primary largely unaffected. This scenario supports the idea that the system experienced significant planetesimal evolution and possible inward transport of solids, consistent with the detection of the debris disk.

\section{Analysis of TESS asteroseismic data} 
\label{sec:osc}

\subsection{Global pulsation properties}
\label{Sec:SeismicCorsaro}
HD~81809 has been photometrically observed by TESS during 27 consecutive days in three different years: 2021, 2023, and 2025 in long cadence at 120\,s sampling \citep{10.17909/t9-nmc8-f686}, and in 2025 also in short cadence at 20\,s \citep{10.17909/t9-st5g-3177}.
%of sector 35 from 2021 February 9 to 2021 March 7, both in short and long cadence mode available on the MAST archive~\footnote{\url{https://archive.stsci.edu/tess/tic_ctl.html}}. 
The two components are blended in the TESS observations, with the primary producing a clear detectable asteroseismic signal because of its brightness.%but the bright primary produces the only detectable asteroseismic signal because of its brightness and perhaps also because the secondary oscillates with a much lower amplitude.

\begin{figure}
    \centering
    \includegraphics[width=\linewidth]{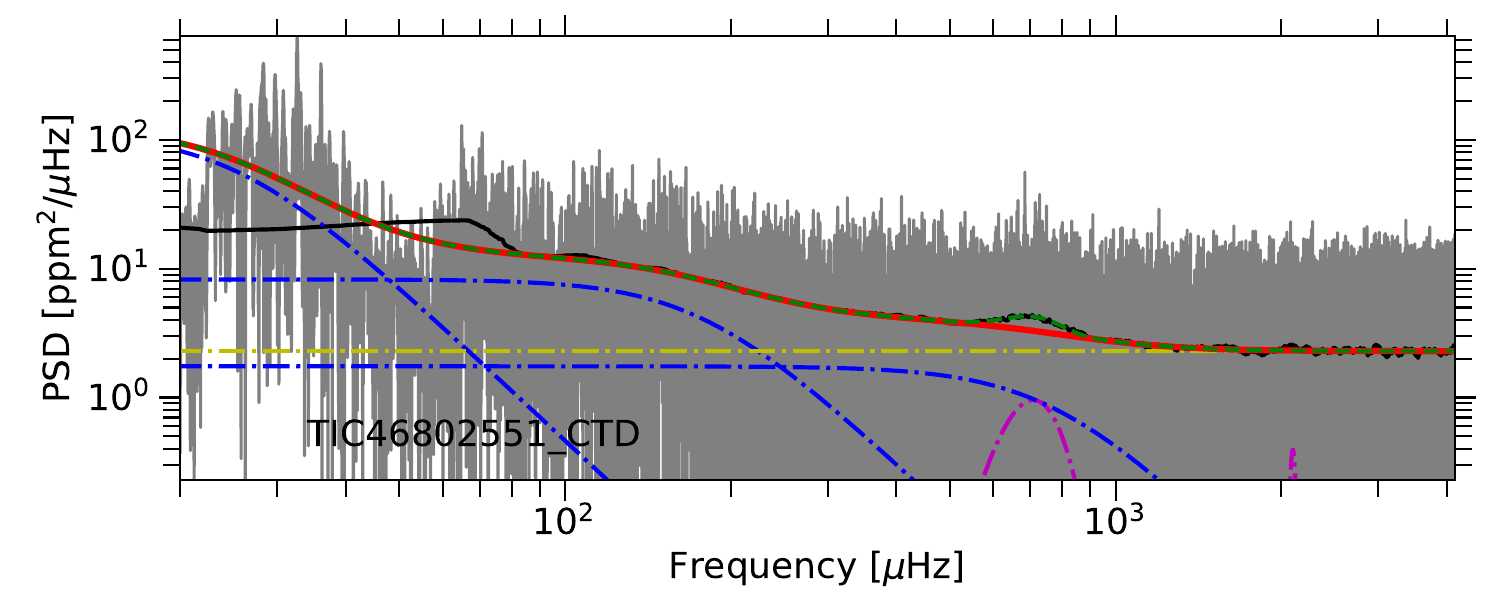}  
    \caption{Power spectral density (gray) for HD~81809 observed by TESS during three different years (2021, 2023, and 2025) for 27 days each in 120\,s. The excess of power (green line) is calculated by subtracting all the background noises which include white and colored noise. The estimated background level is shown in red, while the inclusion of the Gaussian power excess is marked by a green-dashed line. The individual Harvey-like components are plotted in blue, while the level of the white noise is indicated by a dot-dashed yellow line.}
    \label{fig:spectrum}
\end{figure}

Nonetheless, a careful analysis of the seismic light curve using the public tool DIAMONDS developed by \citep{Corsaro14} allowed us to detect an excess of power due to solar-like oscillations not only for the bright primary component, located at $\simeq 700\,\mu$Hz, but also for the faint secondary component, which thanks to the adoption of the concatenated time-series in short cadence, from 2021 to 2025, was found to be detectable around $\simeq 2100\,\mu$Hz. In this way it was possible to identify the global pulsation parameters of both components.
%(Corsaro
%and De Ridder, 2014) 
The resulting global oscillation properties derived for HD~81809 are shown in Table \ref{tab:astero}. The primary component shows the frequency of maximum oscillation power $\nu_\mathrm{max} = 708.74^{+3.23}_{-3.74}\,\mu$Hz, with a corresponding maximum amplitude of oscillation $ A_\mathrm{max} = 2.84 \pm 0.95$\,ppm, and a value for the large frequency separation of $\Delta \nu = 43.32 \pm 3.91\,\mu$Hz derived using the methodology presented in \cite{Corsaro24}, which is also well compatible with the results previously published by \cite{Corsaro24}, who used only 1 sector of TESS short cadence data. For HD~81809\,B we  obtain $\nu_\mathrm{max} = 2098.07_{-2.83}^{+3.07}\,\mu$Hz and $\Delta\nu = 97.75 \pm 4.49\,\mu$Hz, the latter also derived through an auto-correlation function as presented in \cite{Corsaro24}.
The detection of the power excess for the secondary component was performed by means of a Bayesian model comparison, following the approach described by \cite{Corsaro24}, where the resulting background fitting model is shown in Figure~\ref{fig:spectrum}, consisting of three Harvey-like components, a white noise, and two separate Gaussian envelopes.
It is clear that a set of individual oscillation frequencies would have
helped to better constrain this target. For this purpose we also attempted to identify individual oscillation modes through the automated pipeline \textsc{FAMED} \citep{Corsaro20FAMED}, but the high complexity of the oscillation pattern caused by the presence of avoided crossings and the low signal-to-noise ratio of the region of the power excess could not lead us to a solution.

%From the use of seismic scaling relations in \citep{Bellinger2019}, we can derive an estimation of the stellar mass and radius by using $T_{eff}$, $\rm \nu_{max}$ and $\rm \Delta_{\nu}$ introduced above, obtaining for the primary the values $M_{ast}/M_{\odot} = 0.84 \pm 0.032$, $R_{ast} /R_{\odot} = 1.86 \pm 0.011$.

%\newpage
\begin{table}[ht]
\centering
\caption{Asteroseismic global parameters from the analysis of TESS time series.}
\label{tab:astero}
\begin{tabular}{lcc}
\hline
Parameter & HD 81809\,A  & HD 818091\,B \\
\hline
%$R/R_{\odot}$     & $ 2.07 \pm 0.14  $ & $ 1.145 \pm 0.05 $ \\
%$\log g$          & $ 3.77 \pm 0.05  $ & $ 4.22 \pm 0.03  $ \\
$\nu_{\rm max } (\mu \mathrm{Hz}) $ & $ 708.74^{+3.23}_{-3.74}   $ & $ 2098.07^{+3.07}_{-2.83}$ \\
$\Delta \nu (\mu \mathrm{Hz})$& $ 43.32 \pm 3.91   $ & $ 97.75 \pm 4.49   $ \\
$A_{ \rm max}$ (ppm)     & $ 2.84 \pm 0.95  $ & -- \\
\hline
\end{tabular}
\end{table}

\section{Physical parameters through stellar modeling} 
\label{sec:models}
\subsection{Best-fitting procedure }
To characterize the properties of the two stellar components in the HD~81809 system, we employed a global minimization procedure based on classical spectroscopic, asteroseismic, and observed luminosity constraints derived in the previous Sections. The adopted minimization procedure employs the Stellar Parameters Inferred Systematically software \citep[SPInS,][]{Lebreton2020}, which is based on a Markov chain Monte Carlo (MCMC) approach and Bayesian statistics to derive probability distribution functions for the stellar fundamental parameters (age, mass, radius, initial composition). In this context, we used two grids of stellar models for the derivation of the fundamental parameters with SPInS: one computed by means of the Code Liegeois d'Evolution Stellaire \citep[CLES;][]{Scuflaire2008}, and one computed by means of the Modules for Experiments in Stellar Astronomy \citep[MESA;][]{Paxton2013,Paxton2015,Paxton2018,Paxton2019}.

To determine the fundamental parameters and best-fit models of HD~81809A and B, given the remarkably large difference in their chemical composition and Fe content (as shown in Table~\ref{tab:abund}), we decided to perform the modeling of the two components independently. Modeling the two components together in the assumption of a common initial chemical composition and age, would lead to an inferred initial abundance of helium $\rm Y_{i} \sim 0.36$, which is unrealistically large. In this context, two main explanations can be considered to interpret this result: the physics included in the models lacks accuracy, for instance in the computation of the opacity tables or implementation of the transport mechanisms, preventing the correct reproduction of the properties of the two components; the two stars may have formed as binary system, and subsequently accreted metal-rich material from a debris disk (as revealed from the SED), that would have been mixed within the stellar interiors in a different manner, given the diverse structure and evolutionary stage of the two components; it is also possible that the stars formed in chemically different environments, and that they merged into a binary system after formation, due to dynamical interaction. Given the impossibility to select the most probable hypothesis concerning the formation and evolution of the system, we decided to proceed in the modeling by considering one star at the time, and attempt to find physically meaningful solutions.

As first step, for each component we attempted to determine their fundamental parameters ($M, R, Age, ...$) by considering the whole set of observational constraints derived in the above sections, namely $L$ from Table~\ref{tab:params}, $T_{\rm eff}$, $\mathrm{log}~g$, and $\rm [Fe/H]$ from Table~\ref{tab:atmos}, $\rm \nu_{max}$, and $\rm \Delta \nu$ from Table~\ref{tab:astero}, as input for SPInS. Again, also in this case, SPInS failed to converge towards a physically meaningful solution for either component, in any of the considered stellar model grids presented above. For instance, in the modeling of the primary, we found a solution that would fit the observed bolometric luminosity, but overestimating the mass by about $\sim 0.3M_{\odot}$ and the effective temperature by approximately $300~K$ (with respect to the value indicated in Table~\ref{tab:atmos}. While these values for the mass and effective temperature are compatible with the ones found by \cite{Fuhrmann2018}, they result to be in disagreement with the orbital masses deduced in Section \ref{sec:orbit}. A similar issue is encountered when modeling the secondary component, for which the software tends to converge toward solutions with unrealistically advanced ages, and/or unrealistically large $\rm Y_{i}$.

Thus, we decided to exclude the bolometric luminosity from the set of observational constraints, and in the case of the secondary, we applied two modeling approaches:

\begin{itemize}
    \item (B1) Secondary formed with an initial metallicity independent from primary, constrained by observed $\rm [Fe/H] = 0.00 \pm 0.11$;
    \item (B2) Secondary formed with the $\rm [M/H]_{0}$ inferred from modeling the primary, under the hyphotesis of same progenitor cloud.
\end{itemize}

In the following subsections, we provide a detailed description of the optimal models for HD~81809\,A and B obtained with the two aforementioned stellar evolution codes.

\begin{table*}[ht!]
\caption{Stellar fundamental parameters for HD~81809\,A and B from stellar modeling employing two different evolutionary codes. For the secondary we considered two different solutions: B1 using the observed [Fe/H] (see Table \ref{tab:atmos} and B2 using the [M/H]$\rm_0$ deduced bu modeling the primary.
}
\label{tab:models}

\centering
\resizebox{0.8\textwidth}{!}{%
\begin{tabular}{c|cccc}
\hline
& & HD 81809 A & \multicolumn{2}{c}{HD 81809 B} \\ \cline{2-5}
Code & Parameters & & B1 & B2 \\ \hline
& Age (Gyr)                 &  9.75   $\pm$ 1.78    & 12.84   $\pm$ 2.96    & 11.13   $\pm$ 1.51\\
&$\rm [M/H]_0$ (dex)       & -0.3898 $\pm$ 0.3169  & 0.1387  $\pm$  0.2318 &  -0.1390 $\pm$ 0.2149\\
&$M~(\rm M_\odot)$         & 0.87  $\pm$ 0.08  & 0.94 $\pm$ 0.07  & 0.83  $\pm$ 0.01\\
&$R~(\rm R_\odot)$         & 1.95  $\pm$ 0.08  & 1.18 $\pm$ 0.04  & 1.12  $\pm$ 0.03\\
&$T_{\rm eff}$ (K)         & 5619    $\pm$ 125     & 5586   $\pm$ 116      & 5833    $\pm$ 72\\
CLES & $log(g)$ (dex)            & 3.7981  $\pm$ 0.0054  & 4.2682 $\pm$ 0.0454  & 4.2776  $\pm$ 0.0027\\
& $\rm [Fe/H]$ (dex)        & -0.4857 $\pm$ 0.1426  & 0.0510 $\pm$ 0.0995  & -0.2312 $\pm$ 0.0583\\
& $L~(\rm L_\odot)$         & 3.44  $\pm$ 0.43  & 1.2 $\pm$ 0.14 & 1.31  $\pm$ 0.10 \\
& $\rm \Delta\nu\ (\mu HZ)$ & 46.32   $\pm$ 1.05    & 102.43 $\pm$ 1.81    & 106.21  $\pm$ 1.29\\
& $\rm \nu_{max}\ (\mu HZ)$ & 708.58  $\pm$ 3.76    & 2098.04$\pm$ 3.09    & 2097.95  $\pm$ 3.07\\
& $Z_i$                     & 0.0051  $\pm$ 0.0015  & 0.0177 $\pm$ 0.0038  & 0.0092  $\pm$ 0.0011 \\
& $Y_i$                     & 0.3002  $\pm$ 0.0318  & 0.2737 $\pm$ 0.0294 & 0.2866  $\pm$ 0.0243 \\ \hline

& Age (Gyr) & 10.21$\pm$2.71 & 15.77$\pm$4.11 & 10.68$\pm$0.48\\
&$\rm [M/H]_0$ (dex) & -0.4599$\pm$0.1313 & 0.1087$\pm$0.0999 & -0.3712$\pm$0.0484\\
&$M~(\rm M_\odot)$ & 0.88$\pm$0.10 & 0.89$\pm$0.08 & 0.83$\pm$0.01\\
& $R~(\rm R_\odot)$ & 1.97$\pm$0.10 & 1.16$\pm$0.05 & 1.09$\pm$0.01\\
& $T_{\rm eff}$ (K) & 5539$\pm$119 & 5466$\pm$119 & 6044$\pm$41\\
MESA & $\log(g)$ (dex) & 3.7870$\pm$0.0052 & 4.2565$\pm$0.0048 & 4.2781$\pm$0.0016\\
& $\rm [Fe/H]$ (dex) & -0.5140$\pm$0.1382 & 0.0467$\pm$0.1038 & -0.4503$\pm$0.0507\\
& $L~(\rm L_\odot)$& 3.33$\pm$0.51 & 1.09$\pm$0.16& 1.28$\pm$0.03 \\
& $\rm \Delta\nu\ (\mu HZ)$ & 45.62$\pm$1.17 & 102.00$\pm$2.05 & 107.74$\pm$0.62\\
& $\rm \nu_{max}\ (\mu HZ)$ & 708.53$\pm$3.70 & 2097.94$\pm$3.05 & 2097.80$\pm$3.06\\
& $Z_i$ & 0.0047$\pm$0.0015 & 0.0172$\pm$0.0039 & 0.0067$\pm$0.0006 \\
& $Y_i$ & 0.2875$\pm$0.0279 & 0.2753$\pm$0.0254 & 0.2856$\pm$0.0051\\ \hline

\end{tabular}%
}
\end{table*}

\begin{table*}[ht]
\caption{Stellar fundamental parameters for the best fit models of
HD\,81809\,A and B obtained
with two different evolutionary codes.}
\label{tab:best_models}
\begin{center}
\resizebox{\textwidth}{!}{%
\begin{tabular}{lcccccccccccccc}
\hline
Model & Component & $\chi^2$ & $M/M_\odot$ & $R/R_\odot$ & $L/L_\odot$ & $T_\mathrm{eff}$ (K) & $\log(g)$ (dex) & $\Delta\nu$ ($\mu$Hz) & $\nu_\mathrm{max}$ ($\mu$Hz) & $r_{cz}/R_*$ & $X_c$ & $(Z/X)_s$ & Age (Gyr) & $(Z/X)_0$ (dex) \\ \hline
\multirow{3}{*}{CLES} & A & 0.35 & 0.93 & 2.02 & 3.55 & 5570 & 3.79 & 45.40 & 710.85 & 0.57 & 1.61$\times10^{-23}$ & 0.0047 & 9.58 & 0.0064 \\
& B1  & 0.71 & 0.95 & 1.19 & 1.24 & 5580 & 4.27 & 101.67 & 2096.99 & 0.67 & 1.74$\times 10^{-3}$ & 0.015 & 14.31 & 0.023\\
 & B2 & 18.0 & 0.87 & 1.12 & 1.35 & 5886 & 4.28 & 106.22 & 2096.24 & 0.72 & 8.04$\times 10^{-3}$ & 0.0081 & 11.11 & 0.013  \\ \hline
 
\multirow{3}{*}{MESA} & A & 0.23 & 0.96 & 2.07 & 3.649 & 5546 & 3.79 & 44.56 & 708.90 & 0.524  & 9.68$\times10^{-44}$ & 0.0050 & 9.10 & 0.0055 \\
 & B1 & 1.88 & 0.95 & 1.20 & 1.165 & 5477 & 4.26 & 100.31 & 2098.10 & 0.640 & 6.33$\times10^{-4}$ & 0.0194 & 14.53 & 0.0220  \\
& B2 & 3.50 & 0.83 & 1.09 & 1.435 & 6050 & 4.28 & 107.85 & 2098.08 & 0.707  & 3.00$\times10^{-3}$ & 0.0066 & 10.69 & 0.0077 \\
 \hline
\end{tabular}}
\end{center}
\end{table*}

\subsection{Derivation of fundamental stellar parameters based on CLES models}
\label{Sec:mini-CLES}

In this section we provide a brief description of the input physics included in the CLES models for the computation of the grid provided to SPInS. For a more detailed description of the physics included in the code, we refer the interested reader to \cite{Scuflaire2008}. The setup of the input physics employed for the computation of the CLES stellar tracks included: FreeEOS \citep{Irwin2012} equation of state, AGSS09 \citep{Asplund2009} abundances, specifically for the primary component we considered $\alpha$-enhanced ($\rm \langle[\alpha/Fe]\rangle = +0.3$) abundances, and OPAL \citep{IglesiasRogers1996} opacity tables (for $\alpha$-enhanced solar mixture in the case of the primary star). The effects of microscopic diffusion \citep{Thoul1994, Paquette1986} and turbulence at the base of the convective envelope were also taken into account. For the outer boundary conditions we employed \citet{Vernazza1981}. The classical mixing-length theory was applied for convection, with solar calibrated values $\rm \alpha_{MLT} = 1.93$. 

The results obtained with the modeling procedures for HD~81809A and B are indicated on Table \ref{tab:models}, under the flag ``CLES''. The best fit parameters found for the primary component are: $M = (0.87 \pm 0.08) {\rm M}_{\odot}$, Age\,(Gyr) = $\rm 9.75 \pm 1.78$, $R=(1.95 \pm 0.08){\rm R}_{\odot}$, and initial metallicity $\rm [M/X]_0 = -0.3898 \pm 0.3169$. The solution we found is able to fit all the observational constraints, except for the lumnosity, as explained in the above section. The properties of the best-fit stellar model for HD~81809A are indicated in Table~\ref{tab:best_models}.

The two minimization procedures (B1 and B2) applied to the secondary component yield models with significantly different stellar parameters, among which particularly different are the stellar masses. When adopting the observed surface metallicity of HD~81809~B as constraint, labeled as case B1, the model converges towards the observed effective temperature and global seismic parameters. The observed luminosity is not reproduced. However, the inferred mass results $M_B=(0.94\pm0.07)M_{\odot}$, which is formally larger than the mass derived for the primary component. The resulting mass difference between model A and model B1 is $\Delta M=0.07 {\rm M}_{\odot}$ which corresponds to an inconsistency at the $\sim 0.7\sigma$ level. Although this difference can be considered not statistically significant, this  formally implies that the hierarchical configuration of the system is physically implausible, as the secondary component is expected to be less massive than the primary.

In contrast, the minimization assuming the initial $ {\rm[}M/H{\rm ]}_0$ inferred from the modeling of the primary yields a mass of $M_B = (0.83\pm0.01){\rm M}_\odot$, lower than that of the primary and therefore fully consistent with the binary configuration. Nevertheless, this solution overestimates both the effective temperature and the large frequency separation.

\begin{figure}
    \centering
    \includegraphics[width=0.49\linewidth]{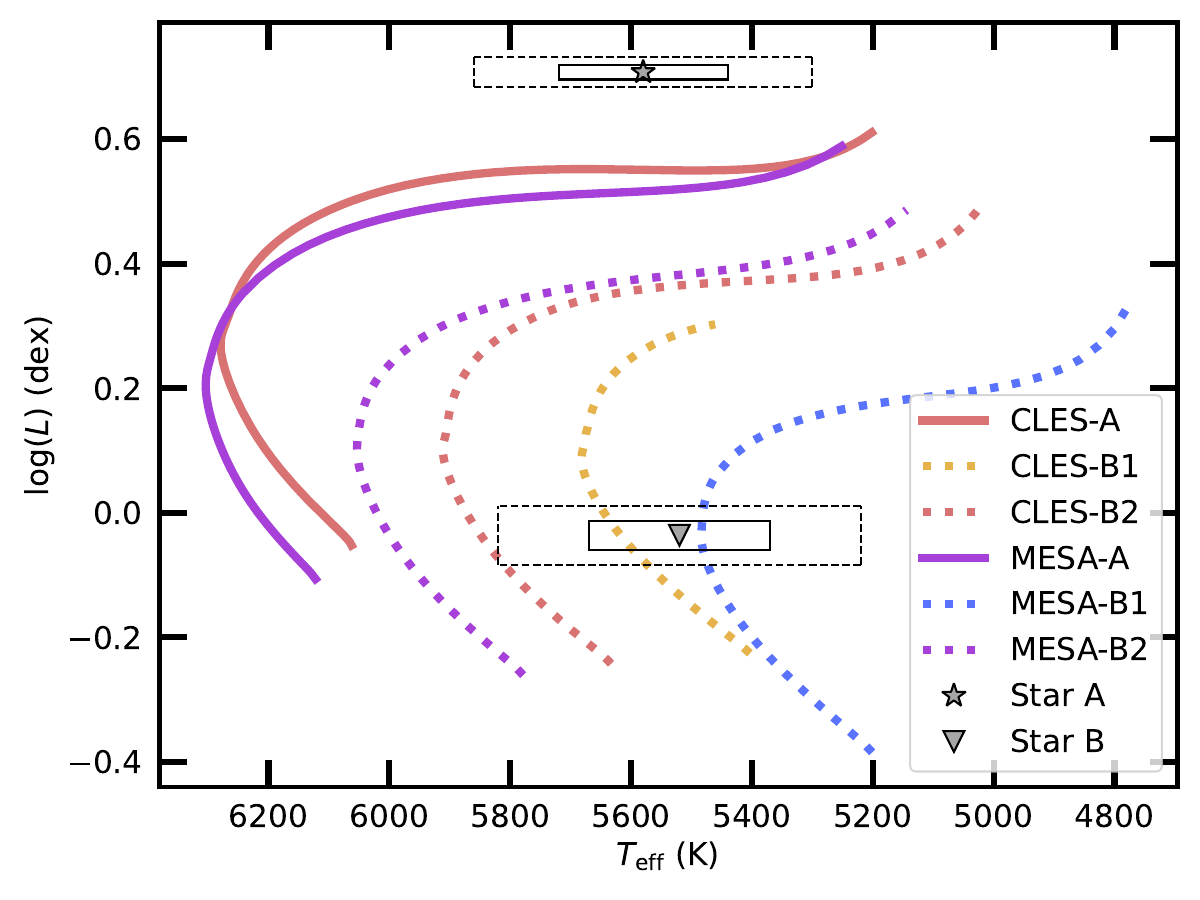}
    \includegraphics[width=0.49\linewidth]{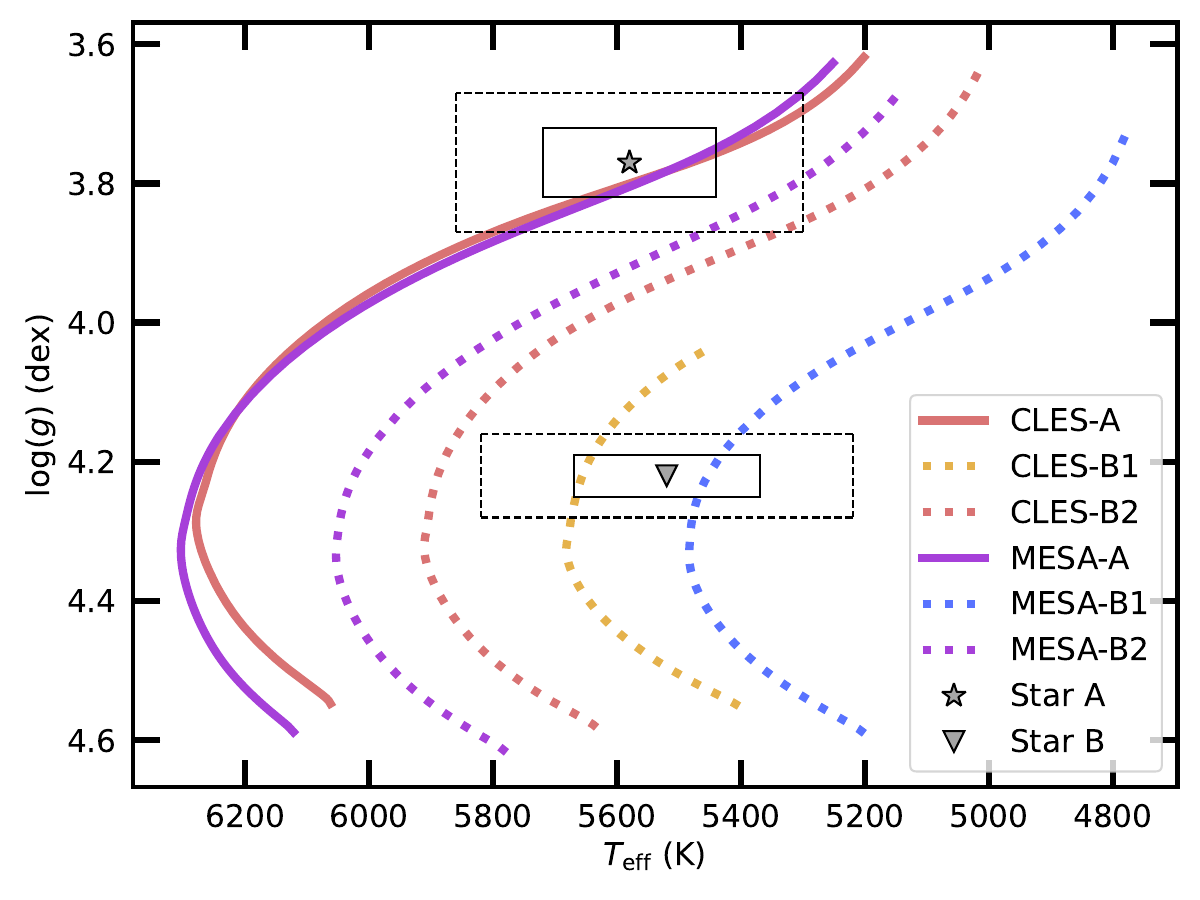}
    \includegraphics[width=0.49\linewidth]{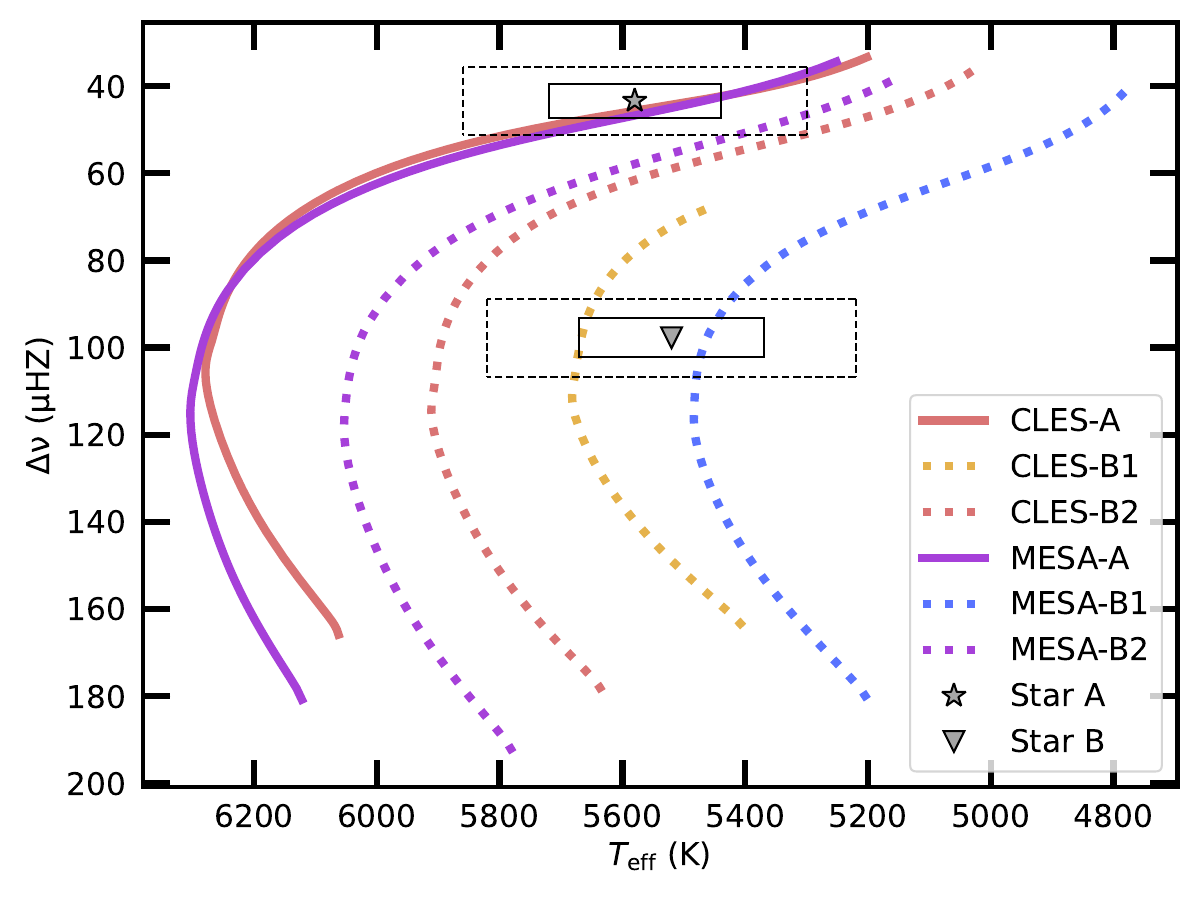}
    \caption{Hertzsprung-Russel (top left panel), Kiel (top right panel) and seismic (bottom panel) diagrams showing the evolutionary tracks of the best-fit stellar models for HD\,81809~A (solid lines)
    and HD\,81809~B (dotted lines) with parameters given in Table \ref{tab:best_models}.
    The star and the triangle shows the correspondent observed  parameters for the two components and the solid and the dashed-lines boxes correspond to the 1 and 2 $\rm \sigma$ errors respectively.}
    
    \label{Fig:HR}
\end{figure}

In Figure~\ref{Fig:HR}, we show the evolutionary tracks obtained from best-fot models of HD\,81809~A, HD\,81809~B1 and HD\,81809~B2.

\begin{figure}
    \centering
    \includegraphics[width=0.49\linewidth]{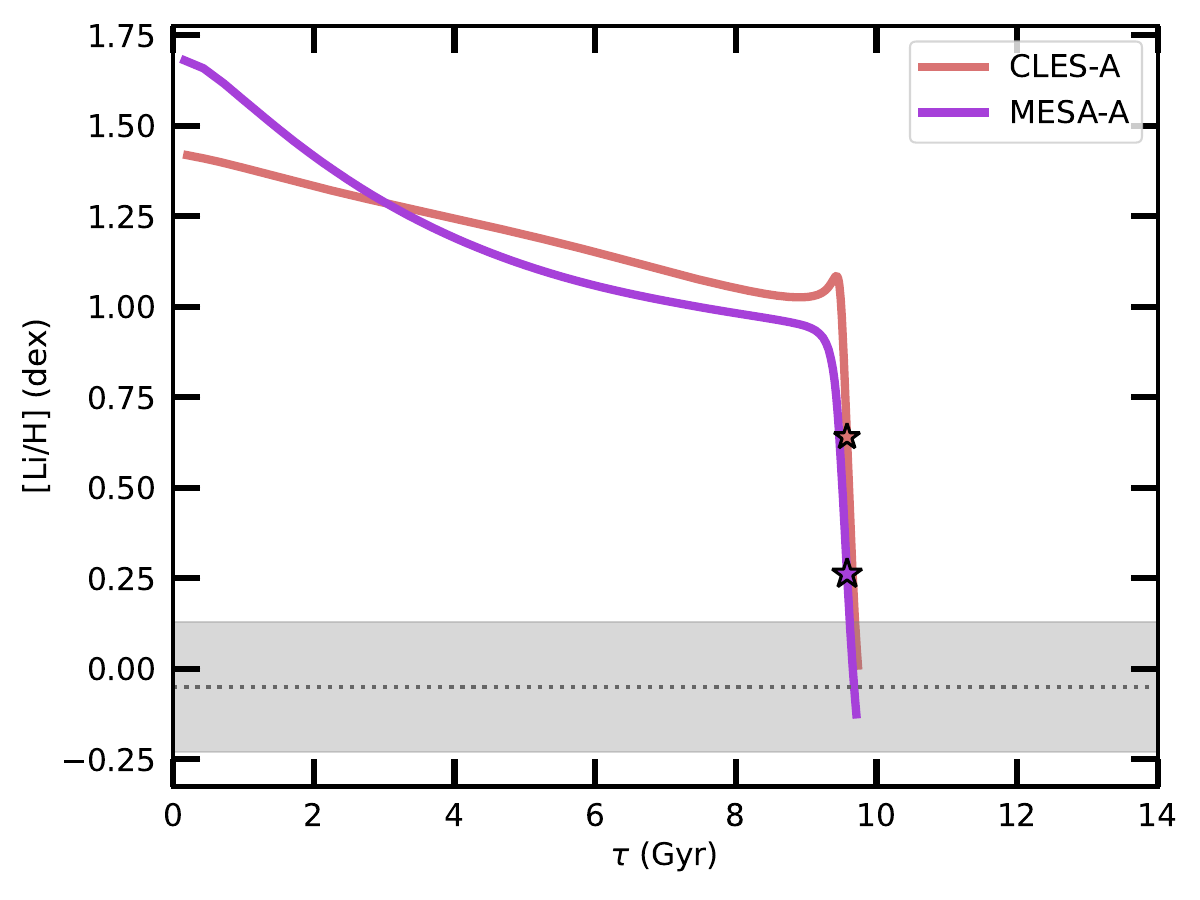}
    \includegraphics[width=0.49\linewidth]{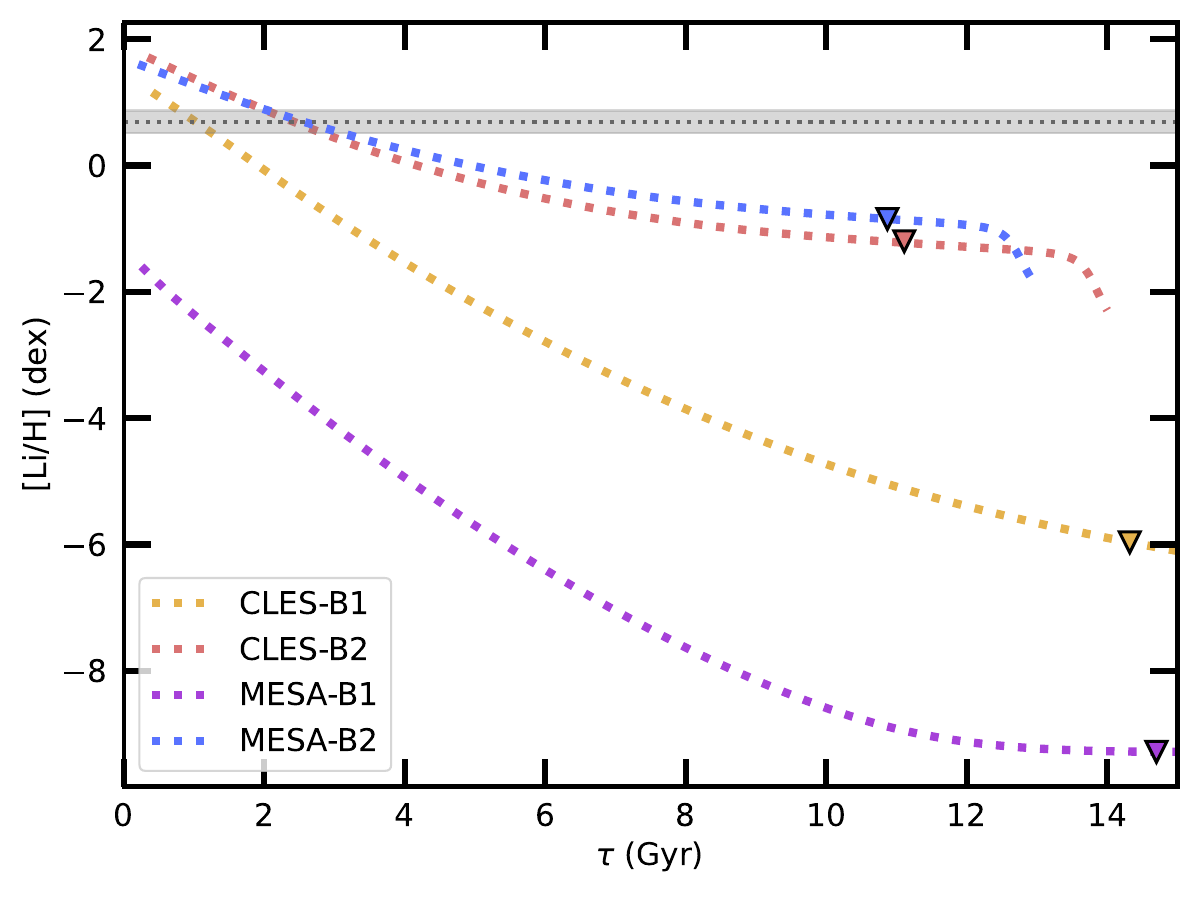}
    \caption{Lithium evolution in CLES and MESA models for HD 81809\,A (left panel) and HD 81809\,B (right panel). The black dotted line shows the observed [Li/H] value, and the gray region shows its uncertainty. The stars and triangles indicate location of the best-fit models of Table \ref{tab:best_models}.}
    %The blue star shows the location of HD81809B, by considering the luminosity derived in Sec.~\ref{sec:fund_params} and the $T_{eff}$ determined through the minimization procedure. 
    \label{Fig:Li_Model}
\end{figure}

\subsection{Fundamental stellar parameters by using SPInS-MESA}

 In order to confirm the obtained results, we performed a characterization of 
 HD\,81809 using stellar models computed by Modules for Experiments in Stellar Astronomy \citep[MESA;][]{Paxton2013,Paxton2015,Paxton2018,Paxton2019}.
  We produced the global minimization by applying SPInS to a
  stellar models grid with
  characteristics similar to those of the grid C of \cite{Moedas2025}, extended to lower metallicity to cover the parameters space of the studied stars. 
  The input physics differs from that of the CLES stellar modeling, because here we adopted the OPAL2005 equation of state \citep{Rogers2002}, and no $\alpha$-enhancement for the primary.

 % We performed three runs, first (Model MESA-1) that considered
 %5 input constraints:
 % the seismic parameters $\Delta_{\nu}$ and $\nu_{max}$, and the spectroscopic parameters $T_{eff}$, $\rm [Fe/H]$ and $\log g$. In the second (Model MESA-2), we also include $L_{SED} $. In the final run (Model MESA-3), we used $L_{bol_A}$ and the $L_{bol_B}  $ the bolometric luminosities of the two components. 
  
%  With MESA it is possible to model the companion in all runs.
  %, even if
%HD\,81809B does not present enough constraints to be correctly fitted.  
%In fact, SPInS allows us to fit both components as binary imposing the prior that both components present the same age and initial chemical composition (same $\rm [Fe/H]_0$ and $Y_0$). Note that we performed the fit of the HD 81809A individually and with the companion, presenting similar results. 

%The results  are presented in Table~\ref{sec:models}
%and can be summarized as follows:
%by using only 5 constraints without luminosity
%we obtained Model MESA-1 with an older and less massive primary star; with 5 constraints and the SED luminosity we obtained Model MESA-2 with a more massive and very young star.
%This is more massive than what was expected, but this is due to the luminosity count the contribution of both stars. 
%The star is brighter than what it is, leading to the preferred models toward more massive stars. 
%By producing a stellar modelling fitting 5 constraints and the bolometric luminosities, we obtained Model MESA-3, obtaining an intermediate picture when comparing the mass and age to the two previews models. 

The results, shown in Table \ref{tab:models} for the primary component are: $M = (0.88 \pm 0.10) {\rm M}_{\odot}$, Age\,(Gyr) = $\rm 10.21 \pm 2.71$, the stellar radius $R=(1.97\pm0.1){\rm R}_{\odot}$ and initial metallicity $\rm [M/X]_0 = -0.4599 \pm 0.1313$. The properties of the best-fit stellar model are indicated in Table~\ref{tab:best_models} with the flag ``MESA''.  The convective envelope is deeper than that of the Sun with a base located at $r_{cz}=0.52R$. As already pointed out for the CLES models, the observed and modeled luminosities differ by about $3.3\sigma$, which is statistically significant.

The two minimization procedures applied to the secondary component yield  models with different stellar parameters (see Table \ref{tab:models}).
 When adopting the observed surface metallicity of HD~81809~B, the model converges toward the observed luminosity, effective temperature and global seismic parameters. However, the inferred mass results $M_B=(0.89\pm0.08)M_{\odot}$, which is formally larger than the mass derived for the primary component,
with a difference between MESA models A and  B1 of 
$\Delta M=0.01 {\rm M}_{\odot}$.
 This indicates that the masses are statistically indistinguishable and hence consistent with the expected binary configuration. The inferred age results $15.77\pm 4.11$ Gyr, with rather large uncertainty that likely overestimates the true error and formally exceeds the age of the Universe. This apparent overestimation likely reflects the large uncertainties in the observational constraints, degeneracies between stellar parameters. Therefore, while the age indicates an old system, the absolute value obtained with MESA should be interpreted with caution.

The minimization assuming the initial $ {\rm[}M/H{\rm ]}_0$ inferred from modeling the primary yields a mass of $M_B = (0.83\pm0.01){\rm M}_\odot$, lower than that of the primary and therefore fully consistent with the binary configuration, an
 Age\,(Gyr) = $\rm 10.68 \pm 0.48$, and a stellar radius $R=(1.09\pm0.01){\rm R}_{\odot}$ (see Table \ref{tab:models}.

\subsection{The modeling results: discrepancies, preferred solutions}  

According to the present modeling obtained with two different evolutionary codes, the system is composed by a main component with a mass $M_A \sim 0.88 M_{\odot} $, $R_A \sim 1.96R_{\odot}$ that has exhausted the hydrogen content in the core, has already passed the turn-off, and is currently on the subgiant branch (see Figure~\ref{Fig:HR}) as indicated by the hydrogen content in the core $X_c$. The convective envelope is deeper than that of the Sun with a base located at $r_{cz} \sim (0.5-0.6)R$. The slightly smaller mass companion is instead in the main-sequence phase.

The results obtained for the two components, as an average of the results obtained with the two codes, indicate that the system is old with an Age$\sim 10$\,Gyr, as expected from the observed high value of [Mg/Fe] of the primary. This result is in disagreement with the one obtained by \citet{Fuhrmann2018} (see Section \ref{sec:intro}).

The Lithium depletion profiles, shown for the best models from CLES and MESA in Figure \ref{Fig:Li_Model}, provide additional information on the age of the stars. For HD\,81809A, the tracks show a very steep Li depletion occurring during the subgiant phase, in a relatively short evolutionary time. The observed stellar value, indicated by the gray region, is consistent with an age $\sim 10$ Gyr. The Li tracks of the secondary (right panel of Figure \ref{Fig:Li_Model}) show that this star due to its low mass, should have lost its initial Li during the main sequence life. At the contrary, the observed abundance of Li shows that it is still relatively Li-rich, with an observed [Li/H]$=0.69$ dex. The possible explanations are: planet engulfment, suppressed depletion due to low rotational mixing or unusual internal structure.

Nevertheless, the comparison between the observed and modeled properties of HD~81809 reveals other notable differences, particularly for the primary component. The observed luminosity, $L/{\rm L}_{\odot} = 5.10 \pm 0.14$, is significantly higher than the one predicted by both CLES ($L/{\rm L}_{\odot}=3.44 \pm 0.43$) and MESA ($L/{\rm L}_{\odot} =3.33 \pm 0.51$) models, corresponding to about $3.5\sigma$ discrepancies. This difference might likely arise from a combination of factors, including uncertainties in the input physics of stellar models (e.g., opacities, convective efficiency, composition, treatment of $\alpha$-enhanced composition), rapid luminosity evolution in the subgiant phase, and potential systematic uncertainties in the observed luminosity. In particular, the non-solar $\alpha$-enhanced composition of HD~81809~A suggests that standard solar-scaled models may not fully capture the stellar structure, indicating caution when interpreting absolute luminosities.

For the secondary component, the modeling considering the observed surface metallicity (B1 solution) yields slightly overestimated masses, exceeding the primary’s mass (with CLES $M_B=(0.94\pm0.07){\rm M}_\odot$, with MESA $M_B=(0.89\pm0.08){ \rm M}_\odot$). This value is physically inconsistent for a hierarchical binary.

Under the assumption of a common initial chemical composition for the system, adopting as initial metallicity the one derived from the primary (B2 solution), the best-fitting produces masses fully consistent with the expected binary hierarchy ($M_B\sim0.83\pm0.01\,M_\odot$ with both the evolutionary codes). Notably, this value is in good agreement with the independently observed dynamical masses derived from the orbital analysis and also with the ratio measurement $M_A/M_B$ given in Table \ref{tab:rv}. Nevertheless, we point out that this solution led to models that do not fully reproduce the observed luminosity, effective temperature, and the asteroseismic large separation.

\section{Analysis of Stellar Magnetic activity} 
\label{sec:magnetic}
HD 81809 was part of the largest and longest observational campaign to monitor the magnetic activity of Sun-like stars, conducted at the Mount Wilson Observatory \citep[MWO,][]{Wilson1968, Wilson1978}. The available dataset, which can be accessed at \url{https://dataverse.harvard.edu/dataverse/mwo_hk_project}, consists of 1879 individual measurements obtained between 1966 and 2001 using the HKP spectrophotometer at the MWO.
The data are expressed in terms of dimensionless S-index, which represents the most widely used stellar activity proxy. The MWO S-index is based on the intensity of emission in the chromospheric Ca II H (393.3 nm) \& K (396.8 nm) lines cores normalized to that two nearby continuum reference bandpasses, as follows:

\begin{equation}
    S_{MWO} = \alpha \frac{N_H + N_K}{N_R + N_V}
\label{S_index_MWO}
\end{equation}
where $N_H$ and $N_K$ are the counts in 1.09 \r{A} triangular bands centered on Ca II H \& K respectively, while $N_R$ and $N_V$ are 20 \r{A} reference bandpasses in the nearby continuum region, and $\alpha$ is a calibration factor \citep[see][for further details]{Vaughan1978, Wilson1978}.

Monitoring the chromospheric activity of this star was subsequently continued with the Solar-Stellar Spectrograph (SSS) at the Lowell Observatory \citep{Hall1995}, resulting in 362 additional measurements over the time interval 1993-2018. Thanks to the combined dataset from the two observatories, it has been possible to study the magnetic activity of this star over nearly 52 years (1966-2018), making it one of the stars whose magnetic activity has been most closely monitored. The presence of a time interval with overlapping data from both MWO and Lowell SSS allows us to inter-calibrate the two datasets. To this aim, we first removed outliers from both datasets, excluding those outside $\pm4\sigma$ of the average, and then applied the relation provided by \citet{Egeland2017_thesis} to calibrate the Lowell SSS S-index to the MWO S-index scale.

\begin{figure}
    \centering
  \includegraphics[width=\linewidth]{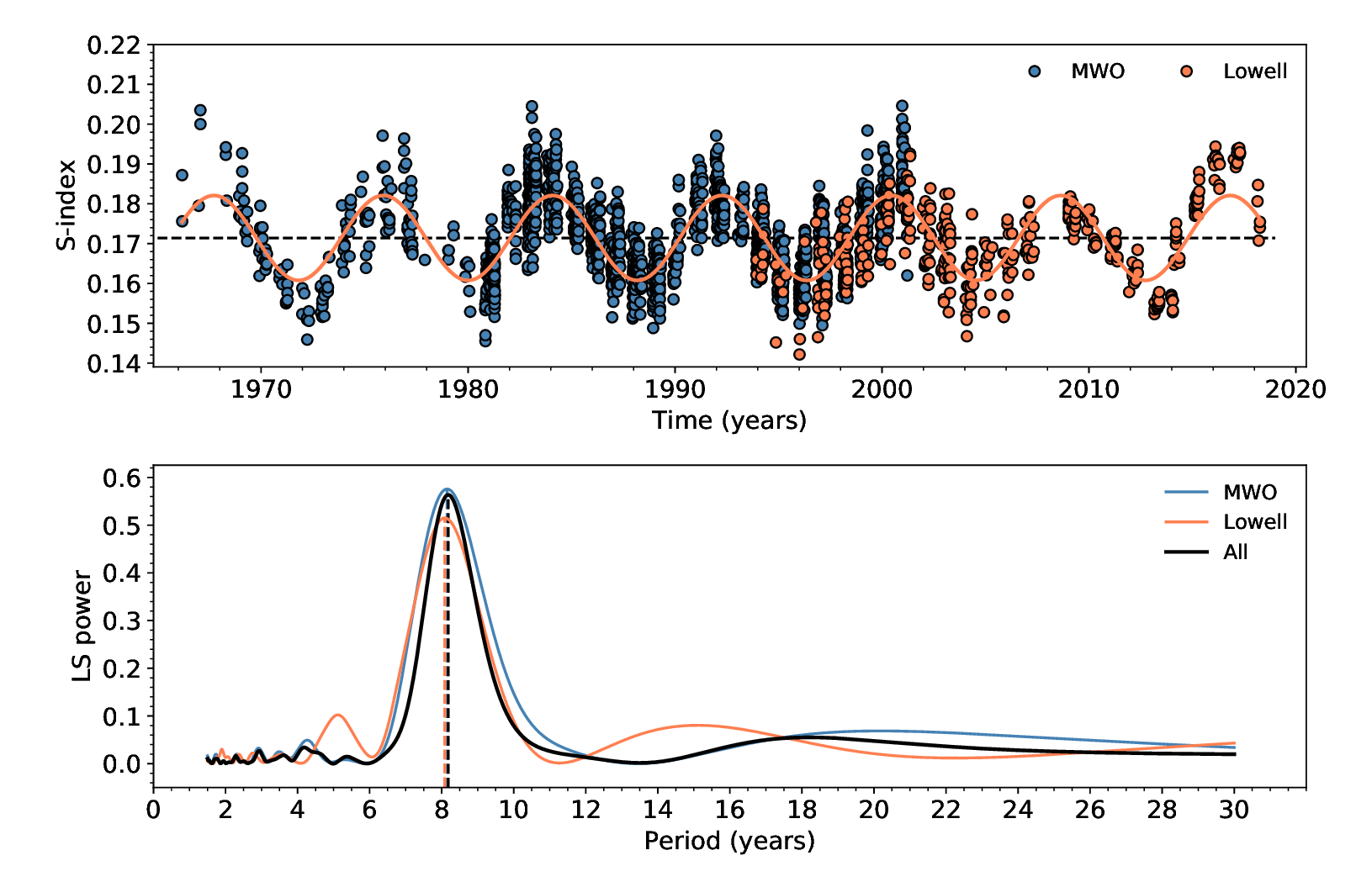}
    \caption{Top: S-index measurements of HD 81809 from MWO (blue points) and Lowell SSS (orange points), with the dashed black line indicating the average S-index value. The orange curve represents the sinusoidal fit to the highest peak of the periodogram, as shown in the bottom panel. Bottom: Lomb-Scargle periodogram of the MWO and Lowell SSS time series, along with the periodogram obtained by merging the two datasets (black line).}
    \label{Fig:S_index}
\end{figure}

The S-index time series plotted in the top panel of Figure \ref{Fig:S_index} indicates that from the MWO data $S_{MWO}=0.172\pm 0.010$, while from the Lowell SSS one (once calibrated to the MWO scale) $S_{Lowell}=0.170\pm0.010$, which confirms the previous estimate by \citet{Egeland2018}.
It is interesting to note that while this star shows higher variability in the S-index ($\sim$ 0.05) - i.e., the difference between the maximum and minimum values - compared to the Sun during a solar cycle ($\simeq 0.02$),
its mean activity level is consistent with the average solar S-index value over the last 10 solar cycles $S=1.01 S_{\odot}$ \citep[see,][for the solar value]{Egeland2017}. 
To analyze the activity cycle of this star, which appears to exhibit clear periodicity based on the S-index measurements, we performed a Lomb-Scargle analysis \citep{Lomb1976, Scargle1982}. The results of this analysis (Figure \ref{Fig:S_index}), performed separately on both the MWO and Lowell time series, indicate a main cycle of 8.15 years for the former and 8.09 years for the latter. 
%When the two time series are merged,
%, the analysis reveals a main cycle of 8.18 years. 
%The results from 
The results are highly consistent, and when the two time series are merged, the analysis indicates the presence of an unambiguous main activity cycle of $\sim 8.18$ years.

The result is in good agreement with the periodicity of $7.3 \pm 1.5$ years obtained by \citet{Orlando2017} for the X-ray luminosity, based on XMM-Newton data. However, it should be noted that the shorter average cycle observed in the X-ray activity may be influenced by the fact that it is derived from only 15 years of data (November 2001-October 2016), a time interval considerably shorter than the 50 years of chromospheric activity. The clear periodical signal retrieved for HD 81809 indicates that the variability is dominated by only one component of the binary. The other component has either flat or a negligible activity with respect to the primary. 

%As noticed by \citet{Egeland2018} the chromospheric magnetic period appears higher than the X-ray cycle period of $7.3\pm0.3$ by \citet{Orlando2017}, based on measurements obtained over 15 years XMM-Newton time series.

It is interesting to show that although the Lomb-Scargle analysis reveals the presence of additional lower amplitude modulations, the dominance of the main cycle places this star in the faculae-dominated activity regime, similar to the Sun. Stars in this regime exhibit photospheric and chromospheric variabilities that are nearly in phase. For HD 81809A, this is further corroborated by \citet{Reinhold2019}, who reported a phase difference of $\Delta\phi = 0.08 \pm 0.01$ between photometric and chromospheric time series.

As shown by decades of observations at Mount Wilson, subgiant stars rarely exhibit activity cycles \citep{Baliunas1995}.
The rotation and cycle period ($P_{cyc}$) of HD~81809 places it on the so-called 'inactive' branch \citep{BohmVitense2007}, which is characterized by a shallower slope in the $P_{cyc}$–$P_{rot}$ relationship.
The fact that this subgiant shows a magnetic activity cycle provides an interesting constraint on its stellar dynamo evolution. According to \citet{Metcalfe2017}, activity cycles are expected to lengthen progressively with increasing rotation period. This behavior can be explained by the fact that as a star evolves from zero age main sequence (ZAMS) and approaches the so-called critical Rossby number \citep{VanSaders2016}, its rotation period remains relatively constant, while the activity cycle becomes longer and weaker before disappearing entirely. 
The results obtained for this subgiant suggest that it may have had a relatively short activity cycle until it reached the so-called critical Rossby number. Beyond this point, the cycle should have then grown longer and weaker at nearly constant Rossby number. When hydrogen core-burning ceased, the core likely contracted and the star became hotter, developing a thinner convection zone. This would have pushed it above the critical Rossby number, resulting in a transition to a 'flat activity' state. However, with the onset of hydrogen shell-burning, the star would have begun to expand and cool, further slowing its rotation through conservation of angular momentum and deepening the outer convection zone. These evolutionary effects can push the Rossby number back below $\rm Ro_{crit}=(0.92\pm0.01)Ro_{\odot}=1.656\pm0.02$ \citep{Metcalfe2024} with $\rm Ro_{\odot}$ from \citet{Rasio1996} so the star can reinvigorate large-scale dynamo action and briefly sustain an activity cycle before ascending the red giant branch as found by \citet[]{Metcalfe2020,Santos2025}.

\section{Rotational and X-ray Luminosity Evolution of HD~81809A}
\label{sec:spi}

In the comprehensive study of the HD~81809 system conducted by \citet{Egeland2018}, the authors concluded that the convolved rotational period and the magnetic activity cycle, determined through the in-phase variability of the S-index and X-ray luminosity observed for HD~81809, can be reasonably attributed to the subgiant component of the system. 

To test the compatibility of the best-fit stellar model of the primary star derived in this work with the observational data for the rotational period and X-ray luminosity, we computed post-processed rotational and X-ray luminosity tracks of HD~81809~A, by providing the CLES evolutionary tracks derived from best-fit stellar modeling in Section~\ref{sec:models}
to a Star-Planet Interaction code \citep{Pezzotti2025}. In this code, the evolution of the stellar surface rotation rate is computed under the assumption of solid body rotation, by accounting for the braking of the stellar surface due to magnetized winds as in \citet{Matt2015, Matt2019}. Given the degeneracy in the rotational history of solar-like low mass stars, we considered initial values of the surface rotation rate ranging between $3.2$ and $18$ times the surface rotation rate of the Sun at solar age ($\Omega_{\odot}=  2.9\times 10^{-6} s^{-1}$), to account for the spread of surface rotation rate observed for stars in young cluster and stellar associations \citep{Gallet2015}.

For initial surface rotation rates in the range $\rm \Omega_{in} < 18~\Omega_{\odot}$, it was assumed a disk-locking timescale of 6 Myr, while for $\rm \Omega_{in} = 18~\Omega_{\odot}$ the
assumed disk-locking timescale was 2 Myr. For values of the stellar Rossby number above the critical value $\rm Ro_{crit} > 0.92~Ro_{\odot}$ \citep{Metcalfe2024}, it is possible to switch into a ``weakened magnetic braking'' regime \citep{VanSaders2016}. In this circumstance, the braking due to magnetized winds is simply turned off.

The local convective turnover timescale at the base of the convective zone is computed as in \citet{Rasio1996, Villaver2009}. The evolution of the stellar X-ray luminosity is computed consistently to the one of the surface rotation rate, by following a recalibration of the $\rm Rx$ vs $\rm Ro$ prescription \citep{Johnstone2021}, as in \citet{Pezzotti2021}, where $\rm Rx$ is the ratio between the X-ray to the bolometric luminosity. For a more detailed description of the physics included in the star-planet interaction code, we refer the interested reader to \citet{Pezzotti2025}.

\begin{figure}[ht]
\centering
\subfloat{\includegraphics[height=0.68\textheight, keepaspectratio]{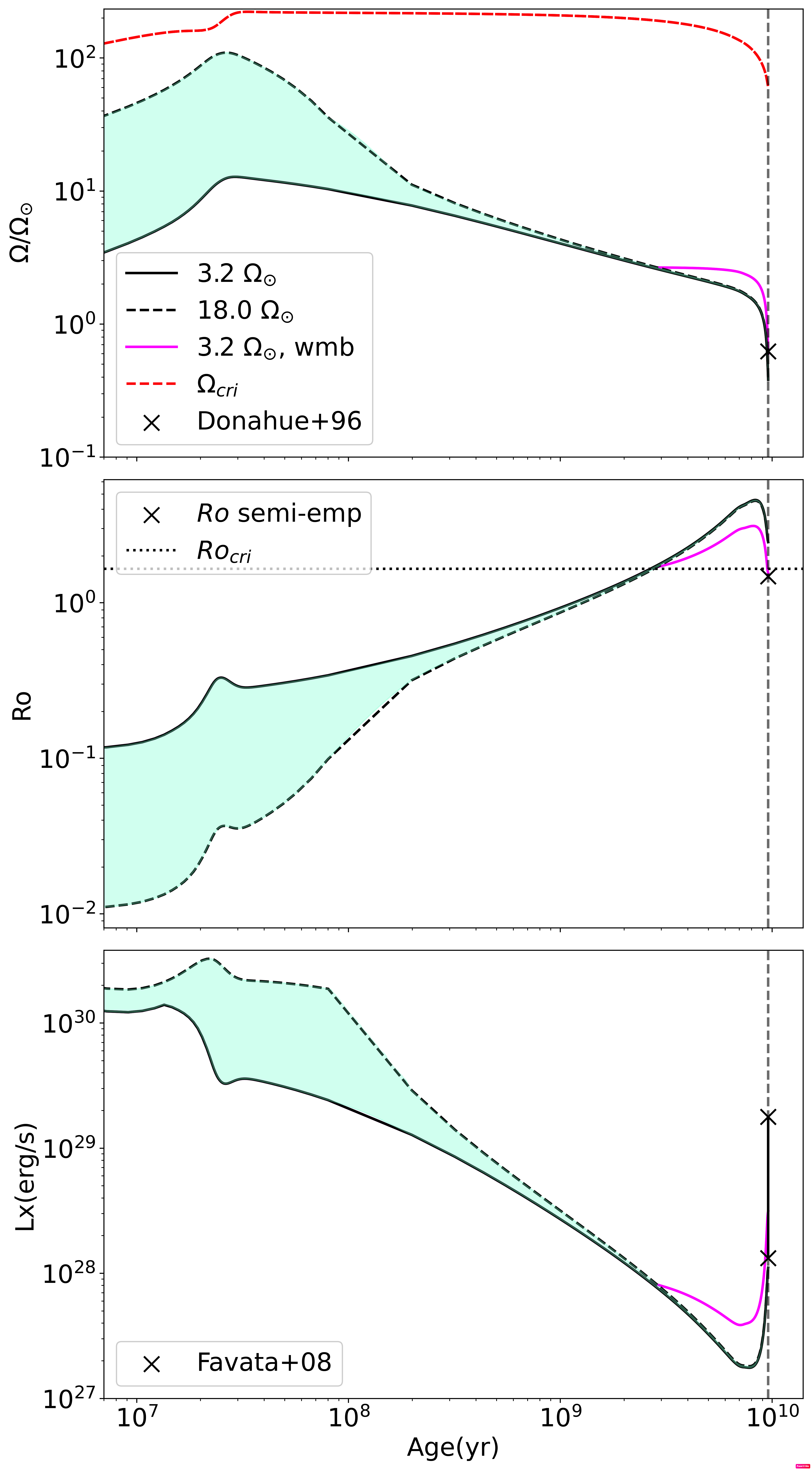}}\hskip 1ex
\subfloat{\includegraphics[height=0.68\textheight, keepaspectratio]{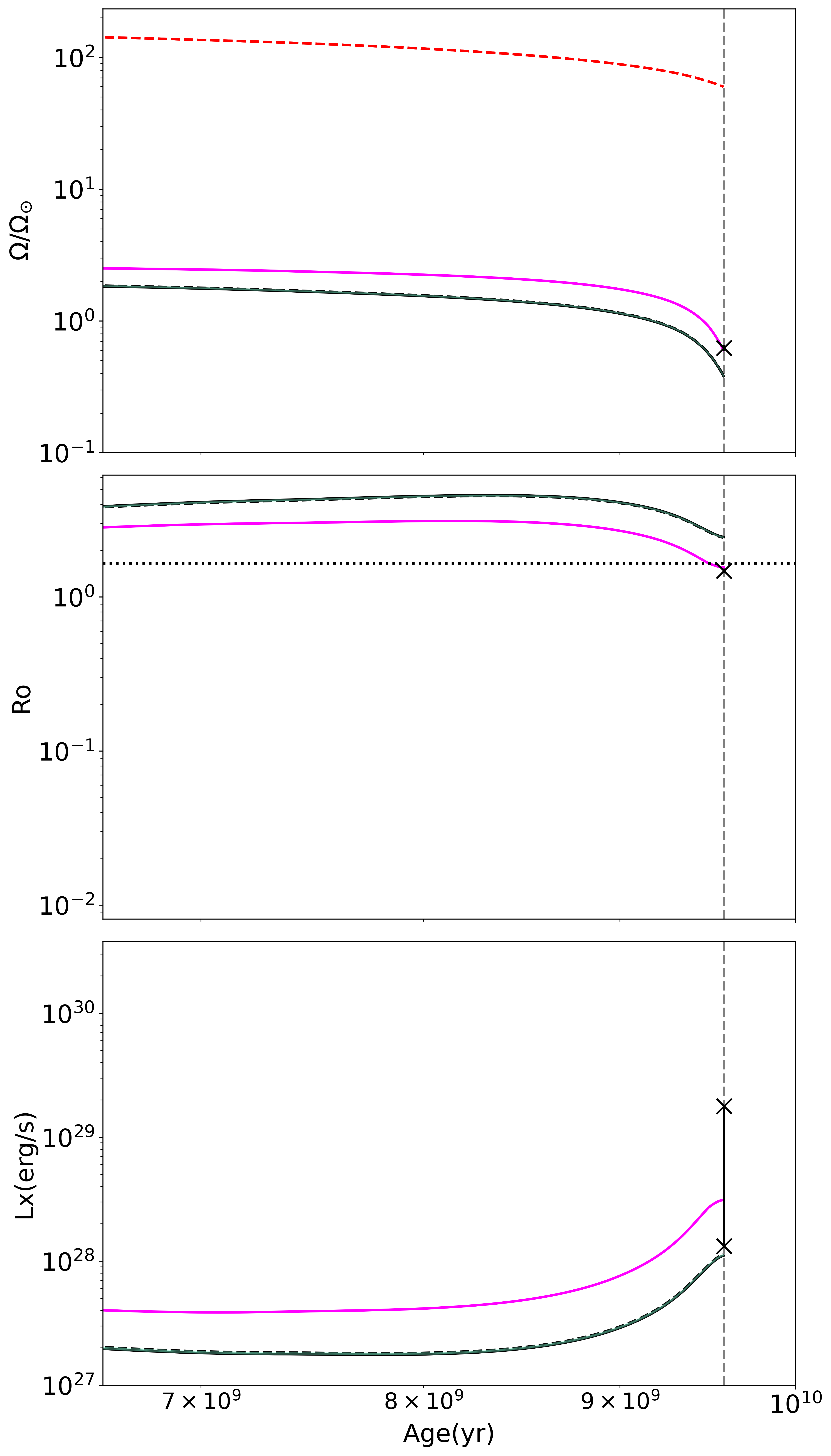}}
\caption{Evolution of the surface rotation rate, Rossby number and X-ray luminosity computed for HD~81809A. In the panels on the right, zoom on the regions of interest are shown. The green shaded area indicates the region of variation of the surface rotation rate across the evolution of the star, for initial values ranging between $3.2$ and $18~\Omega_{\odot}$. The red-dashed line shows the evolution of the critical rotational velocity ($\rm \Omega_{crit}$), defined as the velocity at which the centrifugal acceleration at the equator equals gravity. The purple line is obtained in the hypothesis that the star switched into a weakened magnetic braking at the age of 2\,Gyr, for an initial value of the surface rotation rate $3.2~\Omega_{\odot}$. The black crosses represent the observational values, as indicated in the legend. The observed variation in the X-ray luminosity emitted by the system over a magnetic cycle, with the minimum and maximum indicated by the black crosses, is taken from \citet{Favata2008}.}
\label{Fig:Omega_Ro_X_A_Low_mass}
\end{figure}

In Figure~\ref{Fig:Omega_Ro_X_A_Low_mass} (left panel, from top to bottom) we show the evolution of the surface rotation rate, the stellar Rossby number and the X-ray luminosity as a function of time for the tracks corresponding to the best model of HD~81809\,A. The right panels show a zoom of the region of potential overlap between the tracks and the observational data.

The evolutionary stage at which the onset of the weakened magnetic braking (wmb) occurs is indicated in the middle panel, by the crossing between the tracks and the horizontal-dotted line, which is the critical threshold ($\rm Ro_{crit}$) by \citet{Metcalfe2024}. 
In this figure, it is possible to notice that, at the estimated age of the system ($Age \simeq 9.58$ Gyr), the models accounting for weakened magnetic braking (purple line) result in agreement with the observations of stellar surface rotation \citep{Donahue1993, Donahue1996} and the emitted X-ray luminosity \citep{Favata2008}. %and also quite compatible with the semi-empirical value $\rm Ro=$ (Tab.~\ref{tab:model_rossby_number}) obtained from the best-fit procedure.
As discussed in Section~\ref{sec:magnetic}, 
indeed
a decrease in the Rossby number is visible in the middle panel in correspondence of the end of the MS and the subsequent overall contraction, that could have eventually caused the ``rejuvenation'' of the dynamo at work in the stellar interior, with a reborn magnetic cycle and increase of the $L_x$ emission.

\section{Discussion and Conclusions} 

\label{sec:conclus}

In this work, we have presented a multi-disciplinary characterization of the binary system HD~81809, combining high-resolution spectroscopy, precise radial velocities, GAIA astrometry, TESS asteroseismology, and decades of magnetic activity monitoring. By integrating these diverse datasets with advanced stellar evolutionary modeling (CLES and MESA), we establish HD~81809 as a unique benchmark for stellar physics, particularly regarding the evolution of old, metal-poor stars and the interplay between rotation, accretion, and magnetic activity.

\subsection{Evolutionary status and membership} The joint orbital and asteroseismic analysis confirms that HD~81809 is a long-period binary ($P\simeq 34.5$\,yr) composed of two solar-mass stars. The primary (HD~81809~A, $M\simeq 0.88\mathrm{M_{\odot}}$) has evolved off the main sequence and is currently crossing the subgiant branch, while the secondary (HD~81809~B, $M\simeq 0.83\mathrm{M}_{\odot}$) lays on the main sequence.

The revised atmospheric parameters derived in this study resolve previous inconsistencies regarding the system's age and population membership.
The kinematic properties of the system, combined with the clear $\alpha$-enhancement ($[\alpha/\mathrm{Fe}]\simeq+0.3$) and low metallicity ([Fe/H]$\simeq-0.57$) of the primary component, strongly suggest that HD~81809 belongs to the Galactic thick disk population. This is further corroborated by the present asteroseismic age estimates, which consistently point to an old system with an age of $\simeq 10$\,Gyr. This makes HD~81809 one of the nearest and brightest accessible laboratories for studying the physics of old, metal-poor stars.

The result reconciles the tension found in previous investigations \citep[e.g.,][]{Fuhrmann2018}, where the high [Mg/Fe] ratio and slow rotation suggested an old age that conflicted with younger isochrone fits.

It is important to acknowledge that the current stellar models do not yield a perfect statistical fit to all observational constraints. 
Firstly, both CLES and MESA codes underestimate the bolometric luminosity of the primary by approximately $\sim 3.5\sigma$. 
This discrepancy likely stems from a combination of limitations in the input physics for evolved, $\alpha$-enhanced stars (e.g., opacity tables or mixing length calibrations) and potential systematic effects in the observed photometry due to contamination by the circumstellar debris disk.

Furthermore, another
 significant modeling challenge arises because the secondary star is an "imposter". Its surface composition does not reflect its internal structure, breaking the standard assumptions of stellar evolution models.
Infact,
when models are forced to match the observed solar-like surface metallicity ([Fe/H]$=0.00$), they predict a mass  that is physically impossible, because it is higher than the evolved primary star: the models compensate for the high metallicity by increasing the mass, violating the binary hierarchy.
When models use the true initial metallicity inferred from the primary ([Fe/H]$=-0.57$), they correctly recover the low mass, but fail to match the observed effective temperature and seismic large separation. 
These mismatches highlight the inherent challenge of modeling 'polluted' stars using standard evolutionary tracks, which do not account for the stratification and specific structural changes caused by the accretion of metal-rich material onto a metal-poor envelope.

\subsection{The Chemical Dichotomy and the Accretion Hypothesis} One of the most striking results of this study is the significant chemical discrepancy between the two components. While the primary exhibits a composition typical of old, metal-poor thick disk stars, the secondary displays a solar-like metallicity ([Fe/H]$\simeq 0.0$) and a remarkably high Lithium abundance [Li/H]$\simeq 0.69$ dex above solar).

We interpret this dichotomy as the signature of a pollution event. The presence of a significant infrared excess ($\lambda>30\mu m$) detected in the SED analysis indicates the existence of a cold debris disk ($T_{\rm BB}=96.6$~K) 
in the system. We propose that the secondary star has undergone recent accretion of metal-rich, planetary material. The difference in the observed photospheric abundances can be explained by the internal structure of the two stars. The primary, being a subgiant, possesses a deepening convective envelope that would rapidly dilute any accreted material, effectively erasing the signature of recent pollution and revealing the pristine, metal-poor natal composition.

In contrast, the main-sequence secondary likely retains a thinner surface convection zone (relative to the expanding subgiant). Consequently, accreted metal-rich material and Lithium—likely from the ingestion of rocky planetesimals remained concentrated on the photosphere, creating the illusion of a younger, solar-metallicity star.
 The high Lithium abundance in the secondary supports the ingestion of rocky, planetesimal material, which can replenish surface Li that is otherwise depleted in old stars.

\subsection{Magnetic Rejuvenation in the Subgiant Phase} Our analysis of the 50-year chromospheric activity record (MWO and SSS data) confirms a magnetic cycle of $P_{cycl}\simeq 8.2$-years.

Standard dynamo theory suggests that stars moving off the main sequence should spin down and become magnetically "dead." However, HD~81809~A appears to be an example of "dynamo rejuvenation."

As a star evolved off the main sequence, the contraction of the core and the subsequent expansion of the envelope caused the rotation to slow and the convective zone to deepen. Our evolutionary models and SPI simulations suggest that HD~81809~A has recently evolved back below the critical Rossby number threshold ($Ro<Ro_{crit}$). This structural change has likely reactivated a large-scale $\alpha-\Omega$ dynamo, allowing the star to sustain the coherent, Sun-like activity cycle we observe today. This places HD~81809~A in a crucial, transient evolutionary phase, providing observational evidence for magnetic life after the main sequence.

\subsection{Concluding Remarks}

HD~81809 has proven to be a unique laboratory for binary evolution study, linking several distinct areas of stellar astrophysics: 

\begin{itemize} 
\item It validates evolutionary models for metal-poor, $\alpha$-enhanced stars 
serving as a probe for the history of the Milky Way's thick disk.

\item It provides a rare case study of chemically peculiar binaries, supporting scenarios of planetesimal engulfment and external pollution. 
\item It offers a detailed look at the reactivation of magnetic dynamos in subgiants, constraining the interaction between rotation, convection, and magnetic fields. \end{itemize}

We can conclude that the characterization of HD~81809 provided in this study opens several promising avenues for further investigation. First and foremost, high-resolution spectroscopic monitoring of both components would be essential to disentangle the contribution of the secondary and possibly refine the chemical abundance pattern, including neutron-capture elements, which could offer insight into the early Galactic chemical evolution.

Second, future asteroseismic missions such as PLATO \citep[][]{Rauer2025}, with improved temporal coverage and photometric precision, or perhaps the  ground-based radial velocities from the SONG network \citep{Grundahl2017} could allow the detection of individual oscillation modes,  in order to significantly improve the precision of age and mass estimates.

 Furthermore, a dedicated interferometric campaign, with CHARA \citep[][e.g.,]{Chara, Chara2} or VLTI \citep[][e.g.]{VLTI}, could provide direct radius measurements, offering an independent constraint  particularly in relation to the debated bolometric luminosities.

 Finally, on the magnetic activity front, long-term chromospheric monitoring, possibly combined with new X-ray observations with eROSITA \citep[e.g.,][]{Robrade2023} or Athena \citep[e.g.,][]{Sciortino2021}, could help to further constrain the magnetic cycle characteristics and test dynamo models for evolved, metal-poor subgiants. In particular, HD~81809A offers a unique testbed to study the transition between solar-like dynamos and the weakened magnetic braking regime in evolved stars.
High-resolution spectropolarimetry to map the magnetic topology of the primary would be the ultimate test for dynamo models of evolved, metal-poor stars.

\begin{acknowledgments}
We acknowledge funding for the publication of the present manuscript and for the position of N. M. 
from the research theory grant “Synergic tools for characterizing solar-like stars and habitability conditions of exoplanets  (PI M.P. Di Mauro) under the INAF national call for Fundamental Research 2023.\\
The authors are thankful to Jeffrey C. Hall for providing the Lowell SSS S-index data. A.B. and E.C. are funded by the European Union – NextGenerationEU RRF M4C2 1.1  n: 2022HY2NSX. "CHRONOS: adjusting the clock(s) to unveil the CHRONO-chemo-dynamical Structure of the Galaxy” (PI: S. Cassisi). We acknowledge support
from the research grant “Unveiling the magnetic side of the Stars” (PI A.
Bonanno) funded under the INAF national call for Fundamental Research 2023. CP thanks the Belgian Federal Science Policy Office (BELSPO) for the financial support in the framework of the PRODEX Program of
the European Space Agency (ESA) under contract number 4000141194.\\
This publication makes use of VOSA, developed under the Spanish Virtual Observatory (https://svo.cab.inta-csic.es) project funded by MCIN/AEI/10.13039/501100011033/ through grant PID2020-112949GB-I00.
VOSA has been partially updated by using funding from the European Union's Horizon 2020 Research and Innovation Programme, under Grant Agreement nº 776403 (EXOPLANETS-A)   
PM acknowledges support from UK Science and Technology Facilities Council (STFC) research grant numbers ST/X002047/1 and ST/Y002563/1.

This study uses spectra obtained with the Hermes spectrograph, installed on the Mercator telescope in the Roque de Los Muchachos observatory on the island of La Palma, operated by the KU Leuven university.

This research has made use of the Keck Observatory Archive (KOA), which is operated by the W. M. Keck Observatory and the NASA Exoplanet Science Institute (NExScI), under contract with the National Aeronautics and Space Administration.

Based on observations made with ESO Telescopes at the La Silla Paranal Observatory under programme ID  099.A-9022(A).
\end{acknowledgments}

\appendix

\restartappendixnumbering

\section{Astrometric dataset}
Additional observational data, which complement the results discussed in Section \ref{sec:fund_params}, are listed in Table \ref{tab:rhopa} and Table \ref{tab:lines}.
%%%%%%%%%%%%%%%% Micrometer and speckle measurements
\begin{table}
\caption{Measurements of the angular separation ($\rho$) and position angle ($\theta)$ of HD~81809\,B relative to HD~81809\,A.
The method used for the measurements are indicated as follows: M -- micrometer, S -- speckle interferometry, H -- Hipparcos.
Sources are as follows:
1 -- \citet{1985A+AS...60..333B},
2 -- \citet{1982ApJS...48..273M},
3 -- \citet{1983ApJS...51..309M},
4 -- \citet{1984A+AS...57...31B},
5 -- \citet{1987AJ.....93..688M},
6 -- \citet{1989AJ.....97..510M},
7 -- \citet{1990ApJS...74..275H},
8 -- \citet{1993AJ....106.1639M},
9 -- \citet{2000AJ....119.3084H},
10 -- \citet{1990AJ.....99..965M},
11 -- \citet{1996AJ....111..936H},
12 -- \citet{1997yCat.1239....0E},
13 -- \citet{1994AJ....108.2299H},
14 -- \citet{1993AN....314..303L},
15 -- \citet{1998A+AS..132..237A},
16 -- \citet{1997AJ....114.1639H},
17 -- \citet{1997A+AS..121..405P},
18 -- \citet{1998ApJS..117..587H},
19 -- \citet{1999AJ....118.1395D},
20 -- \citet{2001AJ....121.3224M},
21 -- \citet{2010AJ....139..743T},
22 -- \citet{2012AJ....143...42H},
23 -- \citet{2015AJ....150...50T},
24 -- \citet{2020AJ....160....7T}.
} 
\label{tab:rhopa}

\begin{center}
\begin{tabular}{lrrlc|lrrlc}
\hline
Year & $\rho$ [$^{\prime\prime}$] & $\theta$ [$^{\circ}$]& Source & Method &
Year & $\rho$ [$^{\prime\prime}$] & $\theta$ [$^{\circ}$]& Source & Method \\
\hline
1938.16 &   0.15 &$-41.0$& 1 & M &              1989.30 &  0.412 &$147.2$& 10 & S \\
1941.25 &   0.19 &$-28.3$& 1 & M &              1989.94 &  0.438 &$147.7$& 8 & S \\
1953.03 &   0.30 &$143.6$& 1 & M &              1990.34 &  0.452 &$148.0$& 8 & S \\
1955.18 &   0.37 &$149.2$& 1 & M &              1990.34 &  0.453 &$148.0$& 8 & S \\
1958.04 &   0.36 &$151.6$& 1 & M &              1990.92 &  0.475 &$148.5$& 11 & S \\
1960.21 &   0.43 &$150.9$& 1 & M &              1991.25 &  0.487 &$147.6$& 12 & H \\
1962.21 &   0.42 &$152.5$& 1 & M &              1992.30 &  0.507 &$148.5$& 13 & S \\
1963.27 &   0.44 &$155.7$& 1 & M &              1993.09 &  0.528 &$149.5$& 9 & S \\
1965.06 &   0.40 &$158.4$& 1 & M &              1993.26 &   0.51 &$142.1$& 14 & M \\
1976.28 &   0.29 &$-28.0$& 1 & M &              1994.24 &   0.55 &$151.2$& 15 & M \\
1976.37 &  0.289 &$-26.2$& 2 & S &              1994.31 &  0.545 &$150.3$& 11 & S \\
1977.17 &  0.300 &$-25.0$& 2 & S &              1994.87 &  0.544 &$150.1$& 9 & S \\
1977.34 &  0.295 &$-25.1$& 2 & S &              1995.92 &  0.518 &$150.7$& 16 & S \\
1977.91 &  0.302 &$-28.1$& 2 & S &              1996.16 &   0.40 &$145.6$& 17 & M \\
1977.91 &  0.258 &$-22.1$& 2 & S &              1996.17 &  0.532 &$151.0$& 9 & S \\
1980.15 &  0.188 &$-19.0$& 3 & S &              1996.33 &   0.46 &$151.2$& 18 & M \\
1982.16 &  0.033 &$  1.0$& 4 & S &              1996.87 &  0.513 &$151.1$& 9 & S \\
1984.38 &  0.120 &$134.1$& 5 & S &              1997.30 &  0.470 &$151.6$& 19 & S \\
1987.26 &  0.306 &$144.6$& 6 & S &              1998.91 &  0.392 &$146.5$& 16 & S \\
1988.14 &   0.31 &$144.5$& 7 & M &              1999.16 &  0.446 &$153.4$& 20 & S \\
1988.16 &  0.354 &$146.3$& 8 & S &              2009.26 &  0.264 &$-32.7$& 21 & S \\
1988.25 &  0.360 &$145.8$& 6 & S &              2011.03 & 0.3186 &$-29.6$& 22 & S \\
1988.25 &  0.364 &$146.4$& 9 & S &              2014.18 & 0.2302 &$-24.0$& 23 & S \\
1989.22 &  0.409 &$146.6$& 10 & S &             2019.13 & 0.1255 &$134.6$& 24 & S \\
\hline
\end{tabular}
\end{center}
\end{table}

%\section{Additional material on spectroscopic analysis}
\begin{table}[ht]
\caption{Spectral lines used for deriving the abundances.}
\label{tab:lines}
\centering
\resizebox{0.95\textwidth}{!}{%
\begin{tabular}{lcccclc|lcccclc}
\hline
El  & $\lambda$ & $\log gf$ &  E$_{low}$ &  J$_{low}$ & E$_{up}$ &  J$_{up}$  &  El  & $\lambda$ & $\log gf$ &  E$_{low}$ &  J$_{low}$ & E$_{up}$ &  J$_{up}$ \\
    &  \AA  &           &    eV      &            &    eV    &            &      &  \AA  &           &    eV      &            &    eV    &           \\
\hline
CI   & 4932.050 & -1.658 &   7.685 &  1.0  &  10.198 &   0.0  &   SI   & 4694.113 & -1.713 &   6.524 &  2.0  &   9.165 &   3.0  \\   
CI   & 5052.144 & -1.303 &   7.685 &  1.0  &  10.138 &   2.0  &   SI   & 6052.656 & -0.672 &   7.870 &  3.0  &   9.918 &   4.0  \\   
CI   & 5380.325 & -1.616 &   7.685 &  1.0  &   9.989 &   1.0  &   SI   & 6743.535 & -1.066 &   7.866 &  1.0  &   9.704 &   2.0  \\   
CI   & 6587.620 & -1.003 &   8.537 &  1.0  &  10.419 &   1.0  &   SI   & 6757.153 & -0.352 &   7.870 &  3.0  &   9.704 &   4.0  \\   
CI   & 8335.147 & -0.437 &   7.685 &  1.0  &   9.172 &   0.0  &   SI   & 8694.709 &  0.050 &   7.870 &  3.0  &   9.295 &   4.0  \\   
CI   & 9061.435 & -0.347 &   7.483 &  1.0  &   8.851 &   2.0  &   CaI  & 6161.297 & -1.020 &   2.523 &  2.0  &   4.535 &   2.0  \\   
CI   & 9062.472 & -0.455 &   7.480 &  0.0  &   8.848 &   1.0  &   CaI  & 6162.173 & -0.170 &   1.899 &  2.0  &   3.910 &   1.0  \\   
CI   & 9094.829 &  0.151 &   7.488 &  2.0  &   8.851 &   2.0  &   CaI  & 6166.439 & -0.900 &   2.521 &  1.0  &   4.531 &   0.0  \\   
CI   & 9111.799 & -0.297 &   7.488 &  2.0  &   8.848 &   1.0  &   CaI  & 6169.042 & -0.550 &   2.523 &  2.0  &   4.532 &   1.0  \\      
OI   & 6300.304 & -9.776 &   0.000 &  2.0  &   1.967 &   2.0  &   CaI  & 6462.567 &  0.310 &   2.523 &  2.0  &   4.441 &   3.0  \\   
OI   & 7771.944 &  0.369 &   9.146 &  2.0  &  10.741 &   3.0  &   CaI  & 6471.673 & -8.940 &   5.228 &  2.0  &   7.144 &   1.0  \\   
OI   & 7774.166 &  0.223 &   9.146 &  2.0  &  10.740 &   2.0  &   CaI  & 7148.150 &  0.307 &   2.709 &  2.0  &   4.443 &   2.0  \\   
OI   & 7775.388 &  0.002 &   9.146 &  2.0  &  10.740 &   1.0  &   CaI  & 7326.144 & -5.543 &   5.732 &  3.0  &   7.424 &   3.0  \\   
OI   & 8446.247 & -0.463 &   9.521 &  1.0  &  10.989 &   0.0  &   CaII & 8498.020 & -1.363 &   1.692 &  1.5  &   3.151 &   1.5  \\   
NaI  & 4497.656 & -1.574 &   2.104 &  1.5  &   4.860 &   2.5  &   CaII & 8542.088 & -0.411 &   1.700 &  2.5  &   3.151 &   1.5  \\   
NaI  & 4668.556 & -2.264 &   2.104 &  1.5  &   4.759 &   1.5  &   CaII & 8662.138 & -0.672 &   1.692 &  1.5  &   3.123 &   0.5  \\   
NaI  & 5682.632 & -0.706 &   2.102 &  0.5  &   4.284 &   1.5  &   ScII & 5031.021 & -0.399 &   1.357 &  2.0  &   3.821 &   1.0  \\      
NaI  & 5688.204 & -0.452 &   2.104 &  1.5  &   4.283 &   2.5  &   ScII & 5239.813 & -0.765 &   1.455 &  0.0  &   3.821 &   1.0  \\   
NaI  & 6154.223 & -1.547 &   2.102 &  0.5  &   4.116 &   0.5  &   ScII & 5526.789 &  0.025 &   1.768 &  4.0  &   4.011 &   3.0  \\   
NaI  & 6160.746 & -1.246 &   2.104 &  1.5  &   4.116 &   0.5  &   ScII & 5657.886 & -0.603 &   1.507 &  2.0  &   3.698 &   2.0  \\   
MgI  & 4571.096 & -5.623 &   0.000 &  0.0  &   2.712 &   1.0  &   ScII & 5684.190 & -1.074 &   1.507 &  2.0  &   3.687 &   1.0  \\   
MgI  & 5167.316 & -2.558 &   5.108 &  1.0  &   7.507 &   2.0  &   ScII & 6245.621 & -1.022 &   1.507 &  2.0  &   3.492 &   3.0  \\ 
MgI  & 5172.684 & -0.393 &   2.712 &  1.0  &   5.108 &   1.0  &   TiI  & 5512.524 & -0.400 &   1.460 &  4.0  &   3.709 &   3.0  \\   
MgI  & 5183.604 & -0.167 &   2.717 &  2.0  &   5.108 &   1.0  &   TiI  & 5514.343 & -0.660 &   1.430 &  2.0  &   3.678 &   1.0  \\   
MgI  & 5711.088 & -1.724 &   4.346 &  1.0  &   6.516 &   0.0  &   TiI  & 5514.534 & -0.500 &   1.443 &  3.0  &   3.691 &   2.0  \\   
MgI  & 7691.553 & -0.783 &   5.753 &  2.0  &   7.365 &   3.0  &   TiI  & 5662.152 & -0.053 &   2.318 &  4.0  &   4.507 &   5.0  \\   
MgI  & 8054.231 & -2.238 &   5.933 &  2.0  &   7.472 &   2.0  &   TiI  & 6258.102 & -0.299 &   1.443 &  3.0  &   3.424 &   4.0  \\   
AlI  & 6696.012 & -1.569 &   3.143 &  0.5  &   4.994 &   1.5  &   TiI  & 6261.099 & -0.423 &   1.430 &  2.0  &   3.409 &   3.0  \\   
AlI  & 6698.662 & -1.870 &   3.143 &  0.5  &   4.993 &   0.5  &   TiI  & 7209.436 & -0.430 &   1.460 &  4.0  &   3.179 &   3.0  \\   
AlI  & 7836.108 & -0.534 &   4.022 &  2.5  &   5.603 &   3.5  &   TiI  & 8412.357 & -1.427 &   0.818 &  2.0  &   2.292 &   1.0  \\   
AlI  & 8772.857 & -0.349 &   4.021 &  1.5  &   5.434 &   2.5  &   TiI  & 8434.954 & -0.830 &   0.848 &  5.0  &   2.318 &   4.0  \\   
AlI  & 8773.888 & -0.192 &   4.022 &  2.5  &   5.434 &   3.5  &   TiI  & 8435.652 & -0.967 &   0.836 &  4.0  &   2.305 &   3.0  \\   
SiI  & 5684.484 & -1.733 &   4.954 &  2.0  &   7.134 &   1.0  &   TiII & 4805.093 & -0.941 &   2.061 &  1.5  &   4.641 &   0.5  \\   
SiI  & 5690.425 & -1.772 &   4.930 &  1.0  &   7.108 &   1.0  &   TiII & 5336.786 & -1.600 &   1.582 &  2.5  &   3.904 &   3.5  \\   
SiI  & 5948.541 & -0.780 &   5.082 &  1.0  &   7.166 &   2.0  &   TiII & 5381.022 & -1.970 &   1.566 &  1.5  &   3.869 &   2.5  \\   
SiI  & 6155.134 & -0.755 &   5.619 &  3.0  &   7.633 &   4.0  &   TiII & 5418.768 & -2.130 &   1.582 &  2.5  &   3.869 &   2.5  \\   
SiI  & 6237.319 & -0.975 &   5.614 &  1.0  &   7.601 &   2.0  &   VI   & 4379.190 &  0.580 &   0.301 &  4.5  &   3.131 &   5.5  \\   
SiI  & 6243.815 & -1.244 &   5.616 &  2.0  &   7.601 &   3.0  &   VI   & 5698.516 & -0.120 &   1.064 &  2.5  &   3.239 &   3.5  \\   
SiI  & 6244.466 & -1.091 &   5.616 &  2.0  &   7.601 &   3.0  &   VI   & 5703.555 & -0.210 &   1.051 &  1.5  &   3.224 &   2.5  \\   
SiI  & 7003.569 & -0.939 &   5.964 &  2.0  &   7.734 &   3.0  &   VI   & 6090.194 & -0.070 &   1.081 &  3.5  &   3.116 &   2.5  \\   
SiI  & 7005.880 & -0.741 &   5.984 &  3.0  &   7.753 &   4.0  &   VI   & 6243.098 & -0.940 &   0.301 &  4.5  &   2.286 &   4.5  \\   
SiI  & 7034.901 & -0.624 &   5.871 &  2.0  &   7.633 &   3.0  &   CrI  & 4801.025 & -0.131 &   3.122 &  4.0  &   5.703 &   3.0  \\   
SiI  & 7405.772 & -0.313 &   5.614 &  1.0  &   7.287 &   2.0  &   CrI  & 4876.403 & -2.483 &   4.096 &  2.0  &   6.638 &   2.0  \\   
SiI  & 7409.083 & -0.620 &   5.616 &  2.0  &   7.289 &   3.0  &   CrI  & 5204.511 & -0.198 &   0.941 &  2.0  &   3.323 &   1.0  \\   
SiI  & 7423.496 & -0.176 &   5.619 &  3.0  &   7.289 &   4.0  &   CrI  & 5206.037 &  0.025 &   0.941 &  2.0  &   3.322 &   2.0  \\   
SiI  & 7424.610 & -1.610 &   5.619 &  3.0  &   7.289 &   3.0  &   CrI  & 5208.425 &  0.172 &   0.941 &  2.0  &   3.321 &   3.0  \\   
SiI  & 7932.348 & -0.469 &   5.964 &  2.0  &   7.527 &   3.0  &   CrII & 4824.131 & -0.920 &   3.871 &  4.5  &   6.440 &   4.5  \\
SiI  & 7944.001 & -0.293 &   5.984 &  3.0  &   7.544 &   4.0  &        &          &        &         &       &         &        \\
SiI  & 7970.307 & -1.428 &   5.964 &  2.0  &   7.519 &   2.0  &        &          &        &         &       &         &        \\
\end{tabular}
}

\end{table}  

\begin{table}[]
\caption{continued.}
\addtocounter{table}{-1}
\centering
\resizebox{0.95\textwidth}{!}{%
\begin{tabular}{lcccclc|lcccclc}
\hline
El  & $\lambda$ & $\log gf$ &  E$_{low}$ &  J$_{low}$ & E$_{up}$ &  J$_{up}$  &  El  & $\lambda$ & $\log gf$ &  E$_{low}$ &  J$_{low}$ & E$_{up}$ &  J$_{up}$ \\
    &   \AA  &           &    eV      &            &    eV    &            &      &  \AA  &           &    eV      &            &    eV    &           \\
\hline
MnI  & 4754.035 & -0.080 &   2.282 &  2.5  &   4.889 &   3.5  &  NdII & 5293.163 &  0.100 &   0.823 &  7.5  &   3.165 &   6.5  \\ 
MnI  & 4761.504 & -0.274 &   2.953 &  0.5  &   5.556 &   1.5  &  NdII & 5092.794 & -0.610 &   0.380 &  5.5  &   2.814 &   5.5  \\ 
MnI  & 4762.361 &  0.304 &   2.888 &  3.5  &   5.491 &   4.5  &  NdII & 5688.518 & -0.310 &   0.986 &  6.5  &   3.165 &   6.5  \\ 
MnI  & 4765.856 & -0.086 &   2.941 &  1.5  &   5.542 &   2.5  &  SmII & 4537.941 & -0.480 &   0.485 &  6.5  &   3.216 &   6.5  \\ 
MnI  & 4783.405 &  0.044 &   2.298 &  3.5  &   4.889 &   3.5  &  SmII & 4577.688 & -0.650 &   0.248 &  1.5  &   2.956 &   2.5  \\ 
MnI  & 4823.506 &  0.136 &   2.319 &  4.5  &   4.889 &   3.5  &  SmII & 4642.228 & -0.460 &   0.378 &  5.5  &   3.049 &   5.5  \\ 
MnI  & 5407.440 & -1.743 &   2.143 &  3.5  &   4.435 &   3.5  &  SmII & 4566.202 & -0.590 &   0.333 &  2.5  &   3.048 &   2.5  \\ 
MnI  & 5420.425 & -1.462 &   2.143 &  3.5  &   4.429 &   2.5  &       &          &        &         &       &         &        \\
MnI  & 6013.453 & -0.354 &   3.072 &  1.5  &   5.133 &   2.5  &       &          &        &         &       &         &        \\
MnI  & 6016.632 & -0.181 &   3.073 &  2.5  &   5.133 &   2.5  &       &          &        &         &       &         &        \\
MnI  & 6021.788 & -0.054 &   3.075 &  3.5  &   5.133 &   2.5  &       &          &        &         &       &         &        \\
CoI  & 4118.763 & -0.193 &   1.049 &  2.5  &   4.058 &   3.5  &       &          &        &         &       &         &        \\
CoI  & 5352.020 &  0.125 &   5.892 &  4.5  &   3.576 &   5.5  &       &          &        &         &       &         &        \\
CoI  & 5369.587 & -1.610 &   1.740 &  1.5  &   4.049 &   2.5  &       &          &        &         &       &         &        \\
CoI  & 6450.134 & -1.850 &   2.137 &  3.5  &   4.058 &   3.5  &       &          &        &         &       &         &        \\
CoI  & 6454.958 & -0.139 &   5.552 &  4.5  &   3.632 &   3.5  &       &          &        &         &       &         &        \\
NiI  & 4551.224 & -0.863 &   6.891 &  3.0  &   4.167 &   2.0  &       &          &        &         &       &         &        \\
NiI  & 4815.922 & -1.683 &   3.542 &  2.0  &   6.116 &   2.0  &       &          &        &         &       &         &        \\
NiI  & 5084.089 & -0.084 &   3.678 &  3.0  &   6.116 &   4.0  &       &          &        &         &       &         &        \\
NiI  & 5094.411 & -0.998 &   3.833 &  1.0  &   6.266 &   1.0  &       &          &        &         &       &         &        \\
NiI  & 5099.922 & -0.236 &   3.678 &  3.0  &   6.109 &   3.0  &       &          &        &         &       &         &        \\
NiI  & 5424.635 & -2.632 &   1.951 &  1.0  &   4.236 &   2.0  &       &          &        &         &       &         &        \\
CuI  & 5105.512 & -1.542 &   1.389 &  2.5  &   3.817 &   1.5  &       &          &        &         &       &         &        \\
CuI  & 5218.194 &  0.364 &   3.817 &  1.5  &   6.192 &   2.5  &       &          &        &         &       &         &        \\
CuI  & 5153.237 &  0.116 &   3.786 &  0.5  &   6.191 &   1.5  &       &          &        &         &       &         &        \\
CuI  & 5700.231 & -2.583 &   1.642 &  1.5  &   3.817 &   1.5  &       &          &        &         &       &         &        \\
CuI  & 5782.121 & -1.905 &   1.642 &  1.5  &   3.786 &   0.5  &       &          &        &         &       &         &        \\
CuI  & 8092.623 & -0.131 &   3.817 &  1.5  &   5.348 &   0.5  &       &          &        &         &       &         &        \\
ZnI  & 4680.136 & -0.810 &   4.006 &  0.0  &   6.655 &   1.0  &       &          &        &         &       &         &        \\
ZnI  & 4722.157 & -0.338 &   4.030 &  1.0  &   6.655 &   1.0  &       &          &        &         &       &         &        \\
ZnI  & 4810.532 & -0.125 &   4.078 &  2.0  &   6.655 &   1.0  &       &          &        &         &       &         &        \\
SrII & 4077.709 &  0.167 &   0.000 &  0.5  &   3.040 &   1.5  &       &          &        &         &       &         &        \\
SrII & 4215.519 & -0.145 &   0.000 &  0.5  &   2.940 &   0.5  &       &          &        &         &       &         &        \\
YII  & 4883.682 &  0.070 &   1.084 &  4.0  &   3.622 &   3.0  &       &          &        &         &       &         &        \\
YII  & 4900.119 & -0.090 &   1.033 &  3.0  &   3.562 &   2.0  &       &          &        &         &       &         &        \\
YII  & 5087.418 & -0.170 &   1.084 &  4.0  &   3.520 &   4.0  &       &          &        &         &       &         &        \\
YII  & 5402.773 & -0.360 &   1.839 &  2.0  &   4.133 &   3.0  &       &          &        &         &       &         &        \\
ZrII & 4149.198 & -0.040 &   0.802 &  3.5  &   3.789 &   3.5  &       &          &        &         &       &         &        \\
ZrII & 4496.962 & -0.890 &   0.713 &  2.5  &   3.470 &   2.5  &       &          &        &         &       &         &        \\
BaII & 4554.029 &  0.140 &   0.000 &  0.5  &   2.722 &   1.5  &       &          &        &         &       &         &        \\
BaII & 4934.076 & -0.160 &   0.000 &  0.5  &   2.512 &   0.5  &       &          &        &         &       &         &        \\
BaII & 5853.668 & -0.908 &   0.604 &  1.5  &   2.722 &   1.5  &       &          &        &         &       &         &        \\
BaII & 6141.713 & -0.032 &   0.704 &  2.5  &   2.722 &   1.5  &       &          &        &         &       &         &        \\
BaII & 6496.897 & -0.407 &   0.604 &  1.5  &   2.512 &   0.5  &       &          &        &         &       &         &        \\
LaII & 5114.556 & -1.060 &   0.235 &  1.0  &   2.658 &   1.0  &       &          &        &         &       &         &        \\
LaII & 5290.811 & -1.750 &   0.000 &  2.0  &   2.343 &   2.0  &       &          &        &         &       &         &        \\
CeII & 4222.597 & -0.150 &   0.122 &  4.5  &   3.058 &   4.5  &       &          &        &         &       &         &        \\
CeII & 4562.359 &  0.210 &   0.478 &  3.5  &   3.195 &   4.5  &       &          &        &         &       &         &        \\
\end{tabular}
}

\end{table}

\bibliography{HD81809_biblio}{}

@misc{10.17909/t9-st5g-3177,
doi = {10.17909/T9-ST5G-3177},
url = {http://archive.stsci.edu/doi/resolve/resolve.html?doi=10.17909/t9-st5g-3177},
author = {{TESS Team}},
title = {TESS "Fast" Light Curves - All Sectors},
publisher = {STScI/MAST},
year = {2021}
}

@misc{10.17909/t9-nmc8-f686,
doi = {10.17909/T9-NMC8-F686},
url = {http://archive.stsci.edu/doi/resolve/resolve.html?doi=10.17909/t9-nmc8-f686},
author = {{TESS Team}},
title = {TESS Light Curves - All Sectors},
publisher = {STScI/MAST},
year = {2021}
}

@PROCEEDINGS{Helou1988,
        title = "{Infrared Astronomical Satellite (IRAS) Catalogs and Atlases.Volume 7: The Small Scale Structure Catalog.}",
    booktitle = {Infrared astronomical satellite (IRAS) catalogs and atlases. Volume 7},
         year = 1988,
       editor = {{Helou}, George and {Walker}, D.~W.},
       volume = {7},
        month = jan,
       adsurl = {https://ui.adsabs.harvard.edu/abs/1988iras....7.....H}
}

@ARTICLE{catanzaro2024,
       author = {{Catanzaro}, G. and {Frasca}, A. and {Alonso-Santiago}, J. and {Colombo}, C.},
        title = "{TESS photometry and CAOS spectroscopy of six eclipsing binaries with Am components}",
      journal = {\aap},
         year = 2024,
        month = may,
       volume = {685},
          eid = {A133},
        pages = {A133},
          doi = {10.1051/0004-6361/202449332},
archivePrefix = {arXiv},
       eprint = {2402.16648},
 primaryClass = {astro-ph.SR},
       adsurl = {https://ui.adsabs.harvard.edu/abs/2024A&A...685A.133C}
}

@ARTICLE{2013MSAIS..24..128A,
       author = {{Allard}, F. and {Homeier}, D. and {Freytag}, B. and {Schaffenberger}, W. and {Rajpurohit}, A.~S.},
        title = "{Progress in modeling very low mass stars, brown dwarfs, and planetary mass objects.}",
      journal = {Memorie della Societa Astronomica Italiana Supplementi},
         year = 2013,
        month = jan,
       volume = {24},
        pages = {128},
          doi = {10.48550/arXiv.1302.6559},
archivePrefix = {arXiv},
       eprint = {1302.6559},
 primaryClass = {astro-ph.SR},
       adsurl = {https://ui.adsabs.harvard.edu/abs/2013MSAIS..24..128A}
}

@ARTICLE{gcs2,
       author = {{Holmberg}, J. and {Nordstr{\"o}m}, B. and {Andersen}, J.},
        title = "{The Geneva-Copenhagen survey of the Solar neighbourhood II. New uvby calibrations and rediscussion of stellar ages, the G dwarf problem, age-metallicity diagram, and heating mechanisms of the disk}",
      journal = {\aap},
         year = 2007,
        month = nov,
       volume = {475},
       number = {2},
        pages = {519-537},
          doi = {10.1051/0004-6361:20077221},
archivePrefix = {arXiv},
       eprint = {0707.1891},
 primaryClass = {astro-ph},
       adsurl = {https://ui.adsabs.harvard.edu/abs/2007A&A...475..519H}
}

@ARTICLE{2021A&A...654A..20G,
       author = {{Groenewegen}, M.~A.~T.},
        title = "{The parallax zero-point offset from Gaia EDR3 data}",
      journal = {\aap},
         year = 2021,
        month = oct,
       volume = {654},
          eid = {A20},
        pages = {A20},
          doi = {10.1051/0004-6361/202140862},
archivePrefix = {arXiv},
       eprint = {2106.08128},
 primaryClass = {astro-ph.GA},
       adsurl = {https://ui.adsabs.harvard.edu/abs/2021A&A...654A..20G}
}

@INPROCEEDINGS{Ricker,
       author = {{Ricker}, George R. and {Winn}, Joshua N. and {Vanderspek}, Roland and {Latham}, David W. and {Bakos}, G{\'a}sp{\'a}r. {\'A}. and {Bean}, Jacob L. and {Berta-Thompson}, Zachory K. and {Brown}, Timothy M. and {Buchhave}, Lars and {Butler}, Nathaniel R. and {Butler}, R. Paul and {Chaplin}, William J. and {Charbonneau}, David and {Christensen-Dalsgaard}, J{\o}rgen and {Clampin}, Mark and {Deming}, Drake and {Doty}, John and {De Lee}, Nathan and {Dressing}, Courtney and {Dunham}, E.~W. and {Endl}, Michael and {Fressin}, Francois and {Ge}, Jian and {Henning}, Thomas and {Holman}, Matthew J. and {Howard}, Andrew W. and {Ida}, Shigeru and {Jenkins}, Jon and {Jernigan}, Garrett and {Johnson}, John A. and {Kaltenegger}, Lisa and {Kawai}, Nobuyuki and {Kjeldsen}, Hans and {Laughlin}, Gregory and {Levine}, Alan M. and {Lin}, Douglas and {Lissauer}, Jack J. and {MacQueen}, Phillip and {Marcy}, Geoffrey and {McCullough}, P.~R. and {Morton}, Timothy D. and {Narita}, Norio and {Paegert}, Martin and {Palle}, Enric and {Pepe}, Francesco and {Pepper}, Joshua and {Quirrenbach}, Andreas and {Rinehart}, S.~A. and {Sasselov}, Dimitar and {Sato}, Bun'ei and {Seager}, Sara and {Sozzetti}, Alessandro and {Stassun}, Keivan G. and {Sullivan}, Peter and {Szentgyorgyi}, Andrew and {Torres}, Guillermo and {Udry}, Stephane and {Villasenor}, Joel},
        title = "{Transiting Exoplanet Survey Satellite (TESS)}",
    booktitle = {Space Telescopes and Instrumentation 2014: Optical, Infrared, and Millimeter Wave},
         year = 2014,
       editor = {{Oschmann}, Jacobus M., Jr. and {Clampin}, Mark and {Fazio}, Giovanni G. and {MacEwen}, Howard A.},
       series = {Society of Photo-Optical Instrumentation Engineers (SPIE) Conference Series},
       volume = {9143},
        month = aug,
          eid = {914320},
        pages = {914320},
          doi = {10.1117/12.2063489},
archivePrefix = {arXiv},
       eprint = {1406.0151},
 primaryClass = {astro-ph.EP},
       adsurl = {https://ui.adsabs.harvard.edu/abs/2014SPIE.9143E..20R}
}

@ARTICLE{1990A&AS...83..357B,
       author = {{Bessel}, M.~S.},
        title = "{BVRI photometry of the Gliese catalogue stars.}",
      journal = {\aaps},
         year = 1990,
        month = may,
       volume = {83},
        pages = {357-378},
       adsurl = {https://ui.adsabs.harvard.edu/abs/1990A&AS...83..357B}
}

@ARTICLE{2000A&A...355L..27H,
       author = {{H{\o}g}, E. and {Fabricius}, C. and {Makarov}, V.~V. and {Urban}, S. and {Corbin}, T. and {Wycoff}, G. and {Bastian}, U. and {Schwekendiek}, P. and {Wicenec}, A.},
        title = "{The Tycho-2 catalogue of the 2.5 million brightest stars}",
      journal = {\aap},
         year = 2000,
        month = mar,
       volume = {355},
        pages = {L27-L30},
       adsurl = {https://ui.adsabs.harvard.edu/abs/2000A&A...355L..27H}
}

@ARTICLE{2006AJ....131.1163S,
       author = {{Skrutskie}, M.~F. and {Cutri}, R.~M. and {Stiening}, R. and {Weinberg}, M.~D. and {Schneider}, S. and {Carpenter}, J.~M. and {Beichman}, C. and {Capps}, R. and {Chester}, T. and {Elias}, J. and {Huchra}, J. and {Liebert}, J. and {Lonsdale}, C. and {Monet}, D.~G. and {Price}, S. and {Seitzer}, P. and {Jarrett}, T. and {Kirkpatrick}, J.~D. and {Gizis}, J.~E. and {Howard}, E. and {Evans}, T. and {Fowler}, J. and {Fullmer}, L. and {Hurt}, R. and {Light}, R. and {Kopan}, E.~L. and {Marsh}, K.~A. and {McCallon}, H.~L. and {Tam}, R. and {Van Dyk}, S. and {Wheelock}, S.},
        title = "{The Two Micron All Sky Survey (2MASS)}",
      journal = {\aj},
         year = 2006,
        month = feb,
       volume = {131},
       number = {2},
        pages = {1163-1183},
          doi = {10.1086/498708},
       adsurl = {https://ui.adsabs.harvard.edu/abs/2006AJ....131.1163S}
}

@ARTICLE{1973MNRAS.163..291B,
       author = {{Boksenberg}, A. and {Evans}, R.~G. and {Fowler}, R.~G. and {Gardner}, I.~S.~K. and {Houziaux}, L. and {Humphries}, C.~M. and {Jamar}, C. and {Macau}, D. and {Malaise}, D. and {Monfils}, A. and {Nandy}, K. and {Thompson}, G.~I. and {Wilson}, R. and {Wroe}, H.},
        title = "{The ultra-violet sky-survey telescope in the TD-IA satellite.}",
      journal = {\mnras},
         year = 1973,
        month = jan,
       volume = {163},
        pages = {291},
          doi = {10.1093/mnras/163.3.291},
       adsurl = {https://ui.adsabs.harvard.edu/abs/1973MNRAS.163..291B}
}

@ARTICLE{2012MNRAS.426..903P,
       author = {{Page}, M.~J. and {Brindle}, C. and {Talavera}, A. and {Still}, M. and {Rosen}, S.~R. and {Yershov}, V.~N. and {Ziaeepour}, H. and {Mason}, K.~O. and {Cropper}, M.~S. and {Breeveld}, A.~A. and {Loiseau}, N. and {Mignani}, R. and {Smith}, A. and {Murdin}, P.},
        title = "{The XMM-Newton serendipitous ultraviolet source survey catalogue}",
      journal = {\mnras},
         year = 2012,
        month = oct,
       volume = {426},
       number = {2},
        pages = {903-926},
          doi = {10.1111/j.1365-2966.2012.21706.x},
archivePrefix = {arXiv},
       eprint = {1207.5182},
 primaryClass = {astro-ph.CO},
       adsurl = {https://ui.adsabs.harvard.edu/abs/2012MNRAS.426..903P}
}

@ARTICLE{2010A&A...514A...1I,
       author = {{Ishihara}, D. and {Onaka}, T. and {Kataza}, H. and {Salama}, A. and {Alfageme}, C. and {Cassatella}, A. and {Cox}, N. and {Garc{\'\i}a-Lario}, P. and {Stephenson}, C. and {Cohen}, M. and {Fujishiro}, N. and {Fujiwara}, H. and {Hasegawa}, S. and {Ita}, Y. and {Kim}, W. and {Matsuhara}, H. and {Murakami}, H. and {M{\"u}ller}, T.~G. and {Nakagawa}, T. and {Ohyama}, Y. and {Oyabu}, S. and {Pyo}, J. and {Sakon}, I. and {Shibai}, H. and {Takita}, S. and {Tanab{\'e}}, T. and {Uemizu}, K. and {Ueno}, M. and {Usui}, F. and {Wada}, T. and {Watarai}, H. and {Yamamura}, I. and {Yamauchi}, C.},
        title = "{The AKARI/IRC mid-infrared all-sky survey}",
      journal = {\aap},
         year = 2010,
        month = may,
       volume = {514},
          eid = {A1},
        pages = {A1},
          doi = {10.1051/0004-6361/200913811},
archivePrefix = {arXiv},
       eprint = {1003.0270},
 primaryClass = {astro-ph.IM},
       adsurl = {https://ui.adsabs.harvard.edu/abs/2010A&A...514A...1I}
}

@ARTICLE{2022A&A...661A..89P,
       author = {{Paunzen}, E.},
        title = "{Catalogue of stars measured in the Geneva seven-colour photometric system}",
      journal = {\aap},
         year = 2022,
        month = may,
       volume = {661},
          eid = {A89},
        pages = {A89},
          doi = {10.1051/0004-6361/202142355},
archivePrefix = {arXiv},
       eprint = {2111.04810},
 primaryClass = {astro-ph.SR},
       adsurl = {https://ui.adsabs.harvard.edu/abs/2022A&A...661A..89P}
}

@ARTICLE{2007A&A...474..653V,
       author = {{van Leeuwen}, F.},
        title = "{Validation of the new Hipparcos reduction}",
      journal = {\aap},
         year = 2007,
        month = nov,
       volume = {474},
       number = {2},
        pages = {653-664},
          doi = {10.1051/0004-6361:20078357},
archivePrefix = {arXiv},
       eprint = {0708.1752},
 primaryClass = {astro-ph},
       adsurl = {https://ui.adsabs.harvard.edu/abs/2007A&A...474..653V}
}

@article{teb2,
  author  = {{Maxted}, P.~F.~L. and others},
  year    = {2025},
  title = "{Fundamental effective temperature measurements for eclipsing binary stars – VI. Improved methodology and application to the circumbinary planet host star BEBOP-3}",
  journal = {\mnras,  submitted.}}

@ARTICLE{teb,
       author = {{Miller}, N.~J. and {Maxted}, P.~F.~L. and {Smalley}, B.},
        title = "{Fundamental effective temperature measurements for eclipsing binary stars - I. Development of the method and application to AI Phoenicis}",
      journal = {\mnras},
         year = 2020,
        month = sep,
       volume = {497},
       number = {3},
        pages = {2899-2909},
          doi = {10.1093/mnras/staa2167},
archivePrefix = {arXiv},
       eprint = {2004.04568},
 primaryClass = {astro-ph.SR},
       adsurl = {https://ui.adsabs.harvard.edu/abs/2020MNRAS.497.2899M}
}

@article{Metcalfe2020,
  author       = {Metcalfe, Travis S. and van Saders, Jennifer L. and Basu, Sarbani and Buzasi, Derek and Chaplin, William J. and Egeland, Ricky and Garc{\'\i}a, Rafael A. and Gaulme, Patrick and Huber, Daniel and Reinhold, Timo and et al.},
  title        = {The Evolution of Rotation and Magnetic Activity in 94 Aqr Aa from Asteroseismology with {TESS}},
  journal      = {arXiv e-prints},
  year         = {2020},
  eprint       = {2007.12755},
  archivePrefix= {arXiv},
  primaryClass = {astro-ph.SR}
}

@article{Santos2025,
  author       = {Santos, Angela R.~G. and Metcalfe, Travis S. and Kochukhov, Oleg and Ayres, Tom R. and Gafeira, Ricardo and Campante, Tiago L.},
  title        = {Magnetic braking and dynamo evolution of β Hydri},
  journal      = {Astronomy \& Astrophysics Letters},
  volume       = {698},
  pages        = {L23},
  year         = {2025},
  doi          = {10.1051/0004-6361/202554730}
}

@article{Robrade2023,
  author = {Robrade, J. and Stelzer, B. and Schmitt, J. H. M. M.},
  title = {Coronal X-ray emission of stars in the eROSITA all-sky survey},
  journal = {Astronomy \& Astrophysics},
  year = {2023},
  volume = {669},
  pages = {A155},
  doi = {10.1051/0004-6361/202245369}
}

@article{Sciortino2021,
  author = {Sciortino, S. and Micela, G. and Favata, F. and et al.},
  title = {Stellar activity and flares in the Athena era},
  journal = {Experimental Astronomy},
  year = {2021},
  volume = {51},
  pages = {811–831},
  doi = {10.1007/s10686-021-09725-4}
}

@article{Chara2,
  author = {Gallenne, A. and Mérand, A. and Kervella, P. and et al.},
  title = {Multiplicity of Galactic Cepheids from long-baseline interferometry. I. CHARA/MIRC detection of the companion of V1334 Cygni},
  journal = {Astronomy \& Astrophysics},
  year = {2016},
  volume = {586},
  pages = {A35},
  doi = {10.1051/0004-6361/201527146}
}

@article{VLTI,
  author = {Kervella, Pierre and Mérand, Antoine and Szabados, László and et al.},
  title = {Cepheid distances from infrared long-baseline interferometry. I. VINCI/VLTI observations of seven Galactic Cepheids},
  journal = {Astronomy \& Astrophysics},
  year = {2008},
  volume = {480},
  number = {1},
  pages = {167--179},
  doi = {10.1051/0004-6361:20078408}
}

@article{Chara,
  author = {Mourard, D. and Monnier, J. D. and Ligi, R. and et al.},
  title = {High angular resolution study of the binary star β Lyrae with the CHARA Array and VEGA},
  journal = {Advances in Space Research},
  year = {2015},
  volume = {56},
  number = {9},
  pages = {1764--1774},
  doi = {10.1016/j.asr.2015.07.034}
}

@ARTICLE{Grundahl2017,
       author = {{Grundahl}, F. and {Fredslund Andersen}, M. and {Christensen-Dalsgaard}, J. and {Antoci}, V. and {Kjeldsen}, H. and {Handberg}, R. and {Houdek}, G. and {Bedding}, T.~R. and {Pall{\'e}}, P.~L. and {Jessen-Hansen}, J. and {Silva Aguirre}, V. and {White}, T.~R. and {Frandsen}, S. and {Albrecht}, S. and {Andersen}, M.~I. and {Arentoft}, T. and {Brogaard}, K. and {Chaplin}, W.~J. and {Harps{\o}e}, K. and {J{\o}rgensen}, U.~G. and {Karovicova}, I. and {Karoff}, C. and {Kj{\ae}rgaard Rasmussen}, P. and {Lund}, M.~N. and {Sloth Lundkvist}, M. and {Skottfelt}, J. and {Norup S{\o}rensen}, A. and {Tronsgaard}, R. and {Weiss}, E.},
        title = "{First Results from the Hertzsprung SONG Telescope: Asteroseismology of the G5 Subgiant Star {\ensuremath{\mu}} Herculis}",
      journal = {\apj},
         year = 2017,
        month = feb,
       volume = {836},
       number = {1},
          eid = {142},
        pages = {142},
          doi = {10.3847/1538-4357/836/1/142},
archivePrefix = {arXiv},
       eprint = {1701.03365},
 primaryClass = {astro-ph.SR},
       adsurl = {https://ui.adsabs.harvard.edu/abs/2017ApJ...836..142G}
}

@ARTICLE{Rauer2025,
       author = {{Rauer}, Heike and {Aerts}, Conny and {Cabrera}, Juan and {Deleuil}, Magali and {Erikson}, Anders and {Gizon}, Laurent and {Goupil}, Mariejo and {Heras}, Ana and {Walloschek}, Thomas and {Lorenzo-Alvarez}, Jose and {Marliani}, Filippo and {Martin-Garcia}, C{\'e}sar and {Mas-Hesse}, J. Miguel and {O'Rourke}, Laurence and {Osborn}, Hugh and {Pagano}, Isabella and {Piotto}, Giampaolo and {Pollacco}, Don and {Ragazzoni}, Roberto and {Ramsay}, Gavin and {Udry}, St{\'e}phane and {Appourchaux}, Thierry and {Benz}, Willy and {Brandeker}, Alexis and {G{\"u}del}, Manuel and {Janot-Pacheco}, Eduardo and {Kabath}, Petr and {Kjeldsen}, Hans and {Min}, Michiel and {Santos}, Nuno and {Smith}, Alan and {Suarez}, Juan-Carlos and {Werner}, Stephanie C. and {Aboudan}, Alessio and {Abreu}, Manuel and {Acu{\~n}a}, Lorena and {Adams}, Moritz and {Adibekyan}, Vardan and {Affer}, Laura and {Agneray}, Fran{\c{c}}ois and {Agnor}, Craig and {Aguirre B{\o}rsen-Koch}, Victor and {Ahmed}, Saad and {Aigrain}, Suzanne and {Al-Bahlawan}, Ashraf and {Alcacera Gil}, Ma de los Angeles and {Alei}, Eleonora and {Alencar}, Silvia and {Alexander}, Richard and {Alfonso-Garz{\'o}n}, Julia and {Alibert}, Yann and {Allende Prieto}, Carlos and {Almeida}, Leonardo and {Alonso Sobrino}, Roi and {Altavilla}, Giuseppe and {Althaus}, Christian and {Alvarez Trujillo}, Luis Alonso and {Amarsi}, Anish and {Ammler-von Eiff}, Matthias and {Am{\^o}res}, Eduardo and {Andrade}, Laerte and {Antoniadis-Karnavas}, Alexandros and {Ant{\'o}nio}, Carlos and {Aparicio del Moral}, Beatriz and {Appolloni}, Matteo and {Arena}, Claudio and {Armstrong}, David and {Aroca Aliaga}, Jose and {Asplund}, Martin and {Audenaert}, Jeroen and {Auricchio}, Natalia and {Avelino}, Pedro and {Baeke}, Ann and {Bailli{\'e}}, Kevin and {Balado}, Ana and {Ballber Balaguer{\'o}}, Pau and {Balestra}, Andrea and {Ball}, Warrick and {Ballans}, Herve and {Ballot}, Jerome and {Barban}, Caroline and {Barbary}, Ga{\"e}le and {Barbieri}, Mauro and {Barcel{\'o} Forteza}, Sebasti{\`a} and {Barker}, Adrian and {Barklem}, Paul and {Barnes}, Sydney and {Barrado Navascues}, David and {Barragan}, Oscar and {Baruteau}, Cl{\'e}ment and {Basu}, Sarbani and {Baudin}, Frederic and {Baumeister}, Philipp and {Bayliss}, Daniel and {Bazot}, Michael and {Beck}, Paul G. and {Belkacem}, Kevin and {Bellinger}, Earl and {Benatti}, Serena and {Benomar}, Othman and {B{\'e}rard}, Diane and {Bergemann}, Maria and {Bergomi}, Maria and {Bernardo}, Pierre and {Biazzo}, Katia and {Bignamini}, Andrea and {Bigot}, Lionel and {Billot}, Nicolas and {Binet}, Martin and {Biondi}, David and {Biondi}, Federico and {Birch}, Aaron C. and {Bitsch}, Bertram and {Bluhm Ceballos}, Paz Victoria and {B{\'o}di}, Attila and {Bogn{\'a}r}, Zs{\'o}fia and {Boisse}, Isabelle and {Bolmont}, Emeline and {Bonanno}, Alfio and {Bonavita}, Mariangela and {Bonfanti}, Andrea and {Bonfils}, Xavier and {Bonito}, Rosaria and {Bonomo}, Aldo Stefano and {B{\"o}rner}, Anko and {Boro Saikia}, Sudeshna and {Borreguero Mart{\'\i}n}, Elisa and {Borsa}, Francesco and {Borsato}, Luca and {Bossini}, Diego and {Bouchy}, Francois and {Bou{\'e}}, Gwena{\"e}l and {Boufleur}, Rodrigo and {Boumier}, Patrick and {Bourrier}, Vincent and {Bowman}, Dominic M. and {Bozzo}, Enrico and {Bradley}, Louisa and {Bray}, John and {Bressan}, Alessandro and {Breton}, Sylvain and {Brienza}, Daniele and {Brito}, Ana and {Brogi}, Matteo and {Brown}, Beverly and {Brown}, David J.~A. and {Brun}, Allan Sacha and {Bruno}, Giovanni and {Bruns}, Michael and {Buchhave}, Lars A. and {Bugnet}, Lisa and {Buldgen}, Ga{\"e}l and {Burgess}, Patrick and {Busatta}, Andrea and {Busso}, Giorgia and {Buzasi}, Derek and {Caballero}, Jos{\'e} A. and {Cabral}, Alexandre and {Cabrero Gomez}, Juan-Francisco and {Calderone}, Flavia and {Cameron}, Robert and {Cameron}, Andrew and {Campante}, Tiago and {Campos Gestal}, N{\'e}stor and {Canto Martins}, Bruno Leonardo and {Cara}, Christophe and {Carone}, Ludmila and {Carrasco}, Josep Manel and {Casagrande}, Luca and {Casewell}, Sarah L. and {Cassisi}, Santi and {Castellani}, Marco and {Castro}, Matthieu and {Catala}, Claude and {Catal{\'a}n Fern{\'a}ndez}, Irene and {Catelan}, M{\'a}rcio and {Cegla}, Heather and {Cerruti}, Chiara and {Cessa}, Virginie and {Chadid}, Merieme and {Chaplin}, William and {Charpinet}, Stephane and {Chiappini}, Cristina and {Chiarucci}, Simone and {Chiavassa}, Andrea and {Chinellato}, Simonetta and {Chirulli}, Giovanni and {Christensen-Dalsgaard}, J{\o}rgen and {Church}, Ross and {Claret}, Antonio and {Clarke}, Cathie and {Claudi}, Riccardo and {Clermont}, Lionel and {Coelho}, Hugo and {Coelho}, Joao and {Cogato}, Fabrizio and {Colom{\'e}}, Josep and {Condamin}, Mathieu and {Conde Garc{\'\i}a}, Fernando and {Conseil}, Simon},
        title = "{The PLATO mission}",
      journal = {Experimental Astronomy},
         year = 2025,
        month = jun,
       volume = {59},
       number = {3},
          eid = {26},
        pages = {26},
          doi = {10.1007/s10686-025-09985-9},
archivePrefix = {arXiv},
       eprint = {2406.05447},
 primaryClass = {astro-ph.IM},
       adsurl = {https://ui.adsabs.harvard.edu/abs/2025ExA....59...26R}
}

@article{anguiano2020,
  author  = {Anguiano, Borja and Hayden, M. R. and Recio-Blanco, A. and others},
  title   = {Chemically versus Kinematically Defined Thin and Thick Discs in the Milky Way},
  journal = {Monthly Notices of the Royal Astronomical Society},
  year    = {2020},
  note    = {Defines velocity dispersions and rotational lag for chemically selected thick disk: σ₍R,ϕ,Z₎≈(62,45,41) km/s :contentReference[oaicite:2]{index=2}},
}

@article{vieira2022,
  author  = {Vieira, Katherine and Carraro, Giovanni and Korchagin, Vladimir and Lutsenko, Artem and Girard, Terrence M. and van Altena, William},
  title   = {Milky Way Thin and Thick Disk Kinematics with Gaia EDR3 and RAVE DR5},
  journal = {The Astrophysical Journal},
  year    = {2022},
  volume  = {932},
  pages   = {28},
  doi     = {10.3847/1538-4357/ac6b9b},
  note    = {Analyse kinematics of 278 k RGB stars, deriving velocity dispersions and mean rotations distinguishing thin vs thick disk :contentReference[oaicite:1]{index=1}},
}

@ARTICLE{Kospal2009,
       author = {{K{\'o}sp{\'a}l}, {\'A}gnes and {Ardila}, David R. and {Mo{\'o}r}, Attila and {{\'A}brah{\'a}m}, P{\'e}ter},
        title = "{On the Relationship Between Debris Disks and Planets}",
      journal = {\apjl},
         year = 2009,
        month = aug,
       volume = {700},
       number = {2},
        pages = {L73-L77},
          doi = {10.1088/0004-637X/700/2/L73},
archivePrefix = {arXiv},
       eprint = {0907.0028},
 primaryClass = {astro-ph.SR},
       adsurl = {https://ui.adsabs.harvard.edu/abs/2009ApJ...700L..73K}
}

@ARTICLE{Bryden2009,
       author = {{Bryden}, G. and {Beichman}, C.~A. and {Carpenter}, J.~M. and {Rieke}, G.~H. and {Stapelfeldt}, K.~R. and {Werner}, M.~W. and {Tanner}, A.~M. and {Lawler}, S.~M. and {Wyatt}, M.~C. and {Trilling}, D.~E. and {Su}, K.~Y.~L. and {Blaylock}, M. and {Stansberry}, J.~A.},
        title = "{Planets and Debris Disks: Results from a Spitzer/MIPS Search for Infrared Excess}",
      journal = {\apj},
         year = 2009,
        month = nov,
       volume = {705},
       number = {2},
        pages = {1226-1236},
          doi = {10.1088/0004-637X/705/2/1226},
       adsurl = {https://ui.adsabs.harvard.edu/abs/2009ApJ...705.1226B}
}

@ARTICLE{MoroMartin2007,
       author = {{Moro-Mart{\'\i}n}, Amaya and {Carpenter}, John M. and {Meyer}, Michael R. and {Hillenbrand}, Lynne A. and {Malhotra}, Renu and {Hollenbach}, David and {Najita}, Joan and {Henning}, Thomas and {Kim}, Jinyoung S. and {Bouwman}, Jeroen and {Silverstone}, Murray D. and {Hines}, Dean C. and {Wolf}, Sebastian and {Pascucci}, Ilaria and {Mamajek}, Eric E. and {Lunine}, Jonathan},
        title = "{Are Debris Disks and Massive Planets Correlated?}",
      journal = {\apj},
         year = 2007,
        month = apr,
       volume = {658},
       number = {2},
        pages = {1312-1321},
          doi = {10.1086/511746},
archivePrefix = {arXiv},
       eprint = {astro-ph/0612242},
 primaryClass = {astro-ph},
       adsurl = {https://ui.adsabs.harvard.edu/abs/2007ApJ...658.1312M}
}

@INPROCEEDINGS{Lebreton2014,
       author = {{Lebreton}, Y. and {Goupil}, M.~J. and {Montalb{\'a}n}, J.},
        title = "{How accurate are stellar ages based on stellar models?. II. The impact of asteroseismology}",
    booktitle = {EAS Publications Series},
         year = 2014,
       editor = {{Lebreton}, Y. and {Valls-Gabaud}, D. and {Charbonnel}, C.},
       series = {EAS Publications Series},
       volume = {65},
        month = nov,
        pages = {177-223},
          doi = {10.1051/eas/1465005},
archivePrefix = {arXiv},
       eprint = {1410.5337},
 primaryClass = {astro-ph.SR},
       adsurl = {https://ui.adsabs.harvard.edu/abs/2014EAS....65..177L}
}

@dataset{Soubiran2005,
       author = {{Soubiran}, C. and {Girard}, P.},
        title = "{VizieR Online Data Catalog: Abundances in Milky Way's disk (Soubiran+, 2005)}",
 howpublished = {VizieR On-line Data Catalog: J/A\&A/438/139. Originally published in: 2005A\&A...438..139S},
         year = 2005,
        month = apr,
          eid = {J/A&A/438/139},
          doi = {10.26093/cds/vizier.34380139},
       adsurl = {https://ui.adsabs.harvard.edu/abs/2005yCat..34380139S}
}

@ARTICLE{Pourbaix2004,
       author = {{Pourbaix}, D. and {Tokovinin}, A.~A. and {Batten}, A.~H. and {Fekel}, F.~C. and {Hartkopf}, W.~I. and {Levato}, H. and {Morrell}, N.~I. and {Torres}, G. and {Udry}, S.},
        title = "{S$_{B$^{9}$}$: The ninth catalogue of spectroscopic binary orbits}",
      journal = {\aap},
         year = 2004,
        month = sep,
       volume = {424},
        pages = {727-732},
          doi = {10.1051/0004-6361:20041213},
archivePrefix = {arXiv},
       eprint = {astro-ph/0406573},
 primaryClass = {astro-ph},
       adsurl = {https://ui.adsabs.harvard.edu/abs/2004A&A...424..727P}
}

@ARTICLE{2025RNAAS...9..146M,
       author = {{Maxted}, Pierre F.~L.},
        title = "{Equivalent Width of the Interstellar Na I D$_{1}$ and D$_{2}$ Absorption Lines for Stars with E(B ‑ V) < 0.15}",
      journal = {Research Notes of the American Astronomical Society},
         year = 2025,
        month = jun,
       volume = {9},
       number = {6},
          eid = {146},
        pages = {146},
          doi = {10.3847/2515-5172/ade39b},
       adsurl = {https://ui.adsabs.harvard.edu/abs/2025RNAAS...9..146M}
}

@PHDTHESIS{Donahue1993,
       author = {{Donahue}, Robert Andrew},
        title = "{Surface Differential Rotation in a Sample of Cool Dwarf Stars}",
       school = {New Mexico State University},
         year = 1993,
        month = jan,
       adsurl = {https://ui.adsabs.harvard.edu/abs/1993PhDT.........3D}
}

@ARTICLE{Donahue1996,
       author = {{Donahue}, Robert A. and {Saar}, Steven H. and {Baliunas}, Sallie L.},
        title = "{A Relationship between Mean Rotation Period in Lower Main-Sequence Stars and Its Observed Range}",
      journal = {\apj},
         year = 1996,
        month = jul,
       volume = {466},
        pages = {384},
          doi = {10.1086/177517},
       adsurl = {https://ui.adsabs.harvard.edu/abs/1996ApJ...466..384D}
}

@article{Corsaro14,
	adsurl = {http://adsabs.harvard.edu/abs/2014A%26A...571A..71C},
	archiveprefix = {arXiv},
	author = {{Corsaro}, E. and {De Ridder}, J.},
	doi = {10.1051/0004-6361/201424181},
	eid = {A71},
	eprint = {1408.2515},
	journal = {\aap},
	month = nov,
	pages = {A71},
	primaryclass = {astro-ph.IM},
	title = {{DIAMONDS: A new Bayesian nested sampling tool. Application to peak bagging of solar-like oscillations}},
	volume = 571,
	year = 2014
}

@ARTICLE{Favata2004,
       author = {{Favata}, F. and {Micela}, G. and {Baliunas}, S.~L. and {Schmitt}, J.~H.~M.~M. and {G{\"u}del}, M. and {Harnden}, Jr., F.~R. and {Sciortino}, S. and {Stern}, R.~A.},
        title = "{High-amplitude, long-term X-ray variability in the solar-type star HD 81809: The beginning of an X-ray activity cycle?}",
      journal = {\aap},
         year = 2004,
        month = apr,
       volume = {418},
        pages = {L13-L16},
          doi = {10.1051/0004-6361:20040105},
archivePrefix = {arXiv},
       eprint = {astro-ph/0403142},
 primaryClass = {astro-ph},
       adsurl = {https://ui.adsabs.harvard.edu/abs/2004A&A...418L..13F}
}

@ARTICLE{Orlando2017,
       author = {{Orlando}, S. and {Favata}, F. and {Micela}, G. and {Sciortino}, S. and {Maggio}, A. and {Schmitt}, J.~H.~M.~M. and {Robrade}, J. and {Mittag}, M.},
        title = "{Fifteen years in the high-energy life of the solar-type star HD 81809. XMM-Newton observations of a stellar activity cycle}",
      journal = {\aap},
         year = 2017,
        month = sep,
       volume = {605},
          eid = {A19},
        pages = {A19},
          doi = {10.1051/0004-6361/201731301},
archivePrefix = {arXiv},
       eprint = {1707.06437},
 primaryClass = {astro-ph.SR},
       adsurl = {https://ui.adsabs.harvard.edu/abs/2017A&A...605A..19O}
}

@ARTICLE{Tokovinin,
       author = {{Tokovinin}, Andrei and {Mason}, Brian D. and {Hartkopf}, William I. and {Mendez}, Rene A. and {Horch}, Elliott P.},
        title = "{Speckle Interferometry at SOAR in 2014}",
      journal = {\aj},
         year = 2015,
        month = aug,
       volume = {150},
       number = {2},
          eid = {50},
        pages = {50},
          doi = {10.1088/0004-6256/150/2/50},
archivePrefix = {arXiv},
       eprint = {1506.05718},
 primaryClass = {astro-ph.SR},
       adsurl = {https://ui.adsabs.harvard.edu/abs/2015AJ....150...50T}
}

@ARTICLE{Favata2008,
       author = {{Favata}, F. and {Micela}, G. and {Orlando}, S. and {Schmitt}, J.~H.~M.~M. and {Sciortino}, S. and {Hall}, J.},
        title = "{The X-ray cycle in the solar-type star HD 81809. XMM-Newton observations and implications for the coronal structure}",
      journal = {\aap},
         year = 2008,
        month = nov,
       volume = {490},
       number = {3},
        pages = {1121-1126},
          doi = {10.1051/0004-6361:200809694},
archivePrefix = {arXiv},
       eprint = {0806.2279},
 primaryClass = {astro-ph},
       adsurl = {https://ui.adsabs.harvard.edu/abs/2008A&A...490.1121F}
}

@ARTICLE{Fuhrmann2018,
       author = {{Fuhrmann}, Klaus and {Chini}, Rolf},
        title = "{Fossil Merger of a Population II Star}",
      journal = {\apj},
         year = 2018,
        month = may,
       volume = {858},
       number = {2},
          eid = {103},
        pages = {103},
          doi = {10.3847/1538-4357/aabaff},
       adsurl = {https://ui.adsabs.harvard.edu/abs/2018ApJ...858..103F}
}

@ARTICLE{Egeland2018,
       author = {{Egeland}, Ricky},
        title = "{Deconvolving the HD 81809 Binary: Rotational and Activity Evidence for a Subgiant with a Sun-like Cycle}",
      journal = {\apj},
         year = 2018,
        month = oct,
       volume = {866},
       number = {2},
          eid = {80},
        pages = {80},
          doi = {10.3847/1538-4357/aadf86},
archivePrefix = {arXiv},
       eprint = {1807.10870},
 primaryClass = {astro-ph.SR},
       adsurl = {https://ui.adsabs.harvard.edu/abs/2018ApJ...866...80E}
}

@ARTICLE{Vaughan1978,
       author = {{Vaughan}, A.~H. and {Preston}, G.~W. and {Wilson}, O.~C.},
        title = "{Flux measurements of Ca II and K emission.}",
      journal = {\pasp},
         year = 1978,
        month = jun,
       volume = {90},
        pages = {267-274},
          doi = {10.1086/130324},
       adsurl = {https://ui.adsabs.harvard.edu/abs/1978PASP...90..267V}
}

@ARTICLE{Wilson1978,
       author = {{Wilson}, O.~C.},
        title = "{Chromospheric variations in main-sequence stars.}",
      journal = {\apj},
         year = 1978,
        month = dec,
       volume = {226},
        pages = {379-396},
          doi = {10.1086/156618},
       adsurl = {https://ui.adsabs.harvard.edu/abs/1978ApJ...226..379W}
}

@ARTICLE{Wilson1968,
       author = {{Wilson}, O.~C.},
        title = "{Flux Measurements at the Centers of Stellar H- and K-Lines}",
      journal = {\apj},
         year = 1968,
        month = jul,
       volume = {153},
        pages = {221},
          doi = {10.1086/149652},
       adsurl = {https://ui.adsabs.harvard.edu/abs/1968ApJ...153..221W}
}

@ARTICLE{Hall1995,
       author = {{Hall}, Jeffrey C. and {Lockwood}, G.~W.},
        title = "{The Solar-Stellar Spectrograph: Project Description, Data Calibration, and Initial Results}",
      journal = {\apj},
         year = 1995,
        month = jan,
       volume = {438},
        pages = {404},
          doi = {10.1086/175084},
       adsurl = {https://ui.adsabs.harvard.edu/abs/1995ApJ...438..404H}
}

@phdthesis{Egeland2017_thesis,
    author = {{Egeland}, Ricky},
    title = {LONG-TERM VARIABILITY OF THE SUN IN THE CONTEXT OF
SOLAR-ANALOG STARS},
    school = {Montana State University},
    year = {2017}
}

@ARTICLE{Egeland2017,
       author = {{Egeland}, Ricky and {Soon}, Willie and {Baliunas}, Sallie and {Hall}, Jeffrey C. and {Pevtsov}, Alexei A. and {Bertello}, Luca},
        title = "{The Mount Wilson Observatory S-index of the Sun}",
      journal = {\apj},
         year = 2017,
        month = jan,
       volume = {835},
       number = {1},
          eid = {25},
        pages = {25},
          doi = {10.3847/1538-4357/835/1/25},
archivePrefix = {arXiv},
       eprint = {1611.04540},
 primaryClass = {astro-ph.SR},
       adsurl = {https://ui.adsabs.harvard.edu/abs/2017ApJ...835...25E}
}

@article{grevesse2011chemical,
  title={The chemical composition of the Sun},
  author={Grevesse, Nicolas and Asplund, Martin and Sauval, AJ and Scott, Patrick},
  journal={Canadian Journal of Physics},
  volume={89},
  number={4},
  pages={327--331},
  year={2011},
  publisher={NRC Research Press}
}

@article{kot2000,
	adsurl = {https://ui.adsabs.harvard.edu/abs/2000A\&A...358..587K},
	author = {{Kovtyukh}, V.~V. and {Gorlova}, N.~I.},
	journal = {\aap},
	month = jun,
	pages = {587--592},
	title = {{Precise temperatures of classical Cepheids and yellow supergiants from line-depth ratios}},
	volume = {358},
	year = 2000
}

@inproceedings{kur93,
	adsurl = {https://ui.adsabs.harvard.edu/abs/1993ASPC...44...87K},
	author = {{Kurucz}, R.~L.},
	booktitle = {{IAU Colloq. 138: Peculiar versus Normal Phenomena in A-type and Related Stars}},
	editor = {{Dworetsky}, M.~M. and {Castelli}, F. and {Faraggiana}, R.},
	month = jan,
	pages = {87},
	series = {{Astronomical Society of the Pacific Conference Series}},
	title = {{A New Opacity-Sampling Model Atmosphere Program for Arbitrary Abundances}},
	volume = {44},
	year = 1993
}

@article{kur81,
	adsurl = {https://ui.adsabs.harvard.edu/abs/1981SAOSR.391.....K},
	author = {{Kurucz}, Robert L. and {Avrett}, Eugene H.},
	journal = {SAO Special Report},
	month = may,
	title = {{Solar Spectrum Synthesis. I. A Sample Atlas from 224 to 300 nm}},
	volume = {391},
	year = 1981
}

@article{romaniello2008,
	adsurl = {https://ui.adsabs.harvard.edu/abs/2008A\&A...488..731R},
	archiveprefix = {arXiv},
	author = {{Romaniello}, M. and {Primas}, F. and {Mottini}, M. and {Pedicelli}, S. and {Lemasle}, B. and {Bono}, G. and {Fran{\c c}ois}, P. and {Groenewegen}, M.~A.~T. and {Laney}, C.~D.},
	doi = {10.1051/0004-6361:20065661},
	eprint = {0807.1196},
	journal = {\aap},
	month = sep,
	number = {2},
	pages = {731--747},
	primaryclass = {astro-ph},
	title = {{The influence of chemical composition on the properties of Cepheid stars. II. The iron content}},
	volume = {488},
	year = 2008
}

@ARTICLE{Reinhold2019,
       author = {{Reinhold}, Timo and {Bell}, Keaton J. and {Kuszlewicz}, James and {Hekker}, Saskia and {Shapiro}, Alexander I.},
        title = "{Transition from spot to faculae domination. An alternate explanation for the dearth of intermediate Kepler rotation periods}",
      journal = {\aap},
         year = 2019,
        month = jan,
       volume = {621},
          eid = {A21},
        pages = {A21},
          doi = {10.1051/0004-6361/201833754},
archivePrefix = {arXiv},
       eprint = {1810.11250},
 primaryClass = {astro-ph.SR},
       adsurl = {https://ui.adsabs.harvard.edu/abs/2019A&A...621A..21R}
}

@ARTICLE{Lomb1976,
       author = {{Lomb}, N.~R.},
        title = "{Least-Squares Frequency Analysis of Unequally Spaced Data}",
      journal = {\apss},
         year = 1976,
        month = feb,
       volume = {39},
       number = {2},
        pages = {447-462},
          doi = {10.1007/BF00648343},
       adsurl = {https://ui.adsabs.harvard.edu/abs/1976Ap&SS..39..447L}
}

@ARTICLE{Scargle1982,
       author = {{Scargle}, J.~D.},
        title = "{Studies in astronomical time series analysis. II. Statistical aspects of spectral analysis of unevenly spaced data.}",
      journal = {\apj},
         year = 1982,
        month = dec,
       volume = {263},
        pages = {835-853},
          doi = {10.1086/160554},
       adsurl = {https://ui.adsabs.harvard.edu/abs/1982ApJ...263..835S}
}

@ARTICLE{kaufer99,
       author = {{Kaufer}, A. and {Stahl}, O. and {Tubbesing}, S. and {N{\o}rregaard}, P. and {Avila}, G. and {Francois}, P. and {Pasquini}, L. and {Pizzella}, A.},
        title = "{Commissioning FEROS, the new high-resolution spectrograph at La-Silla.}",
      journal = {The Messenger},
         year = 1999,
        month = mar,
       volume = {95},
        pages = {8-12},
       adsurl = {https://ui.adsabs.harvard.edu/abs/1999Msngr..95....8K}
}

@ARTICLE{Lebreton2020,
       author = {{Lebreton}, Y. and {Reese}, D.~R.},
        title = "{SPInS, a pipeline for massive stellar parameter inference. A public Python tool to age-date, weigh, size up stars, and more}",
      journal = {\aap},
         year = 2020,
        month = oct,
       volume = {642},
          eid = {A88},
        pages = {A88},
          doi = {10.1051/0004-6361/202038602},
archivePrefix = {arXiv},
       eprint = {2009.00037},
 primaryClass = {astro-ph.SR},
       adsurl = {https://ui.adsabs.harvard.edu/abs/2020A&A...642A..88L}
}

@ARTICLE{Scuflaire2008,
       author = {{Scuflaire}, R. and {Th{\'e}ado}, S. and {Montalb{\'a}n}, J. and
         {Miglio}, A. and {Bourge}, P. -O. and {Godart}, M. and {Thoul}, A. and
         {Noels}, A.},
        title = "{CL{\'E}S, Code Li{\'e}geois d'{\'E}volution Stellaire}",
      journal = {\apss},
         year = 2008,
        month = aug,
       volume = {316},
       number = {1-4},
        pages = {83-91},
          doi = {10.1007/s10509-007-9650-1},
archivePrefix = {arXiv},
       eprint = {0712.3471},
 primaryClass = {astro-ph},
       adsurl = {https://ui.adsabs.harvard.edu/abs/2008Ap&SS.316...83S}
}

@ARTICLE{BailerJones2021,
       author = {{Bailer-Jones}, C.~A.~L. and {Rybizki}, J. and {Fouesneau}, M. and {Demleitner}, M. and {Andrae}, R.},
        title = "{Estimating Distances from Parallaxes. V. Geometric and Photogeometric Distances to 1.47 Billion Stars in Gaia Early Data Release 3}",
      journal = {\aj},
         year = 2021,
        month = mar,
       volume = {161},
       number = {3},
          eid = {147},
        pages = {147},
          doi = {10.3847/1538-3881/abd806},
archivePrefix = {arXiv},
       eprint = {2012.05220},
 primaryClass = {astro-ph.SR},
       adsurl = {https://ui.adsabs.harvard.edu/abs/2021AJ....161..147B}
}

@ARTICLE{Green2018,
       author = {{Green}, Gregory M. and {Schlafly}, Edward F. and {Finkbeiner}, Douglas and {Rix}, Hans-Walter and {Martin}, Nicolas and {Burgett}, William and {Draper}, Peter W. and {Flewelling}, Heather and {Hodapp}, Klaus and {Kaiser}, Nicholas and {Kudritzki}, Rolf-Peter and {Magnier}, Eugene A. and {Metcalfe}, Nigel and {Tonry}, John L. and {Wainscoat}, Richard and {Waters}, Christopher},
        title = "{Galactic reddening in 3D from stellar photometry - an improved map}",
      journal = {\mnras},
         year = 2018,
        month = jul,
       volume = {478},
       number = {1},
        pages = {651-666},
          doi = {10.1093/mnras/sty1008},
archivePrefix = {arXiv},
       eprint = {1801.03555},
 primaryClass = {astro-ph.GA},
       adsurl = {https://ui.adsabs.harvard.edu/abs/2018MNRAS.478..651G}
}

@ARTICLE{Paxton2013,
       author = {{Paxton}, Bill and {Cantiello}, Matteo and {Arras}, Phil and {Bildsten}, Lars and {Brown}, Edward F. and {Dotter}, Aaron and {Mankovich}, Christopher and {Montgomery}, M.~H. and {Stello}, Dennis and {Timmes}, F.~X. and {Townsend}, Richard},
        title = "{Modules for Experiments in Stellar Astrophysics (MESA): Planets, Oscillations, Rotation, and Massive Stars}",
      journal = {\apjs},
         year = 2013,
        month = sep,
       volume = {208},
       number = {1},
          eid = {4},
        pages = {4},
          doi = {10.1088/0067-0049/208/1/4},
archivePrefix = {arXiv},
}

@ARTICLE{Paxton2015,
       author = {{Paxton}, Bill and {Marchant}, Pablo and {Schwab}, Josiah and {Bauer}, Evan B. and {Bildsten}, Lars and {Cantiello}, Matteo and {Dessart}, Luc and {Farmer}, R. and {Hu}, H. and {Langer}, N. and {Townsend}, R.~H.~D. and {Townsley}, Dean M. and {Timmes}, F.~X.},
        title = "{Modules for Experiments in Stellar Astrophysics (MESA): Binaries, Pulsations, and Explosions}",
      journal = {\apjs},
         year = 2015,
        month = sep,
       volume = {220},
       number = {1},
          eid = {15},
        pages = {15},
          doi = {10.1088/0067-0049/220/1/15},
archivePrefix = {arXiv},
       eprint = {1506.03146},
 primaryClass = {astro-ph.SR},
       adsurl = {https://ui.adsabs.harvard.edu/abs/2015ApJS..220...15P}
}

@ARTICLE{Paxton2018,
       author = {{Paxton}, Bill and {Schwab}, Josiah and {Bauer}, Evan B. and {Bildsten}, Lars and {Blinnikov}, Sergei and {Duffell}, Paul and {Farmer}, R. and {Goldberg}, Jared A. and {Marchant}, Pablo and {Sorokina}, Elena and {Thoul}, Anne and {Townsend}, Richard H.~D. and {Timmes}, F.~X.},
        title = "{Modules for Experiments in Stellar Astrophysics (MESA): Convective Boundaries, Element Diffusion, and Massive Star Explosions}",
      journal = {\apjs},
         year = 2018,
        month = feb,
       volume = {234},
       number = {2},
          eid = {34},
        pages = {34},
          doi = {10.3847/1538-4365/aaa5a8},
archivePrefix = {arXiv},
       eprint = {1710.08424},
 primaryClass = {astro-ph.SR},
       adsurl = {https://ui.adsabs.harvard.edu/abs/2018ApJS..234...34P}
}

@ARTICLE{Paxton2019,
       author = {{Paxton}, Bill and {Smolec}, R. and {Schwab}, Josiah and {Gautschy}, A. and {Bildsten}, Lars and {Cantiello}, Matteo and {Dotter}, Aaron and {Farmer}, R. and {Goldberg}, Jared A. and {Jermyn}, Adam S. and {Kanbur}, S.~M. and {Marchant}, Pablo and {Thoul}, Anne and {Townsend}, Richard H.~D. and {Wolf}, William M. and {Zhang}, Michael and {Timmes}, F.~X.},
        title = "{Modules for Experiments in Stellar Astrophysics (MESA): Pulsating Variable Stars, Rotation, Convective Boundaries, and Energy Conservation}",
      journal = {\apjs},
         year = 2019,
        month = jul,
       volume = {243},
       number = {1},
          eid = {10},
        pages = {10},
          doi = {10.3847/1538-4365/ab2241},
archivePrefix = {arXiv},
       eprint = {1903.01426},
 primaryClass = {astro-ph.SR},
       adsurl = {https://ui.adsabs.harvard.edu/abs/2019ApJS..243...10P}
}

@ARTICLE{Moedas2025,
       author = {{Moedas}, Nuno and {Deal}, Morgan and {Bossini}, Diego},
        title = "{Impact of radiative accelerations on the stellar characterization of FGK-type stars using spectroscopic and seismic constraints}",
      journal = {\aap},
         year = 2025,
        month = mar,
       volume = {695},
          eid = {A9},
        pages = {A9},
          doi = {10.1051/0004-6361/202453130},
archivePrefix = {arXiv},
       eprint = {2502.05025},
 primaryClass = {astro-ph.SR},
       adsurl = {https://ui.adsabs.harvard.edu/abs/2025A&A...695A...9M}
}

@software{Irwin2012,
       author = {{Irwin}, Alan W.},
        title = "{FreeEOS: Equation of State for stellar interiors calculations}",
 howpublished = {Astrophysics Source Code Library, record ascl:1211.002},
         year = 2012,
        month = nov,
          eid = {ascl:1211.002},
       adsurl = {https://ui.adsabs.harvard.edu/abs/2012ascl.soft11002I}
}

@ARTICLE{Asplund2009,
       author = {{Asplund}, Martin and {Grevesse}, Nicolas and {Sauval}, A. Jacques and {Scott}, Pat},
        title = "{The Chemical Composition of the Sun}",
      journal = {\araa},
         year = 2009,
        month = sep,
       volume = {47},
       number = {1},
        pages = {481-522},
          doi = {10.1146/annurev.astro.46.060407.145222},
archivePrefix = {arXiv},
       eprint = {0909.0948},
 primaryClass = {astro-ph.SR},
       adsurl = {https://ui.adsabs.harvard.edu/abs/2009ARA&A..47..481A}
}

@ARTICLE{IglesiasRogers1996,
       author = {{Iglesias}, Carlos A. and {Rogers}, Forrest J.},
        title = "{Updated Opal Opacities}",
      journal = {\apj},
         year = 1996,
        month = jun,
       volume = {464},
        pages = {943},
          doi = {10.1086/177381},
       adsurl = {https://ui.adsabs.harvard.edu/abs/1996ApJ...464..943I}
}

@ARTICLE{Vernazza1981,
       author = {{Vernazza}, J.~E. and {Avrett}, E.~H. and {Loeser}, R.},
        title = "{Structure of the solar chromosphere. III. Models of the EUV brightness components of the quiet sun.}",
      journal = {\apjs},
         year = 1981,
        month = apr,
       volume = {45},
        pages = {635-725},
          doi = {10.1086/190731},
       adsurl = {https://ui.adsabs.harvard.edu/abs/1981ApJS...45..635V}
}

@ARTICLE{Paquette1986,
       author = {{Paquette}, C. and {Pelletier}, C. and {Fontaine}, G. and {Michaud}, G.},
        title = "{Diffusion Coefficients for Stellar Plasmas}",
      journal = {\apjs},
         year = 1986,
        month = may,
       volume = {61},
        pages = {177},
          doi = {10.1086/191111},
       adsurl = {https://ui.adsabs.harvard.edu/abs/1986ApJS...61..177P}
}

@ARTICLE{Thoul1994,
       author = {{Thoul}, Anne A. and {Bahcall}, John N. and {Loeb}, Abraham},
        title = "{Element Diffusion in the Solar Interior}",
      journal = {\apj},
         year = 1994,
        month = feb,
       volume = {421},
        pages = {828},
          doi = {10.1086/173695},
archivePrefix = {arXiv},
       eprint = {astro-ph/9304005},
 primaryClass = {astro-ph},
       adsurl = {https://ui.adsabs.harvard.edu/abs/1994ApJ...421..828T}
}

@ARTICLE{BohmVitense2007,
       author = {{B{\"o}hm-Vitense}, Erika},
        title = "{Chromospheric Activity in G and K Main-Sequence Stars, and What It Tells Us about Stellar Dynamos}",
      journal = {\apj},
         year = 2007,
        month = mar,
       volume = {657},
       number = {1},
        pages = {486-493},
          doi = {10.1086/510482},
       adsurl = {https://ui.adsabs.harvard.edu/abs/2007ApJ...657..486B}
}

@ARTICLE{Baliunas1995,
       author = {{Baliunas}, S.~L. and {Donahue}, R.~A. and {Soon}, W.~H. and {Horne}, J.~H. and {Frazer}, J. and {Woodard-Eklund}, L. and {Bradford}, M. and {Rao}, L.~M. and {Wilson}, O.~C. and {Zhang}, Q. and {Bennett}, W. and {Briggs}, J. and {Carroll}, S.~M. and {Duncan}, D.~K. and {Figueroa}, D. and {Lanning}, H.~H. and {Misch}, T. and {Mueller}, J. and {Noyes}, R.~W. and {Poppe}, D. and {Porter}, A.~C. and {Robinson}, C.~R. and {Russell}, J. and {Shelton}, J.~C. and {Soyumer}, T. and {Vaughan}, A.~H. and {Whitney}, J.~H.},
        title = "{Chromospheric Variations in Main-Sequence Stars. II.}",
      journal = {\apj},
         year = 1995,
        month = jan,
       volume = {438},
        pages = {269},
          doi = {10.1086/175072},
       adsurl = {https://ui.adsabs.harvard.edu/abs/1995ApJ...438..269B}
}

@ARTICLE{Metcalfe2017,
       author = {{Metcalfe}, Travis S. and {van Saders}, Jennifer},
        title = "{Magnetic Evolution and the Disappearance of Sun-Like Activity Cycles}",
      journal = {\solphys},
         year = 2017,
        month = sep,
       volume = {292},
       number = {9},
          eid = {126},
        pages = {126},
          doi = {10.1007/s11207-017-1157-5},
archivePrefix = {arXiv},
       eprint = {1705.09668},
 primaryClass = {astro-ph.SR},
       adsurl = {https://ui.adsabs.harvard.edu/abs/2017SoPh..292..126M}
}

@ARTICLE{VanSaders2016,
       author = {{van Saders}, Jennifer L. and {Ceillier}, Tugdual and {Metcalfe}, Travis S. and {Silva Aguirre}, Victor and {Pinsonneault}, Marc H. and {Garc{\'\i}a}, Rafael A. and {Mathur}, Savita and {Davies}, Guy R.},
        title = "{Weakened magnetic braking as the origin of anomalously rapid rotation in old field stars}",
      journal = {\nat},
         year = 2016,
        month = jan,
       volume = {529},
       number = {7585},
        pages = {181-184},
          doi = {10.1038/nature16168},
archivePrefix = {arXiv},
       eprint = {1601.02631},
 primaryClass = {astro-ph.SR},
       adsurl = {https://ui.adsabs.harvard.edu/abs/2016Natur.529..181V}
}

@ARTICLE{Pezzotti2025,
       author = {{Pezzotti}, C. and {Buldgen}, G. and {Magaudda}, E. and {Farnir}, M. and {Van Grootel}, V. and {Bellotti}, S. and {Poppenhaeger}, K.},
        title = "{Planetary inward migration as the potential cause of GJ 504's fast rotation and bright X-ray luminosity: New constraints from eROSITA}",
      journal = {\aap},
         year = 2025,
        month = feb,
       volume = {694},
          eid = {A179},
        pages = {A179},
          doi = {10.1051/0004-6361/202452580},
archivePrefix = {arXiv},
       eprint = {2501.07402},
 primaryClass = {astro-ph.EP},
       adsurl = {https://ui.adsabs.harvard.edu/abs/2025A&A...694A.179P}
}

@ARTICLE{Matt2015,
       author = {{Matt}, Sean P. and {Brun}, A. Sacha and {Baraffe}, Isabelle and
         {Bouvier}, J{\'e}r{\^o}me and {Chabrier}, Gilles},
        title = "{The Mass-dependence of Angular Momentum Evolution in Sun-like Stars}",
      journal = {\apjl},
         year = 2015,
        month = jan,
       volume = {799},
       number = {2},
          eid = {L23},
        pages = {L23},
          doi = {10.1088/2041-8205/799/2/L23},
archivePrefix = {arXiv},
       eprint = {1412.4786},
 primaryClass = {astro-ph.SR},
       adsurl = {https://ui.adsabs.harvard.edu/abs/2015ApJ...799L..23M}
}

@ARTICLE{Matt2019,
       author = {{Matt}, Sean P. and {Brun}, A. Sacha and {Baraffe}, Isabelle and
         {Bouvier}, J{\'e}r{\^o}me and {Chabrier}, Gilles},
        title = "{Erratum: {\textquotedblleft}The Mass-dependence of Angular Momentum Evolution in Sun-like Stars{\textquotedblright} (<A href=``http://doi.org/10.1088/2041-8205/799/2/l23''>2015, ApJL, 799, L23</A>)}",
      journal = {\apjl},
         year = 2019,
        month = jan,
       volume = {870},
       number = {2},
          eid = {L27},
        pages = {L27},
          doi = {10.3847/2041-8213/aafa1b},
       adsurl = {https://ui.adsabs.harvard.edu/abs/2019ApJ...870L..27M}
}

@ARTICLE{Johnstone2021,
       author = {{Johnstone}, C.~P. and {Bartel}, M. and {G{\"u}del}, M.},
        title = "{The active lives of stars: A complete description of the rotation and XUV evolution of F, G, K, and M dwarfs}",
      journal = {\aap},
         year = 2021,
        month = may,
       volume = {649},
          eid = {A96},
        pages = {A96},
          doi = {10.1051/0004-6361/202038407},
archivePrefix = {arXiv},
       eprint = {2009.07695},
 primaryClass = {astro-ph.SR},
       adsurl = {https://ui.adsabs.harvard.edu/abs/2021A&A...649A..96J}
}

@ARTICLE{Metcalfe2024,
       author = {{Metcalfe}, Travis S. and {Strassmeier}, Klaus G. and {Ilyin}, Ilya V. and {Buzasi}, Derek and {Kochukhov}, Oleg and {Ayres}, Thomas R. and {Basu}, Sarbani and {Chontos}, Ashley and {Finley}, Adam J. and {See}, Victor and {Stassun}, Keivan G. and {van Saders}, Jennifer L. and {Sepulveda}, Aldo G. and {Ricker}, George R.},
        title = "{Weakened Magnetic Braking in the Exoplanet Host Star 51 Peg}",
      journal = {\apjl},
         year = 2024,
        month = jan,
       volume = {960},
       number = {1},
          eid = {L6},
        pages = {L6},
          doi = {10.3847/2041-8213/ad0a95},
archivePrefix = {arXiv},
       eprint = {2401.01944},
 primaryClass = {astro-ph.EP},
       adsurl = {https://ui.adsabs.harvard.edu/abs/2024ApJ...960L...6M}
}

@ARTICLE{Pezzotti2021,
       author = {{Pezzotti}, C. and {Eggenberger}, P. and {Buldgen}, G. and {Meynet}, G. and {Bourrier}, V. and {Mordasini}, C.},
        title = "{Revisiting Kepler-444. II. Rotational, orbital, and high-energy fluxes evolution of the system}",
      journal = {\aap},
         year = 2021,
        month = jun,
       volume = {650},
          eid = {A108},
        pages = {A108},
          doi = {10.1051/0004-6361/202039652},
archivePrefix = {arXiv},
       eprint = {2104.06061},
 primaryClass = {astro-ph.EP},
       adsurl = {https://ui.adsabs.harvard.edu/abs/2021A&A...650A.108P}
}

@ARTICLE{Rasio1996,
       author = {{Rasio}, F.~A. and {Tout}, C.~A. and {Lubow}, S.~H. and {Livio}, M.},
        title = "{Tidal Decay of Close Planetary Orbits}",
      journal = {\apj},
         year = 1996,
        month = oct,
       volume = {470},
        pages = {1187},
          doi = {10.1086/177941},
archivePrefix = {arXiv},
       eprint = {astro-ph/9605059},
 primaryClass = {astro-ph},
       adsurl = {https://ui.adsabs.harvard.edu/abs/1996ApJ...470.1187R}
}

@ARTICLE{Villaver2009,
       author = {{Villaver}, Eva and {Livio}, Mario},
        title = "{The Orbital Evolution of Gas Giant Planets Around Giant Stars}",
      journal = {\apjl},
         year = 2009,
        month = nov,
       volume = {705},
       number = {1},
        pages = {L81-L85},
          doi = {10.1088/0004-637X/705/1/L81},
archivePrefix = {arXiv},
       eprint = {0910.2396},
 primaryClass = {astro-ph.EP},
       adsurl = {https://ui.adsabs.harvard.edu/abs/2009ApJ...705L..81V}
}

@ARTICLE{Rogers2002,
       author = {{Rogers}, F.~J. and {Nayfonov}, A.},
        title = "{Updated and Expanded OPAL Equation-of-State Tables: Implications for Helioseismology}",
      journal = {\apj},
         year = 2002,
        month = sep,
       volume = {576},
       number = {2},
        pages = {1064-1074},
          doi = {10.1086/341894},
       adsurl = {https://ui.adsabs.harvard.edu/abs/2002ApJ...576.1064R}
}

@article{Corsaro24,
	adsurl = {https://ui.adsabs.harvard.edu/abs/2024A&A...683A.161C},
	archiveprefix = {arXiv},
	author = {{Corsaro}, E. and {Bonanno}, A. and {Kayhan}, C. and {Di Mauro}, M.~P. and {Reda}, R. and {Giovannelli}, L.},
	doi = {10.1051/0004-6361/202348403},
	eid = {A161},
	eprint = {2401.16226},
	journal = {\aap},
	month = mar,
	pages = {A161},
	primaryclass = {astro-ph.SR},
	title = {{A new catalog of magnetically active solar-like oscillators}},
	volume = {683},
	year = 2024
}

@article{Corsaro20FAMED,
	adsurl = {https://ui.adsabs.harvard.edu/abs/2020A&A...640A.130C},
	archiveprefix = {arXiv},
	author = {{Corsaro}, E. and {McKeever}, J.~M. and {Kuszlewicz}, J.~S.},
	doi = {10.1051/0004-6361/202037930},
	eid = {A130},
	eprint = {2006.08245},
	journal = {\aap},
	month = aug,
	pages = {A130},
	primaryclass = {astro-ph.IM},
	title = {{Fast and Automated Peak Bagging with DIAMONDS (FAMED)}},
	volume = {640},
	year = 2020
}

@ARTICLE{2010AJ....139..743T,
       author = {{Tokovinin}, Andrei and {Mason}, Brian D. and {Hartkopf}, William I.},
        title = "{Speckle Interferometry at the Blanco and SOAR Telescopes in 2008 and 2009}",
      journal = {\aj},
         year = 2010,
        month = feb,
       volume = {139},
       number = {2},
        pages = {743-756},
          doi = {10.1088/0004-6256/139/2/743},
archivePrefix = {arXiv},
       eprint = {0911.5718},
 primaryClass = {astro-ph.SR},
       adsurl = {https://ui.adsabs.harvard.edu/abs/2010AJ....139..743T}
}

@ARTICLE{1987AJ.....93..688M,
       author = {{McAlister}, Harold A. and {Hartkopf}, William I. and {Hutter}, Donald J. and {Franz}, Otto G.},
        title = "{ICCD Speckle Observations of Binary Stars. II. Measurements During 1982 -1985 from the Kitt Peak 4 CM Telescope}",
      journal = {\aj},
         year = 1987,
        month = mar,
       volume = {93},
        pages = {688},
          doi = {10.1086/114353},
       adsurl = {https://ui.adsabs.harvard.edu/abs/1987AJ.....93..688M}
}

@ARTICLE{2015AJ....150...50T,
       author = {{Tokovinin}, Andrei and {Mason}, Brian D. and {Hartkopf}, William I. and {Mendez}, Rene A. and {Horch}, Elliott P.},
        title = "{Speckle Interferometry at SOAR in 2014}",
      journal = {\aj},
         year = 2015,
        month = aug,
       volume = {150},
       number = {2},
          eid = {50},
        pages = {50},
          doi = {10.1088/0004-6256/150/2/50},
archivePrefix = {arXiv},
       eprint = {1506.05718},
 primaryClass = {astro-ph.SR},
       adsurl = {https://ui.adsabs.harvard.edu/abs/2015AJ....150...50T}
}

@ARTICLE{1989AJ.....97..510M,
       author = {{McAlister}, Harold A. and {Hartkopf}, William I. and {Sowell}, James R. and {Dombrowski}, Edmund G. and {Franz}, Otto G.},
        title = "{ICCD Speckle Observations of Binary Stars. IV. Measurements During 1986-1988 from the Kitt Peak 4-m Telescope}",
      journal = {\aj},
         year = 1989,
        month = feb,
       volume = {97},
        pages = {510},
          doi = {10.1086/115001},
       adsurl = {https://ui.adsabs.harvard.edu/abs/1989AJ.....97..510M}
}

@ARTICLE{2020AJ....160....7T,
       author = {{Tokovinin}, Andrei and {Mason}, Brian D. and {Mendez}, Rene A. and {Costa}, Edgardo and {Horch}, Elliott P.},
        title = "{Speckle Interferometry at SOAR in 2019}",
      journal = {\aj},
         year = 2020,
        month = jul,
       volume = {160},
       number = {1},
          eid = {7},
        pages = {7},
          doi = {10.3847/1538-3881/ab91c1},
archivePrefix = {arXiv},
       eprint = {2005.05305},
 primaryClass = {astro-ph.SR},
       adsurl = {https://ui.adsabs.harvard.edu/abs/2020AJ....160....7T}
}

@ARTICLE{2012AJ....143...42H,
       author = {{Hartkopf}, William I. and {Tokovinin}, Andrei and {Mason}, Brian D.},
        title = "{Speckle Interferometry at SOAR in 2010 and 2011: Measures, Orbits, and Rectilinear Fits}",
      journal = {\aj},
         year = 2012,
        month = feb,
       volume = {143},
       number = {2},
          eid = {42},
        pages = {42},
          doi = {10.1088/0004-6256/143/2/42},
       adsurl = {https://ui.adsabs.harvard.edu/abs/2012AJ....143...42H}
}

@ARTICLE{1993AN....314..303L,
       author = {{Ling}, Josefina F. and {Lanchares}, V.},
        title = "{Micrometer measurements of visual double stars at Calar Alto}",
      journal = {Astronomische Nachrichten},
         year = 1993,
        month = jul,
       volume = {314},
       number = {4},
        pages = {303-305},
          doi = {10.1002/asna.2113140409},
       adsurl = {https://ui.adsabs.harvard.edu/abs/1993AN....314..303L}
}

@ARTICLE{1993AJ....106.1639M,
       author = {{McAlister}, Harold A. and {Mason}, Brian D. and {Hartkopf}, William I. and {Shara}, Michael M.},
        title = "{ICCD Speckle Observations of Binary Stars. X. A Further Survey for Duplicity Among the Bright Stars}",
      journal = {\aj},
         year = 1993,
        month = oct,
       volume = {106},
        pages = {1639},
          doi = {10.1086/116753},
       adsurl = {https://ui.adsabs.harvard.edu/abs/1993AJ....106.1639M}
}

@ARTICLE{1990AJ.....99..965M,
       author = {{McAlister}, Harold and {Hartkopf}, William I. and {Franz}, Otto G.},
        title = "{ICCD Speckle Observations of Binary Stars. V. Measurements During 1988-1989 from the Kitt Peak and the Cerro Tololo 4 M Telescopes}",
      journal = {\aj},
         year = 1990,
        month = mar,
       volume = {99},
        pages = {965},
          doi = {10.1086/115387},
       adsurl = {https://ui.adsabs.harvard.edu/abs/1990AJ.....99..965M}
}

@ARTICLE{2000AJ....119.3084H,
       author = {{Hartkopf}, William I. and {Mason}, Brian D. and {McAlister}, Harold A. and {Roberts}, Jr., Lewis C. and {Turner}, Nils H. and {ten Brummelaar}, Theo A. and {Prieto}, Cristina M. and {Ling}, Josefina F. and {Franz}, Otto G.},
        title = "{ICCD Speckle Observations of Binary Stars. XXIII. Measurements during 1982-1997 from Six Telescopes, with 14 New Orbits}",
      journal = {\aj},
         year = 2000,
        month = jun,
       volume = {119},
       number = {6},
        pages = {3084-3111},
          doi = {10.1086/301402},
       adsurl = {https://ui.adsabs.harvard.edu/abs/2000AJ....119.3084H}
}

@ARTICLE{1983ApJS...51..309M,
       author = {{McAlister}, H.~A. and {Hendry}, E.~M. and {Hartkopf}, W.~I. and {Campbell}, B.~G. and {Fekel}, F.~C.},
        title = "{Speckle interferometric measurements of binary stars. VIII.}",
      journal = {\apjs},
         year = 1983,
        month = mar,
       volume = {51},
        pages = {309-320},
          doi = {10.1086/190851},
       adsurl = {https://ui.adsabs.harvard.edu/abs/1983ApJS...51..309M}
}

@ARTICLE{1997AJ....114.1639H,
       author = {{Hartkopf}, William I. and {McAlister}, Harold A. and {Mason}, Brian D. and {ten Brummelaar}, Theo and {Roberts}, Jr., Lewis C. and {Turner}, Nils H. and {Wilson}, John W.},
        title = "{ICCD Speckle Observations of Binary Stars. XVII. Measurements During 1993-1995 From the Mount Wilson 2.5-M Telescope.}",
      journal = {\aj},
         year = 1997,
        month = oct,
       volume = {114},
        pages = {1639},
          doi = {10.1086/118594},
       adsurl = {https://ui.adsabs.harvard.edu/abs/1997AJ....114.1639H}
}

@ARTICLE{2001AJ....121.3224M,
       author = {{Mason}, Brian D. and {Hartkopf}, William I. and {Holdenried}, Ellis R. and {Rafferty}, Theodore J.},
        title = "{Speckle Interferometry of New and Problem Hipparcos Binaries. II. Observations Obtained in 1998-1999 from McDonald Observatory}",
      journal = {\aj},
         year = 2001,
        month = jun,
       volume = {121},
       number = {6},
        pages = {3224-3234},
          doi = {10.1086/321096},
       adsurl = {https://ui.adsabs.harvard.edu/abs/2001AJ....121.3224M}
}

@ARTICLE{1996AJ....111..936H,
       author = {{Hartkopf}, William I. and {Mason}, Brian D. and {McAlister}, Harold A. and {Turner}, Nils H. and {Barry}, Donald J. and {Franz}, Otto G. and {Prieto}, Christina M.},
        title = "{ICCD Speckle Observations of Binary Stars. XIII. Measurements During 1989- 1994 From the Cerro Tololo 4 M Telescope}",
      journal = {\aj},
         year = 1996,
        month = feb,
       volume = {111},
        pages = {936},
          doi = {10.1086/117841},
       adsurl = {https://ui.adsabs.harvard.edu/abs/1996AJ....111..936H}
}

@ARTICLE{1994AJ....108.2299H,
       author = {{Hartkopf}, William I. and {McAlister}, Harold A. and {Mason}, Brian D. and {Barry}, Donald J. and {Turner}, Nils H. and {Fu}, Hsieh-Hai},
        title = "{ICCD Speckle Observations of Binary Stars. XI. Measurements During 1991-1993 From the Kitt Peak 4m Telescope}",
      journal = {\aj},
         year = 1994,
        month = dec,
       volume = {108},
        pages = {2299},
          doi = {10.1086/117242},
       adsurl = {https://ui.adsabs.harvard.edu/abs/1994AJ....108.2299H}
}

@ARTICLE{1982ApJS...48..273M,
       author = {{McAlister}, H.~A. and {Hendry}, E.~M.},
        title = "{Speckle interferometric measurements of binary stars. VI.}",
      journal = {\apjs},
         year = 1982,
        month = mar,
       volume = {48},
        pages = {273-278},
          doi = {10.1086/190778},
       adsurl = {https://ui.adsabs.harvard.edu/abs/1982ApJS...48..273M}
}

@ARTICLE{1998ApJS..117..587H,
       author = {{Heintz}, W.~D.},
        title = "{Observations of Double Stars. XVIII.}",
      journal = {\apjs},
         year = 1998,
        month = jul,
       volume = {117},
       number = {2},
        pages = {587-598},
          doi = {10.1086/313127},
       adsurl = {https://ui.adsabs.harvard.edu/abs/1998ApJS..117..587H}
}

@ARTICLE{1984A+AS...57...31B,
       author = {{Balega}, Iu. and {Bonneau}, D. and {Foy}, R.},
        title = "{Speckle interferometric measurements of binary stars. II.}",
      journal = {\aaps},
         year = 1984,
        month = jul,
       volume = {57},
        pages = {31-36},
       adsurl = {https://ui.adsabs.harvard.edu/abs/1984A&AS...57...31B}
}

@ARTICLE{1990ApJS...74..275H,
       author = {{Heintz}, W.~D.},
        title = "{Observations of Double Stars and New Pairs. XIV.}",
      journal = {\apjs},
         year = 1990,
        month = sep,
       volume = {74},
        pages = {275},
          doi = {10.1086/191499},
       adsurl = {https://ui.adsabs.harvard.edu/abs/1990ApJS...74..275H}
}

@ARTICLE{1998A+AS..132..237A,
       author = {{Alzner}, A.},
        title = "{Measurements of double stars 1993.67 - 1998.13}",
      journal = {\aaps},
         year = 1998,
        month = oct,
       volume = {132},
        pages = {237-252},
          doi = {10.1051/aas:1998291},
       adsurl = {https://ui.adsabs.harvard.edu/abs/1998A&AS..132..237A}
}

@ARTICLE{1999AJ....118.1395D,
       author = {{Douglass}, Geoffrey G. and {Mason}, Brian D. and {Germain}, Marvin E. and {Worley}, Charles E.},
        title = "{Speckle Interferometry at the US Naval Observatory. IV.}",
      journal = {\aj},
         year = 1999,
        month = sep,
       volume = {118},
       number = {3},
        pages = {1395-1405},
          doi = {10.1086/301022},
       adsurl = {https://ui.adsabs.harvard.edu/abs/1999AJ....118.1395D}
}

@ARTICLE{1985A+AS...60..333B,
       author = {{Baize}, P.},
        title = "{Elements orbitaux de quinze etoiles doubles visuelles.}",
      journal = {\aaps},
         year = 1985,
        month = may,
       volume = {60},
        pages = {333-337},
       adsurl = {https://ui.adsabs.harvard.edu/abs/1985A&AS...60..333B}
}

@ARTICLE{1997A+AS..121..405P,
       author = {{Prieto}, C.},
        title = "{Micrometer measurements of southern double stars made at The Observatory of Llano del Hato at Merida (Venezuela)}",
      journal = {\aaps},
         year = 1997,
        month = mar,
       volume = {121},
        pages = {405-406},
          doi = {10.1051/aas:1997121},
       adsurl = {https://ui.adsabs.harvard.edu/abs/1997A&AS..121..405P}
}

@dataset{1997yCat.1239....0E,
       author = {{ESA}},
        title = "{VizieR Online Data Catalog: The Hipparcos and Tycho Catalogues (ESA 1997)}",
 howpublished = {VizieR On-line Data Catalog: I/239.  Originally published in: 1997HIP...C......0E},
         year = 1997,
        month = feb,
          eid = {I/239},
       adsurl = {https://ui.adsabs.harvard.edu/abs/1997yCat.1239....0E}
}

@ARTICLE{2024A&A...681A.107R,
       author = {{Royer}, P. and {Merle}, T. and {Dsilva}, K. and {Sekaran}, S. and {Van Winckel}, H. and {Fr{\'e}mat}, Y. and {Van der Swaelmen}, M. and {Gebruers}, S. and {Tkachenko}, A. and {Laverick}, M. and {Dirickx}, M. and {Raskin}, G. and {Hensberge}, H. and {Abdul-Masih}, M. and {Acke}, B. and {Alonso}, M.~L. and {Bandhu Mahato}, S. and {Beck}, P.~G. and {Behara}, N. and {Bloemen}, S. and {Buysschaert}, B. and {Cox}, N. and {Debosscher}, J. and {De Cat}, P. and {Degroote}, P. and {De Nutte}, R. and {De Smedt}, K. and {de Vries}, B. and {Dumortier}, L. and {Escorza}, A. and {Exter}, K. and {Goriely}, S. and {Gorlova}, N. and {Hillen}, M. and {Homan}, W. and {Jorissen}, A. and {Kamath}, D. and {Karjalainen}, M. and {Karjalainen}, R. and {Lampens}, P. and {Lobel}, A. and {Lombaert}, R. and {Marcos-Arenal}, P. and {Menu}, J. and {Merges}, F. and {Moravveji}, E. and {Nemeth}, P. and {Neyskens}, P. and {Ostensen}, R. and {P{\'a}pics}, P.~I. and {Perez}, J. and {Prins}, S. and {Royer}, S. and {Samadi-Ghadim}, A. and {Sana}, H. and {Sans Fuentes}, A. and {Scaringi}, S. and {Schmid}, V. and {Siess}, L. and {Siopis}, C. and {Smolders}, K. and {S{\'o}dor}, {\'A}. and {Thoul}, A. and {Triana}, S. and {Vandenbussche}, B. and {Van de Sande}, M. and {Van De Steene}, G. and {Van Eck}, S. and {van Hoof}, P.~A.~M. and {Van Marle}, A.~J. and {Van Reeth}, T. and {Vermeylen}, L. and {Volpi}, D. and {Vos}, J. and {Waelkens}, C.},
        title = "{MELCHIORS. The Mercator Library of High Resolution Stellar Spectroscopy}",
      journal = {\aap},
         year = 2024,
        month = jan,
       volume = {681},
          eid = {A107},
        pages = {A107},
          doi = {10.1051/0004-6361/202346847},
archivePrefix = {arXiv},
       eprint = {2311.02705},
 primaryClass = {astro-ph.SR},
       adsurl = {https://ui.adsabs.harvard.edu/abs/2024A&A...681A.107R}
}

@ARTICLE{2013PASP..125..306F,
       author = {{Foreman-Mackey}, Daniel and {Hogg}, David W. and {Lang}, Dustin and {Goodman}, Jonathan},
        title = "{emcee: The MCMC Hammer}",
      journal = {\pasp},
         year = 2013,
        month = mar,
       volume = {125},
       number = {925},
        pages = {306},
          doi = {10.1086/670067},
archivePrefix = {arXiv},
       eprint = {1202.3665},
 primaryClass = {astro-ph.IM},
       adsurl = {https://ui.adsabs.harvard.edu/abs/2013PASP..125..306F}
}

@ARTICLE{Gallet2015,
       author = {{Gallet}, F. and {Bouvier}, J.},
        title = "{Improved angular momentum evolution model for solar-like stars. II. Exploring the mass dependence}",
      journal = {\aap},
         year = 2015,
        month = may,
       volume = {577},
          eid = {A98},
        pages = {A98},
          doi = {10.1051/0004-6361/201525660},
archivePrefix = {arXiv},
       eprint = {1502.05801},
 primaryClass = {astro-ph.SR},
       adsurl = {https://ui.adsabs.harvard.edu/abs/2015A&A...577A..98G}
}
\bibliographystyle{aasjournal}

\end{document}